# Case Study Of GIPSY and MARF

Project Milestone 4

Ajay Kumar Thakur
Master of Software Engineering
Concordia University
Montreal, Canada
thakur.ajay777@gmail.com

Biswajit Banik
Masters of Software Engineering
Concordia University
Montreal, Canada
biswajitbanik59@gmail.com

Pankaj Kumar Pant
Master of Software Engineering
Concordia University
Montreal, Canada
Itp.pankaj@gmail.com

Dhanashree Sankini
Master of Software Engineering
Concordia University
Montreal, Canada
dhana.sankini@gmail.com

Dipesh Walia
Master of Software Engineering
Concordia University
Montreal, Canada
dipeshwalia2013@gmail.com

Renuka Milkoori
Master of Software Engineering
Concordia University
Montreal, Canada
renuka.milkoori@gmail.com

***Abstract- Metrics are used mainly to predict software engineering efforts such as maintenance effort, error Prone ness, and error rate. This document emphasis on experimental study based on two open source systems namely MARF and GIPSY. With the help of various research papers we were able to analyze and give priorities to various metrics that are implemented with JDeodrant. LOGISCOPE and McCabe tools are used to identify problematic classes with help of Kiviat graph and average Cyclomatic Complexity that further are implemented with highest priority metric with JDeodrant. To obtain accurate results we collected data using different tools. The analysis of the two systems is done as a conclusion of study using different tools.***

***Keywords: problematic class analysis MARF, GIPSY, LOGICSCOPE, JDeodoant, McCabe, MARFCAT, Software Measurement Analysis, QMOOD***

## I. INTRODUCTION

Software metric in general terms can be defined as a measure of a property of a software. To detect possible flaws in terms of efficiency of the code, we can assess if the quality of the project can be improved with respect to time. Two systems MARF and GIPSY were considered to perform the experimental study. Section A deals with proper research on both the systems. This was carried out to understand the systems better, skimming through different research papers. Different implementations, features of MARF and GIPSY were then merged. Section B compares the systems on the results obtained from the tools namely McCabe and LOGISCOPE. Firstly for each given OSS code case studies measurement data on Maintainability is collected. This is done basing on the software metrics like MOOD, QMOOD so on and so forth. The maintainability of the code in the case studies is then compared.

We then identify the weak classes for both the systems. A rank is assigned to the software metrics based on the poor and fair category reported in the class. Also in case of worst quality code (in each case study), classes, which are characterized as fair or poor are listed. The results were then visualized using the Kiviat Graph (provided by LOGISCOPE) for the two selected classes, comparing the quality of the two classes. There were few recommendations that were discussed among the team which are listed later in the same section. Secondly, results obtained using the McCabe tool are drafted. Metrics like Average Cyclomatic Complexity,Essential Complexity ,Module Design Complexity etc. are calculated .Basing on the results ,a comparative study of both the tools is drafted.

The further sections explain about the implementation of the metrics mentioned in the earlier section. We then match the metric results of both the classes with the results obtained using JDeodrant. The vulnerability of the classes are noted using the log files using another tool called MARFCAT. Analysis and interpretation of the measured data is done according to the results obtained using the tools.

## II. BACKGROUND

This document emphasis on different research papers based on two open source systems namely MARF and GIPSY. MARF is an open-source research platform which includes collection of voice/sound/speech/text and natural language processing (NLP) algorithms written in Java, General Intensional Programming System (GIPSY) is designed to support intensional programming languages built upon intensional logic and their imperative counter-parts for the intensional execution model.

The list of team members to the corresponding research paper read and summarized is mentioned in Table II (MARF) and in Table III (GIPSY) in appendix.

*A. OSS Case Studies*

*1) MARF*

Modular Audio Recognition Framework (MARF) is an open source collection of pattern recognition API's. In simple words MARF is designed in such a way that the generality and the default settings are taken care of. There were many implementations related to MARF including the applications used.

The figure below describes how the data flow and the transformation between different stages involved in MARF, Commonly known as MARF pipeline. MARF pipeline consist of four steps and group together with same kind of algorithms:

1. *Sample loading*
2. *Preprocessing,*
3. *Feature extraction and*
4. *Training/Classification.*

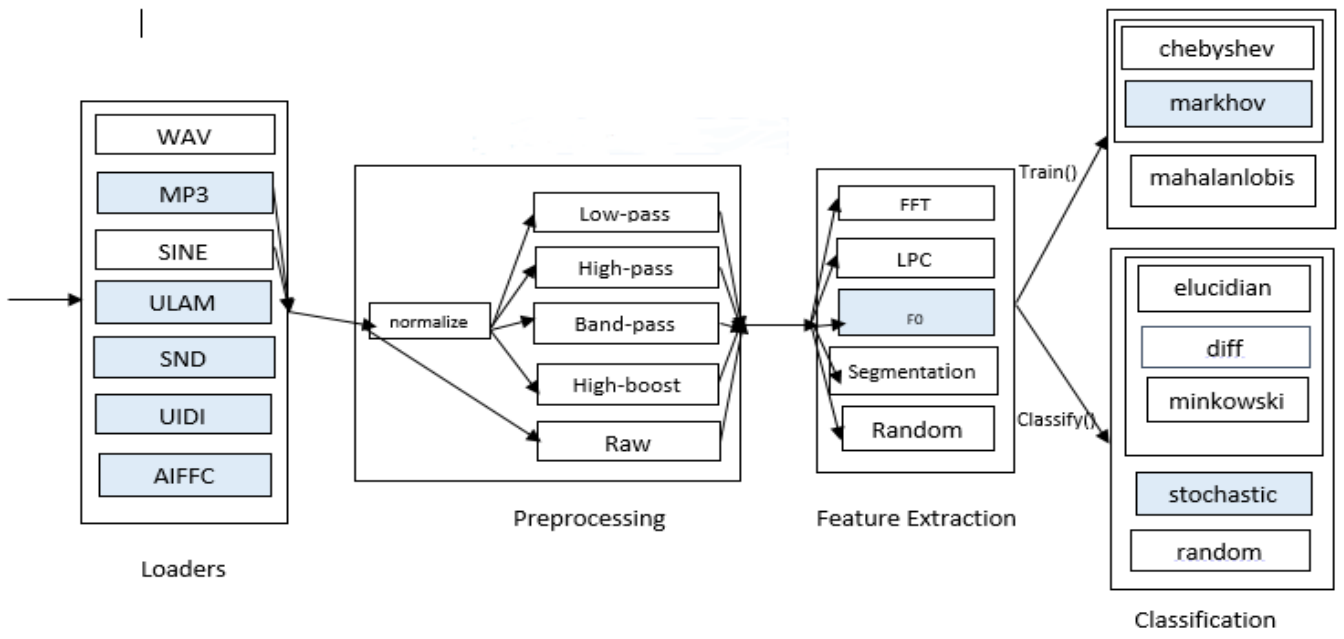

*Figure 1 MARF Architecture [2]*

Pipelines are capable to run as distributed nodes as well as a front-end. There is no backup is require for basic stage implementation and front end .It just require communication with JAVA [17], CORBA [18] and XML-RPC Web Services [19]. In order to test and employ MARF functionality there are various applications. The main object test on MARF [20] is textual, or imagery data are possible pattern-recognition, high-volume processing of recorded audio and phone conversations.

Different Algorithms are implemented on MARF and some are under process of implementation. Incomplete list of algorithms are:

- Fast Fourier transform (FFT) [23]
- Linear predictive coding (LPC)
- Various distance classifiers (Chebyshev, Euclidean, Minkowski [24], Mahalanobis [25], Diff, and Hamming [26]) (4) Cosine similarity measure [27]
- A number of math-related tools, for matrix and vector processing.
- Artificial neural network (classification).
- Zipf's Law-based classifier [25].
- General probability classifier.
- Continuous Fraction Expansion (CFE)-based filters [8].

The methodology used in MARF is the pipeline architecture that has been discussed earlier. This is done using its plugin architecture. MARF is used in many different applications having different goals.

In its initials days MARF was a single-threaded pipeline audio recognition system framework but now it turns into more general autonomic distributed system designed to work in heterogeneous environment which supports voice, speech, Natural Language, pattern recognition, machine learning, comparative study of multiple algorithms etc. MARF uses various algorithms which includes CYK natural language parsing, n-gram based language models, zipf's law, artificial neural networks, back propagation and various signal processing techniques. Speaker gender identification is one of the many application which uses MARF framework, MARF allows any permutation of algorithms listed in each stage that together comprise a configuration.

In general there are two phases

1. *Training (train())*
2. *Recognition, i.e. recognize()*

Train is used on some random data while recognize uses actual sample for identification process. Distributed and Autonomic MARF is used for the purpose of distributed systems, web services, security & forensics and sample Cryptolysis application [40, 41]. Comparing MARF with other tools like GATE is not fair as, MARF and its NLP framework is developed at later stage and not initially. Future development in MARF includes, use of API to get its pipeline and algorithmic implementations, also to develop swt plugin for eclipse [42] for MARF to make scripting more visual.

Firstly, it was used in a SpeakerIdentApp which is a text-independent speaker application derived from Modular Audio Recognition Framework (MARF)'s API. This application mainly focuses on speech processing tasks analyzing outcomes used by of median clusters rather than using default mean clusters. Prime focus of study is identification of speaker's as of whom they are, their gender, and accent by using mean and median clusters and then differentiate their results so selection of best algorithm can become clear. For speaker identification, median clusters are used instead of mean by replacing a method getMeanVector with getMedianVector in code base. Speaker identification application is useful in applications including safety, national security, recorded conversation forensic analysis, as well as conference assistance applications.

Secondly, MARF is also implemented in case of writer identification. This is done to identify the hand written copies "visually" by scanning the whole document as one. This is addressed as "inexpensive "as the CPU cycles are referred with respect to the preliminary identification techniques implemented. This particular experiment carried out proves that there is lot more MARF can do apart from just "audio".

Pipeline architecture is used in this case too but with few modifications. These modifications were done using the extensible plugins. Few of the uses include exams verification in case of fraud claims sorting out of hand-written mail consuming minimum time.

Let us consider DEFT 2010 as a system and analyze it using MARF. This challenge describes the two tracks in identification. Track 1 is to find decade of publication which can be reusable of publication and track2 is geographical location which can be partially tested. The statistics of each task is given back like train the system and test many algorithm combinations to know the best and the worst.

There are two pipelines approach. One is Classical MARF pipeline approach which has many algorithms selected at run-time and allows any combinations of algorithm. The statistical NLP processing is done by combining pipeline with NLP components. The most important things are statistical analysis [31], recognition using machine learning, language models and signal processing techniques. The other approach is NLP MARF pipeline approach consists set of options and parameters. And here the input stream is broken into characters and create n number of gram models. Language models are computed and tested for each part of article from training data. Variable tunable parameters are the multiple combinations of variable parameters used to know the best combination through implementations. The DEFT 2010 specific option is to select a combination and do the data processing. In Sample loading an interpretation have two types 1) each character is translated to a waveform and converted into audio wave signal with 2 bytes per amplitude value 2) or by using tokenization. In preprocessing the objects are filtered and normalized so that are silence and noise is removed by applying low-pass FFT filter.

In feature extraction module the preprocessed data is taken and produce feature length vector x and the result is stored in mean and median clusters [30] of training models. Then the fixed length vector is taken by classification module and then store them as a mean in default or in cluster and these clusters generally save processing and storage space. Different choice of cluster can provide different algorithm combination. Both pipelines have Zipf's Law [32] class and it's very slow in Classical pipeline but in NLP Zipf's Law should gather all discriminative terms from training to rank them. And here the distance and similarity is measured by measuring top N rank set and ranked Zipf's Law.

Using track 1 data in track 2 can improve the testing. The training set for each part of article does not increase by combining titles into texts for each part of article. Journals identify the place and which is not vice versa. Mostly location was considered as primary, But by assuming journal as primary class from which the location can be identified.

Let us have a look at making the properties of MARF making them automatic using ASSL which is a Distributed MARF Case Study. Pattern recognition techniques are widely used in computer science like image, sound and voice recognition have been derived from brain. The principle of autonomic computing [28] is used to solve specific problems of distributed pattern recognition like availability, performance, security etc. All this issues can be managed automatically with help of ASSL (Autonomic System Speciation Language) [21] and DMARF (Distributed Modular Audio Recognition Framework). First, ASSL framework is used to develop self-management features then it integrate into DMARF. These features enhance DMARF with an autonomic middleware that manages the four stages of the frameworks pattern recognition pipeline. Moreover, DMARF framework helps us in signal processing, pattern analysis and natural language processing.

The Autonomic System Specification Language (ASSL) [21] uses three major abstraction tiers to solve problem of formal specification and code generation of autonomic systems (ASs) within a framework. A first tier is AS -which describes the autonomic rules with help of service level objectives and self-management features, topology and global actions and metric applied in these rules. Second tier is AS Interaction Protocol (ASIP) tier –which composed with channel, communication functions and messages. It basically deals with communication between AEs. Third tier is Autonomic Elements +AE tier- in which set of individual AEs describe with their own behavior. This tier integrates with AE friends (a list of AEs forming a circle of trust), AE actions and events, AE rules (SLO and self-management features), an AE interaction protocol (AEIP) and AE metrics. For that it require two things-self managing behaviors intended to control the managed elements associated with an AE and communication interface. For self-managing behavior self-management policies are required. These policies are specified with ASSL constructs term fluent and mappings [21]. AS enter with fluent activating events and mapping connect with particular actions.

DMARF can be autonomic with help of atomicity to the DMARF behavior. A special manager is required to add in each DMARF stage it makes the DMARF to capable of self-management. Self-healing of ADMARF- ASSL is used by node replacement and node recovery services. Self-healing algorithm is spread over AS and AE tiers where metrics, actions and events and interface functions are used to achieve self-healing behavior of ADMARF. Self –protecting policy is also spread on AS and AE tiers. Moreover, two interaction protocols is used – a public (ASIP tier) and a private (AEIP tier), which are responsible for secure interaction used by both DMARF nodes and external entities to communicate. In order to achieve self ASSL self-optimization model for DMARF there are two major functional is required that is Training set classification data replication and Dynamic communication protocol selection. In self-optimizing algorithm a self-optimization behavior takes place any time when ADMARF enters in classification stage where it synchronize their latest cached results. Each stage node strives to adapt current available communication protocol before proceeding with the problem computation. In AS tier specification for self-optimization a system level policy, action and events are describe with special constructs called fluent and mapping.

Fluent –inClassificationStage is used and further mapped to an AS-level run-GlobalOptimization action. In AE tier specification for self-optimization- single node adapt to most efficient communication protocol.Fluent –inCPAdaptation is triggered to adapt when ADMARF enter in specification stage and fluent is mapped to an adaptCP action to perform the needed adaptation.

Therefore, algorithm is devised with ASSL for the pipelined stages of the DMARF's pattern recognition pipeline. Autonomic feature is described for self-protecting, and self-optimizing and self-managing policies for self-healing in ADMARF. It is notice that ADMARF system can function in autonomous environment those are on Internet, pattern recognition team can rely more on availability of system that execute on multiple days and robotic spacecraft can also use this system.

Lastly let us discuss about the security hardening of scientific distributed demand driven and pipelined computing systems using MARF. Here Distributed Modular Audio Recognition Framework (DMARF) describes scientific computational aspects and case studies with respect to the security aspects. Here there is threat on the nodes, the network that can drop, inject, alter data and can have malicious code injection. In this problem area the security risks those are related to incorrect computation results and cache. Those malware can spread using the system and attack as vectors. Thus it can be a threat for the system that contains confidential data. In the solution statement there is concrete distributive system, understanding the needs and requirements. The Java Data security Framework (JDSF), proxy certificates and other solutions can be useful.

DMARF is generally based on the classical MARF. The MARF is generally an open source research platform, pattern recognition, signal processing and NLP algorithm that is written in Java. MARF can be useful because it can run distributive over the network, run stand alone or can act as a library application. The main working principle in MARF is consists of pipeline stages that communicate with each other to get the data in a chaining approach.

The difference between demand driven and pipelined is that GIPSY generally useful for demand driven execution model. General Education Engine (GEE) plays an important role here. It helps to send procedural demand to a network demand store and from there it can be picked up by an observing worker or by any other nodes. After that the result is stored back to the warehouse to execute program again. DMARF is useful for process which implements a pipelined or chained way to connecting some of its distributive nodes. But DMARF does not by itself represented as a demand driven model of computation but it allows applications to connect with each other through nodes. For security issues the four main criteria includes confidentiality, integrity, authentication and availability

*2) GIPSY*

The GIPSY mainly has three modular sub systems like General Intentional programming Language Compiler (GIPC), the General Education Engine (GEE) and the Runtime Programming Environment (RIPE). All of its components are designed in a modular manner so that it can accept the eventual replacements of each of its components. It can handle replacement of components at both run time and compile time. Thus it improves the overall efficiency of the system.

General Intentional Programming Language Compiler (GIPC) – Similar like other programming language [33] [34] there are many versions of lucid which mainly depends on the set of types, constants, and data operations similar like basic algebra. It can also include conditional expressions, intentional navigation and intentional query.

The GIPSY architecture and program compilation- Gipsy programs are compiled in a two stage process. In the first stage the intentional part of the GIPSY program is translated into C. After translation into C, it is compiled in the standard way. The source code mainly consists of two parts. The lucid part generally defines the intentional data dependencies between variables and the sequential part defines the granular sequential computation units. The lucid part is compiled into an Intentional Data Dependency Structure (IDS). This part generally interpreted at run time by GEE. These are generated by the given communication layer definition such as IPC, COBRA or the WOS. This kind of modular design allows sequential threads those are written in different languages.

General Education Engine (GEE): The GIPSY mainly uses demand driven model of computation which takes place only if there is an explicit demand for it. For every demand there is a procedure call generates which is either computed locally or remotely. It mainly consists of two parts, Intentional Demand Propagator (IDP) and Intentional value Warehouse (IVW). The main task for IDP is to generate and propagate demands according to data dependency structure that are generated by GIPC. Intentional value Warehouse (IVW) is the second part of GEE, which acts like cache.

The GEE mainly uses the data flow context tags to build a store of values those are already computed. Runtime Interactive Programming Environment (RIPE) is mainly useful for visual runtime programming environment which display the data flow diagram according to the GIPSY program.

Autonomic GIPSY is augmenting the self-managing capability in General Intentional Programming System (GIPSY). The desired result of autonomous computing composes of goal-driven self-protection, self-healing, self-optimization and self-configuration [15] [16].

As an introduction, Autonomic computing (AC) is research led by IBM, whose focus is to make complex computing system smarter and easily manageable called Autonomic Systems (ASs). Moreover, each tier can be duplicated and distributed to assist increased volume and system redundancy. As GIPSY architecture permits high scalability but do not have self-management capacity. So to overcome this very requirement there is need to make GIPSY a self-adaptive and autonomous computing system.

GMT (GIPSY Manager Tier), instance of this tier are GIPSY managers (GMs), which registers GIPSY Nodes (GNs) to GIPSY instances and their allocations. GM provides interface to user to register GNs and GMs are peer to peer instances. GIPSY Node (GN) is computer that registered in GIPSY network as a host of one or more GIPSY tier instances. AGIPSY foundation is a model build with ASSL. The ASSL comprises three main tiers:

- *AS Tier: Specifies an AS (Autonomic System) in term of Service-level objectives (AS SLO), which is higher form of behavioral specification that focuses the objective of system, such as performance.*

- *AS Interaction Protocol: This specifies an AS-Level interaction Protocol (ASIP), a public communication interface express as message that exchanged between Autonomic Elements (AEs).*
- *AE Tier: AEs manages their own behavior and relationship with other AEs. AGIPSY Architecture and Behavior AGIPSY comprises autonomous GNs (GIPSY Nodes). Each GNs are autonomous and has control over its actions and state. Control provided by node Manager (NM). The GNs running a GMT instance are global autonomic managers, which monitor and manage the work of the entire system.*

Some of the autonomic features are also described as sub section stated as, GNs can recover from many types if failures. The self-maintenance can also be achieved and defined by self-configuration, self-optimization, self-healing and self-protection

GIPSY AE is GIPSY node and autonomic properties are achieved using Node Manager (NM). There are four distinct control-loop components- Monitor, Simulator, Decision Maker and Executor. In addition there are two controllers: channel controller and sensor controller. Channels are means of communication among NMs and channel Controller is responsible for sending and receiving messages over the channels. Sensors are used to measure parameter of GIPSY tier instance and Effectors are kind of “manageability interface” used by executor to control managed GIPSY tier instance. Sensor Controller is responsible for controlling the sensors

GIPSY AE trade-offs includes

- *Performance Trade-off: To monitor and control is performance overhead.*
- *Scalability-Complexity Trade-off: Scalability is also important issue while designing AGIPSY architecture.*

The GIPSY mainly has three modular sub systems like General Intentional programming Language Compiler (GIPC), the General Education Engine (GEE) and the Runtime Programming Environment (RIPE). All of its components are designed in a modular manner so that it can accept the eventual replacements of each of its components. It can handle replacement of components at both run time and compile time. Thus it improves the overall efficiency of the system.

General Intentional Programming Language Compiler (GIPC) is similar to other programming language that also includes conditional expressions, intentional navigation and intentional query. There is a Multi-Tier Architecture in the GIPSY Environment. Execution is the idea of generating, propagating and computing demands and their results is the main concept of demand driven computation. Different types of demands and their syntax.

- *Intensional demands: {GEERid, programId, context}*
- *Procedural demands: {GEERid, programId, Object params[], context, [code]}*
- *Resource demands: { resourceTypeId, resourceId}*
- *System demands: {destinationTierId, systemDemandTypeId, Object params[]}*

In multi-tier architecture of GIPSY, manual interaction can be done by three interfaces that is DWT, DGT and DST .It is also notice that GMT plays an important role in the network management and START/STOP mechanism of nodes. During the run time implementation, a command line UI is used for interaction with GMT, which the user manually bootstraps, and controls the nodes. Moreover, set of configuring files with appropriate settings and properties for each tier type are required. To start a network there are following set of steps that required performing: first, process of gipsy node should be created that start GMT tier. When a GMT is start then

GIPSY nodes automatically register and allocate the DST [22]. It is notice that registration of DST enables the GMT to receive demands from further node and tier allocations. In second step, user will create node locally on computer and register the node to existing GMT with the help of register command.

To evaluate heterogeneous programs for a distributed multi-tier demand driven GIPSY system provides a framework which contains various modules and has a lot of configurable component so it requires automation solution for configuring and managing GIPSY deployment components.

The system should provide an integrated tool that allows the user to: generate a GIPSY network and configure its components; able to save GIPSY network configuration; able to start and stop GIPSY and should be register with GMT; able to allocate/de-allocate with tiers; able to identify and manipulate their properties during run time; increase the overall usability of GIPSY system; visualization of nodes and tiers at run time; able to give means and semantics for scheduling, validation, and visual mapping to lucid programs.

In order to achieve all above aspects a graph-based graphical user interface is develop that enables the set of user interface in which user can directly use the distributed GIPSY during run time. The main objective of this solution is to increase the usability at run time and overall control of GIPSY network should be accessible to user with less manual details. A bootstrap to GIPSY network is required before to this work in which user manually execute shear number of commands and scripts. For graphical design -GMT component is used that allow the user to manage and operate the GIPSY network which translate simple graphical interaction with complex message passing between development components.

This solution help user to create, control and configure a GIPSY network through graph based interface. All connected graph nodes are described as GIPSY tiers, which read the properties from files and store the configured objects in graph nodes. User can identify new tier to network graph as color assigned to tier is associated to the node and tier is assigned to node.

Therefore, the presented tool is effective solution to assist management automation of GIPSY [29] artifacts that are distributed across multiple machines forming an overlay network. This solution is based upon graph programing and visualization to represent a GIPSY network. User can easily load information at run time and network can be create, configure and save to file. It is also notice that GIPSY network can easily bootstrapped and manageable. This solution allows the user to identify their status and properties of nodes and tiers of GIPSY at run time.

GIPSY framework helps in compilation and executions of programs written in intentional programming languages. GIPL is Generic intentional programming language which all family variants of intentional programming language use [Lucid 39, 38, 37, 35, 36]. GIPL uses Kripke possible world's semantics for his semantic markup, kripke's semantics works around the notion of context, To define context calculus as a first class atomic value is the language, which is yet not defined in GIPSY is the purpose of this paper, which is the purpose of this paper. To provide the solution, integrate the context calculus extension of GIPL syntax and semantics as first class atomic value.

Just like java GIPL is kind of intermediate language, uses to run non Neumann model based computation. GIPSY also helps you in filling gap between hybrid elements, it also check the elements in both run time and at compile time. GIPSYIdentifier for identifier & GIPSYFunction for operator and function checking.

Adding recursive function in GISPY semantics is in to do list for future.

#### *3) SUMMARY*

MARF is arranged into a modular and extensible framework facilitating addition of new algorithms, which is predominantly written in java. The main motive and design approach of MARF is to provide a tool to compare algorithms in a homogeneous environment to allow dynamic module selection.

GIPSY is a multi-language intensional programming environment that provides a distributed multi-tiered architecture to provide maximum scalability. The GIPSY is an ongoing effort for the development of a flexible and adaptable multi-lingual programming language development framework aimed at the investigation on the lucid family of intensional programming languages. The framework approach adopted is aimed at providing the possibility of easily developing compiler components for other languages of intensional nature and to execute them on a language-independent run-time system.

### *B. BASIC METRICS*

#### *1) Metrics definition*

Quantitative measure of some of the given properties of software system which further help the project team in obtaining various goal related to software quality assurance, code optimization, budget planning etc. Below is the some calculation related to direct Measure such as Line of Code (LOC), Number of files in given project, Number of classes etc.

#### *2) Methodologies Used*

*a)* Metrics 1.3.6 [Eclipse Puglin] : "Provide metrics calculation and dependency analyzer plugin for the Eclipse platform measure various metrics with average and standard deviation and detect cycles in package and type dependencies and graph them" [43].

Lines of Code: a) TLOC: Total lines of code counts non-blank and non-comment lines in a compilation unit. Useful if interested in computed KLOC [43]. b) MLOC: Method lines of code will count and sum up the non-blank and non-comment lines inside method bodies [43]

Number of Classes: Total number of classes in the selected scope [43]

*b)* SLOCCount: "A set of tools for counting physical Source Lines of Code (SLOC) in a large number of languages of a potentially large set of programs" [44]

*c)* Linux Commands: a) *T*o count files (even files without an extension) at the root of the current directory, use:

*ls -l | grep ^- | wc –l*

b) To count files (even files without an extension) recursively from the root of the current directory, use:

*ls -lR | grep ^- | wc –l*

*d)* ohloh.net (Ohcount): Ohcount is the source code line counter that powers Ohloh. [45]

*3) Metric Table (Result and Comparitive study of different tools)*

As the definition of LOC and related properties of software varies tool to tool, below is the comparative analysis of MARF and GIPSY properties with respect to the different tools used.

| PROPERTIES / TOOLS | Number of languages (MARF) | (GIPSY) | Lines of text (MARF) | (GIPSY) | Line of classes (MARF) | (GIPSY) | Number of files MARF | GIPSY |
|---|---|---|---|---|---|---|---|---|
| Metrics 1.3.6 | - | - | 24548 | 104073 | 199 | 540 | - | - |
| SLOCCount | JAVA (88.45%) SH (5.41%) PERL (4.67%) XML (1.47%) | JAVA (97.53%) ANSIC (1.62%) SH (0.45%) XML (0.36%) HASKELL (0.04%) | (SLOC) 27,751 | (SLOC) 100,973 | - | - | - | - |
| Linux Commands | - | - | - | - | - | - | 771 | 3510 |
| ohloh.net (Ohcount) | Java (72.0%) TeX/LaTeX ( 13.2%) XML (4.5%) Make (3.8%) Perl (3.0%) HTML (1.5%) shell script (1.1%) DOS batch script (0.4%) MetaFont (0.3%) CSS (0.2%) | - | 136,248 (Total Lines) 79,932 (Code Line) 37,160 (Comment Line) | - | - | - | - | - |

*Figure 2 comparative analyses of MARF and GIPSY properties*

All the results snapshot are included in APPENDIX section

*4) METRICS*

*a) Towards a simplied implementation of object- oriented design metrics*

The purpose of this research paper [46] is to introduce new language SAIL (a simple interpreted language) designed to write simple yet effective object oriented Metrics. Before moving on to SAIL author discussed about the two existing models namely structure based and repository based approaches, also he identified some key implementation process i.e. navigation, selection, set arithmetic & property aggregation for calculation of object oriented design metrics and their respective limitations.

Before moving on to limitations of the above mentioned models, let's see how these model works. Structure based approach uses data structures as for Meta model representation, author uses one of his existing research [47] and MEMORIA [48] as Meta to model and showed simplified model (only aggregation relationship is shown) as shown in figure 6.

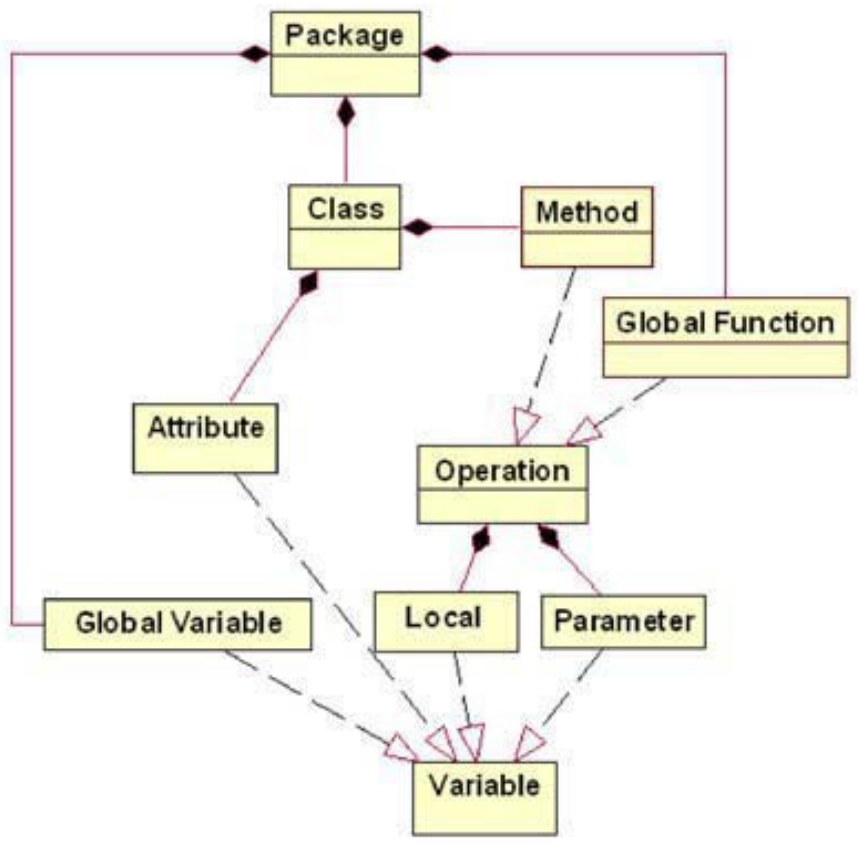

*Figure 3 Structure based approach [46]*

In Repository based approach [figure 7], Meta model can be compare with relational database system with uses foreign keys and queries for performing aforementioned properties. Below is the five key mechanism for implementation of object oriented design matrices and how they works in above mention systems.

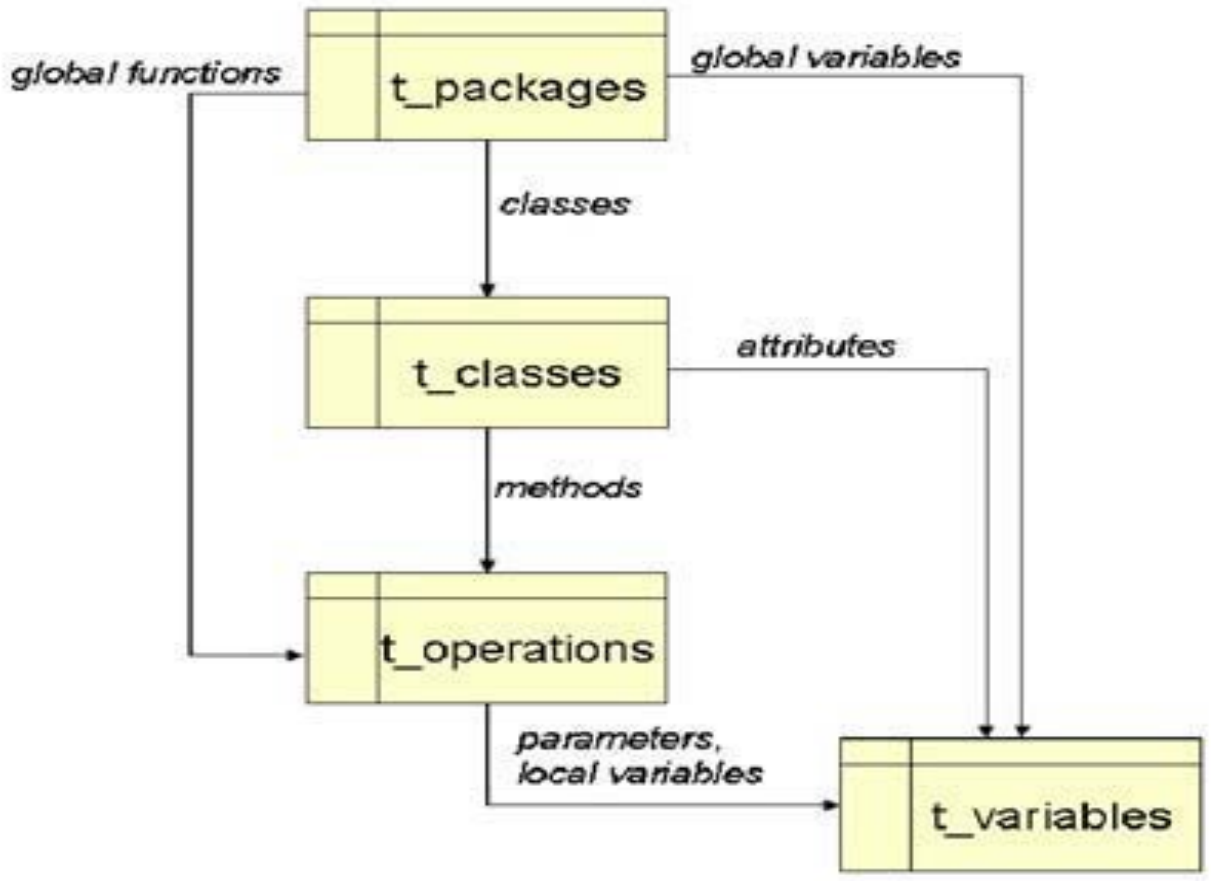

*Figure 4 Repository based approach [46]*

Group building and property computing are the two basic categories for any analysis, former is accumulation Meta models that are related with dedicated rule (includes navigation, selection, set arithmetic and filtering) while later is assigning new non elementary property to entity.

In structure based approach, for filtering

```
Package myPack;
Iterator it = sysPackages.iterator();
while(it.hasNext()){
myPack=(Package)it.next();
if(myPack.getName().equals("my.package"))
break;}
```

[46]

i.e. complexity will increase with respect to structure depthness and same is true for the navigation mechanism. For selection and arithmetic mechanism, Meta model required to calculate lot of function and hence more complex analysis.

In case of repository based approach there is the use of primary key and foreign keys as a part of SQL languages.

```
select * from t_packages
where f_name="my.package"
```
[46]

Below is the comparative analysis of repository based and structure based system in all 5 key elements

| | Filtering | Navigation | Selection | Set Arithmetics | Property Aggregation |
|---|---|---|---|---|---|
| Structure-Based Approach | ✗ | ✗ | ✗ | ~ | ~ |
| Repository-Based Approach | ✓ | ✗ | ✓ | ✗ | ✓ |

*Figure 5 Structured Based vs Repository based approach [46]*

Filtering, selection and Property aggregation is easy to calculate with repository based system. However for SAIL language, which use predefined set of data structure which may include elementary, structured type or collection. For filtering

```
Package myPack;
myPack = select (*) from sysPackages
where name="my.package";
```
[46]

Compare with repository based structure SAIL uses modulatory to lower its complexity. The use of SELECT statement made things very simpler and easy to understand, same applies to other key mechanism as well. Following is the brief comparison with respect to size.

| | Java | SQL | SAIL |
|---|---|---|---|
| LOC | 1650 | 1073 | 1461 |
| Size | 37580 | 48057 | 31012 |

*Figure 6 LOC and Size analysis [46]*

So does true in case of TCC and CM result metric.

| Size | Java | SQL | SAIL |
|---|---|---|---|
| TCC | 2016 | 1346 | 893 |
| CM | 3600 | 1152 | 933 |

*Figure 7 size in bytes TCC and CM[46]*

Lastly, there is brief summary of approaches used based on mentions models, for structural model Moose [49], smalltalk, etc but Repository based model are not limited to rational database and sql, audit c/c++ [50] is commercial tool for auditing use repository based approach.

Comparing SAIL (procedural language which uses an SQL like SELECT statement [46]) with Embedded SQL, OQL, GQOL. Embedded SQL improves modularity and also it does not allow loops and related decision controls. The major difference between SAIL and Embedded SQL like ( PL/SQL ) lies in data model and corresponding manipulation of data, in embedded SQL the analyzed code is saved in rational database, and thus for navigation there is need for complex select statements. In contrast SAIL uses data structures no burden for imperative statements and all. Comparing SAIL with OQL (an object-oriented query language), both uses SELECT statements, within navigation OQL cannot move using only 1 select statement to two level content. There is also have GOQL is like OQL query language used for querying graphs [51].

Future of sail involves, integration of query, representation of the data model in SAIL, simple manipulation of collections [46]

*b) A validation of object-oriented design metrics as quality indicators [52]*

Below is the some points about validation of different types of design metrics that can be used as a quality indicators. First of all these software metrics are needed to serve as different kinds of activities like giving information and guidelines to managers, scheduling and allocating activities or resources for software development process. Metrics [53] can also helpful to identify where the resources must be allocated and help to take decision. The main aim is to identify those modules which are fault prone and thus one can focus on those modules during testing or verification rather than the entire module. This action eventually saves cost and time of testing the whole module. There are different kinds of product metrics like number of lines of code, McCabe complexity metric etc.

In these days most of the companies use Object–oriented technology [54][55] to improve its monitoring, controlling, and improving the way of development and maintain software. For this reason metrics must be defined and validated for using in software industry. Metrics may be correct in terms of measurement theory or sometimes measure may not be satisfactory in terms of theoretical aspects but it is helpful to work in practice.

In this paper the study was based on to find that the proposed metrics can able to predict or detect the faulty classes. Acceptance testing was the method to find the faulty classes through these design metrics. The result was only 36 percent faults were detected and 84 percent classes contained less than three faults. These metrics are not language independent and thus there is need to change some of the Chidamber and Kemerer's metrics [56][57] so that they can suitable with specifications of C++. Those metrics are as follows-

Weighted Methods per Class (WMC) - it measures the individual class complexity.

Depth of Inheritance Tree of a class (DIT)- it defines the maximum depth of the inheritance graph of each class. The benefit with C++ is that it allows multiple inheritance.

Number of Children of a Class (N0C) - it measures the number of direct children of each class.

Coupling between Object classes (CB0) - it provides the number of classes to which a class is coupled.

Response For a Class (RFC) - it measures the number of responses that can occur response to a message received by an object of that class.

Lack of Cohesion on Methods (LC0M)- it is the number of member function without shared instance variable minus member function with instance variable. The result if comes to negative the metric is set to zero

For validating the above mentioned metrics their relationship [58] with fault occurrence must be validated too. To test each metrics hypothesis HWMC, HDIT, HNOC, HCBO, HRFC, HLCOM are used separately.

From the proposed example in this paper get this data charts that shows

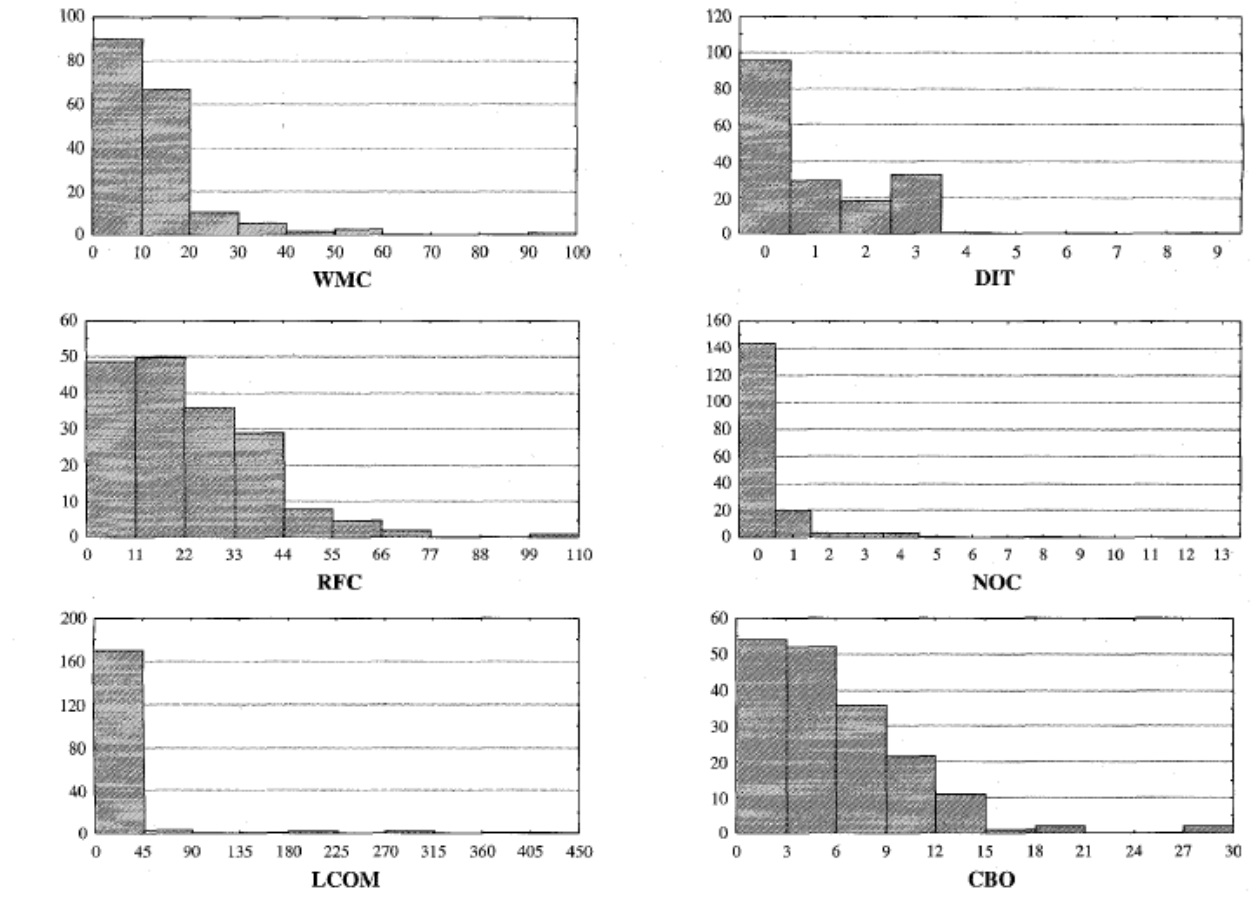

Fig. 1. Distribution of the analyzed OO metrics. The X axes represents the values of the metric. The Y axes represents the number of class.

*Figure 8[52]*

*c) A unified framework for coupling measurement in object-oriented systems. Empirical Software Engineering,*

This paper [59] to an increased understanding of the state-of-the-art: A mechanism is provided for comparing measures and their potential use, integrating existing measures which examine the same concepts in different ways, and facilitating more rigorous decision making regarding the definition of new measures and the selection of existing measures for a specific goal of measurement. In addition, our review of the state-of-the-art highlights that many measures are not defined in a fully operational form, and relatively few of them are based on explicit empirical models, as recommended by measurement theory.

Software measurement has evolved in past few years which has led to many research developing software. The Object Oriented development and its techniques are being used widely and coupling measurement has occurred in this branch of study.

According to this paper there is a bit of empirical study which led to these new measures. According to few researchers it is quite difficult to relate one measure to another including the part which says which measure is related to what application. Making it difficult for the practitioners to obtain a clear picture of the state-of-the-art in order to select or define measures for object-oriented systems. This is done using different activities. Firstly a standardized terminology is being used to express measures. This ensures that all the measures are consistent and operational. Secondly to obtain a structured synthesis a review is done on the existing frameworks in order to interpret the coupling measurement in object oriented systems.

Thirdly a Unified framework on the reviews that were obtained on existing frameworks is used to categorize existing measures accordingly.

*d) A Hierarchical Model for Object-Oriented Design Quality Assessment [60]*

A hierarchical model for object-oriented design has been developed and validate with help of various object-oriented design framework system because improve model of quality is always welcome as assessment of software quality is tedious job. It is notice that earlier models or frameworks are not able to evaluate the overall quality of software. A new model has been developing which has lower level design metrics that can easily describe with help of design characteristics and quality attributes.

The basic idea of methodology to develop hierarchical quality model for object-oriented design (QMOOD) has been taken from Dromey's generic quality model methodology that involves various step as show in Fig 12

It has four levels ($L_1$, $L_2$, $L_3$ and $L_4$) and three mapping links ($L_{12}$, $L_{23}$ ,and $L_{34}$).

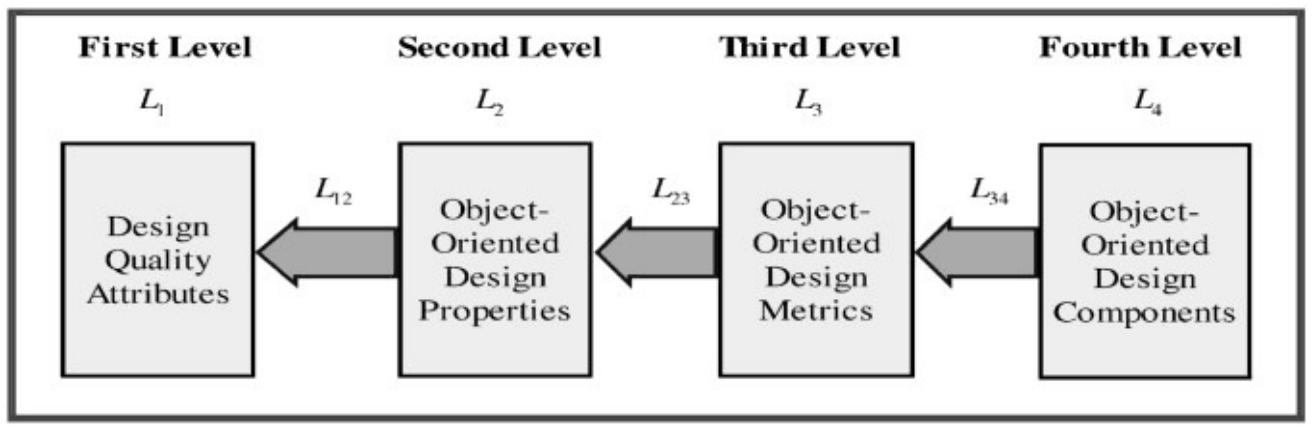

*Figure 9 Levels and links in QMOOD [60]..*

In Level ($L_1$) of model has identify various quality attributes that is functionality, effectiveness, understandability, extendibility, reusability and flexibility and it can changed according to different objectives and goal. In Level ($L_2$) of model has identified design properties where abstraction, encapsulation, coupling, cohesion, complexity, and design size are used for design quality characteristics in both structural as well as object oriented development. Whereas messaging, composition, inheritance, polymorphism, and class hierarchies plays an important role in quality of object oriented design. In Level ($L_3$) of model describe and identity various design metrics- DSC (Design Size in Classes), NOH (Number of Hierarchies), ANA (Average Number of Ancestor), DAM (Data Access Metric) , DCC (Direct Class Coupling), CAM (Cohesion Among Method of Class), MOA (Measure of Aggregation), MFA (Measure of Functional Abstraction), NOP (Number of Polymorphic Methods), CIS (Class Interface Size), NOM (Number of Methods).In Level ($L_4$) various design components are identify and define the architecture of an object-oriented design are objects, classes and relationship between objects and classes. It also identifies the quality carrying properties of component. In Fig1 -$L_{34}$ shows the mapping between quality components and design properties and with help of this Table 1 has been construct that describes the positive () and negative () influence on quality attributes where as $L_{23}$ assigning design metrics to design properties.

| | Reusability | Flexibility | Understandability | Functionality | Extendibility | Effectiveness |
|---|---|---|---|---|---|---|
| Design Size | ↑ | | | ↑ | | |
| Hierarchies | | | | ↑ | | |
| Abstraction | | | | | ↑ | ↑ |
| Encapsulation | | ↑ | ↑ | | | ↑ |
| Coupling | | | | | | |
| Cohesion | ↑ | | ↑ | ↑ | | |
| Composition | | ↑ | | | | ↑ |
| Inheritance | | | | | ↑ | ↑ |
| Polymorphism | | ↑ | | ↑ | ↑ | ↑ |
| Messaging | ↑ | | | ↑ | | |
| Complexity | | | | | | |

*Figure 10 Quality Attributes Design Property Relationships [60]*

In order to give weighting- $L_{12}$ is used between design properties to quality attributes. And with help of Table 1 attribute is weighted proportionally so that calculated values of all quality attributes have same range.

\

| Quality Attribute | Index Computation Equation |
|---|---|
| Reusability | -0.25 * Coupling + 0.25 * Cohesion + 0.5 * Messaging + 0.5 * Design Size |
| Flexibility | 0.25 * Encapsulation - 0.25 * Coupling + 0.5 * Composition + 0.5 * Polymorphism |
| Understandability | -0.33 * Abstraction + 0.33 * Encapsulation - 0.33 * Coupling + 0.33 * Cohesion - 0.33 * Polymorphism - 0.33 * Complexity - 0.33 * Design Size |
| Functionality | 0.12 * Cohesion + 0.22 * Polymorphism + 0.22 Messaging + 0.22 * Design Size + 0.22 * Hierarchies |
| Extendibility | 0.5 * Abstraction - 0.5 * Coupling + 0.5 * Inheritance + 0.5 * Polymorphism |
| Effectiveness | 0.2 * Abstraction + 0.2 * Encapsulation + 0.2 * Composition + 0.2 * Inheritance + 0.2 * Polymorphism |

*Figure 11 Computation Formulas for Quality Attributes[60]*

The validation of QMOOD model has done in two ways - Individual attribute validation and Overall quality. In order to validate all calculate values for each attribute a validation suite is been select and calculate expected result values of reusable, flexible, extendible, and effective frameworks with help of MFC (Microsoft Foundation Classes) and OWL (Objects Window library). Different version of MFC and OWL are used to compare and analyze quality individual attributes. On another hand, medium size C++ project has been taken to validate overall quality in which 13 independent evaluator is used to study [60] quality of project in validation suite. Moreover, [60] project also evaluated by QMOOD ++ tool. In order to rank project designs TQI (Total Quality index) is used and Spearman's rank correlation coefficient test is used for result comparison of evaluators.

Therefore, this model has ability to estimate overall design quality by using various functional equivalent projects where model has significant correlation between quality characteristics that determine by independent evaluators. The main attribute of this model is that it can be easily modified according to their relations and weights. It gives facility of practical quality assessment tool, which can be used in variety of demands. This gives an indication that model can effectively use in order to monitor the quality of software.

*e) Measurement of Cohesion and Coupling in OO Analysis Model based on Crosscutting Concerns [61]*

Software quality is an integral and vital part concerned while developing a software system. To achieve this there is a fundamental software engineering principle known as Separation of Concerns, both functional and nonfunctional, achieved through implementing software quality patterns Low Coupling and High Cohesion throughout whole software development cycle. The main concern of this paper is to measure to control coupling and cohesion of Object Oriented (OO) Analysis based on crosscutting concerns. Proposed cohesion measure is New but coupling measure is adopted from existing OO design. Although object-orientation approach in achieving separation of concern, both functional and nonfunctional, is effective but certain properties cannot be directly mapped from problem domain to solution domain, so they cannot be localized to single modular units. These properties are studied under crosscutting concerns (or aspects). Aspect-Oriented Programming (AOP) [65] is term used to describe technologies and approaches adopted that supports explicit capture of crosscutting concern and implementation of functional component is carried out separately. Aspect-Oriented Software Development (AOSD) extended AOP to provide support for separation, identification, representation and composition of crosscutting concerns with facility of mechanism to trace them throughout software development. This paper focuses on extending this AOSD framework by introducing measurement at analysis level to identify early crosscutting implication in system.

*Figure 12 Role of OO Analysis Model Measurement in AOSD [61]*

AOSD and Crosscutting Concerns

As per current approach, AOSD is used to handle non function requirements at early stage of development process. But proposed framework support capturing, analysis and design of both functional and nonfunctional requirements. Main target were as follows

- To eliminate gap between functional and nonfunctional requirements.
- To identify and resolve conflicts among crosscutting concerns.
- Smooth transition from requirement phase to analysis and design phase.

This paper extends work of [64] by augmenting measurement to analysis level to identify early crosscutting implications in system. The goal is to assist stakeholders to analyze system description (structural and behavioral) into highly cohesive, loosely coupled structure, so that makes it easy to model into modules. This analysis is done at first stage- requirement analysis, when behavior of system is elicited and modeled into different use cases and visible parts of software and relations with real-world objects structured in domain model. This is explained in following diagram.

Cohesion means degree to which task performed by single module are functionally related. A module is highly cohesive if elements of that particular module exhibits high degree of semantic relatedness. It also states that each element in module should be essential for that module to achieve its purpose. There has been different approach to measure cohesion, LCOM [62], Its AOP counterpart [67], OO cohesion measure [68]. The LCOM metric measure structural cohesion rather than semantic cohesion. Our goal is to make mechanism that deal with semantic relationship between elements of component and a single, overall abstraction (single, well defined purpose) application to analysis model.

Coupling is interdependency among components in a system that is responsible for its nature, also it extends the relationship between elements in the software system. Low coupling pattern increases prediction and controlling of scope of the system. Moreover complexity can be reduced with lowest coupling between classes, which in turn increases modularity and encapsulation. A good example of system level OO design measure of coupling is MOOD [63], inclusive of Coupling Factor (CF). Our aim is to obtain feedback on coupling level in analysis model (coupling inherent in problem domain).

Software requirement analysis is most critical and crucial activity in development process, its objective is to gather requirements (textual description) and model into OO Analysis Model. Analysis model is representation of specification in elements of OO analysis model, such as conceptual classes and relationships that forms domain model, Use case and their relations extend or include. The Unified Process (UP) captures analysis model as per boundary description, which is use case diagram, restricted class diagram (represent real world concept) and their associations and interaction diagram.

Both coupling and cohesion are way to measure quality of partitioning used in analysis model, which is described below

Cohesion

Each use case is set of scenarios, which is set of paths and conditions that are for interest for system analysis. Each scenario defines expected behavior of the system. It may contain sub goals to fulfill the requirements. If a use case is included in a certain use case and same use case is extended in another use case. Then that use case is considered to be crosscutting both the use cases. For that cohesion of use case (local level) and cohesion of use case model (global level) needed to measure.

Measurement Method

A scenario is defined as

$\Sigma = (SE, \angle SE, SO, MEO, MET)$

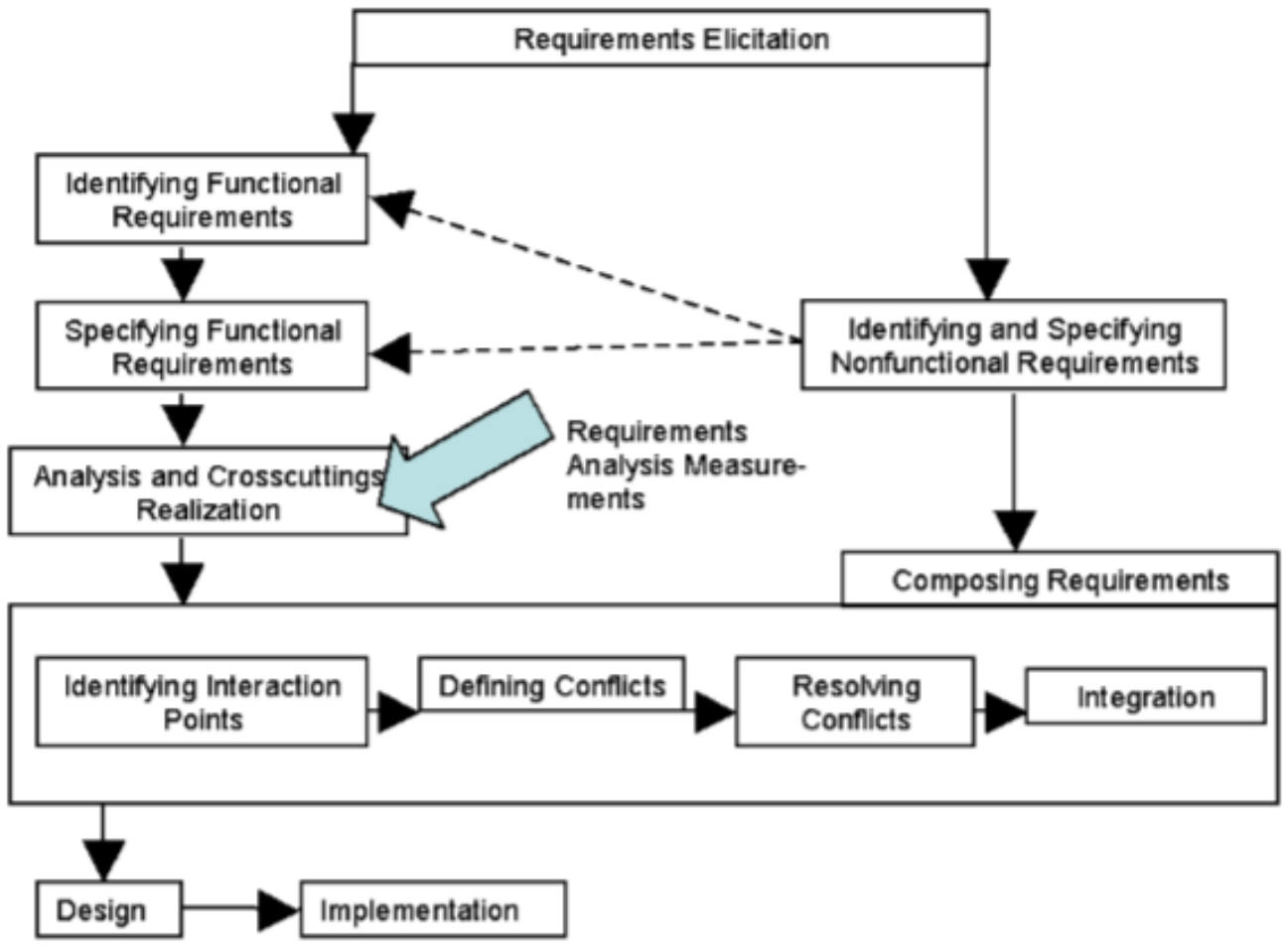

Where,

SE = all the environmental (Input/Output) events in the scenario

$\angle$SE = order imposed on the events in time

SO = set of domain concepts participating in the scenario

MEO = mapping from SE to the pairs of objects that exchange events

MET = mapping from SE to the time axis

Cohesion measurement in use case

As use case is set of scenarios. Similar scenario (operates on common exchanged messages) increases the cohesion level and those operating in disjoint messages lower the cohesion.

Therefore, cohesion level in use case is defined as

$CL_UC = |Q|/|P|$

Where,

Q = set of the similar pairs of scenarios of one use-case

P = set of all pairs of scenarios of the same use-case

The range of CL_UC is [0..1] with 1 with highest cohesion and 0 as lack of cohesion.

Cohesion measurement in use case model

Crosscutting corresponds to common to at least two scenarios belonging to different use cases. Let there be two use cases U1 and U2 presented by scenarios Σ1 and Σ2 respectively. Therefore cohesion level in use case model is defined as

CL_UCM = 1-|QM|/|PM
Where,
QM = set of the pairs of similar scenarios belonging to different use cases
PM = set of all pairs of scenarios (same condition apply)
The range of CL_UCM is [0..1] with 1 with highest cohesion (no common pairs) and 0 as lack of cohesion (all pairs are crosscutted)

The coupling is adopted from MOOD coupling factor [4] to specify the level of coupling in domain model due to association between classes. Coupling Factor (CF) is defined as
CF = ∑TCi=1 ∑TCj=1 client (Ci, Cj) / (TC2 – TC)
Where,

1, iff Cc=> Cs ^ Cc ≠ Cs client (Ci, Cj) =
0, otherwise

Cc=>Cs represent relationship between client class (Cc) and supplier class (Cs). The range of this value is [0..1], where 0 is lack of coupling and 1 is highest level of coupling. The case study in concern is taken as Web-based invoicing system, system can receive multiple orders / cancellation at same time, moreover multiple teller can access system for order process to change status of product from "Pending" to "Invoiced", if quantity available else order wait in queue.

Let's take two use cases "Place Order" and "View Order" with alternative scenarios (authentication failure, wrong product number, etc.) and by applying CL_UC results that "Place Order" use case as highly cohesive, while applying CL_UCM in these two use cases, results found that there is crosscutting in these two use cases. So the partial domain model is

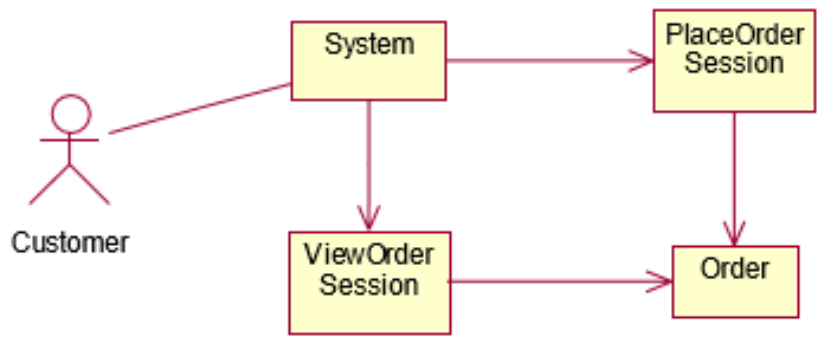

*Figure 13 Partial Domain Model for PlaceOrder and ViewOrder Use Cases [61]*

This paper specifically focuses on measurement to control coupling and cohesion of OO Analysis Model based on crosscutting concern. This is basically extension of the work in [65] to identify early crosscutting implications in a software system.

*f) An evaluation of the MOOD set of objectoriented software metrics*

This paper [66] explains about the MOOD metrics which has six metrics for object oriented design. Based on the object oriented quality properties like encapsulation, inheritance, coupling and polymorphism the six metrics is measured from a measurement theory viewpoint [68], [69], [70], [71]. Then by taking three application domains empirical data is collected and analyzed.

The strength of this investigation centers on the consideration of a number of criteria [67] for valid metrics are to measure an attribute, other entities which are distinguished from one another should follow, and the direct metrics include: A valid metric must obey the representation condition [66]; that is, it must preserve all intuitive notions about the attribute and the way in which the metric distinguishes between entities. Each unit of an attribute contributing to a valid metric is equivalent. Different entities can have the same attribute value.

There is difference between direct measurement and indirect measurement. In direct measurement an attribute does not depend upon another attribute and in indirect measurement other attributes are used. There are internal and external attributes of a product or process as well. Generally managers focus on external attributes for reliability and maintainability. OO metrics are based on internal which is related external attributes. According to the framework of Kitchenham et al. [67], indirect metrics should exhibit the following properties in addition to those listed earlier:

The metric should be based on an explicitly defined model of the relationship between certain attributes

The model must be dimensionally consistent.

3. The metric must not exhibit any unexpected discontinuities.

4. The metric must use units and scale types correctly.

For each quality properties the direct and indirect metric properties apply

Theoretical value of quality properties of MOOD metrics include Encapsulation, Inheritance, Coupling and Polymorphism.

Encapsulation: It includes Method Hiding Factor (MHF) and Attribute Hiding Factor (AHF). Here the use of information hiding concept is defined in terms of methods and attributes to other code. MHF consists the percentage of hidden methods and in AHF consists percentage of hidden attributes. Here MHF and AHF meets all criteria's of direct metrics and if the metrics discontinuity is accounted in single class systems indirect metrics also meets all the criteria's

Inheritance: It includes Method Inheritance Factor (MIF) and attribute Inheritance Factor (AIF). Here number of inherited methods and attributes are considered. Considering each metric related to direct metric for both MIF and AIF like difference in programs and inheritance have different MIF values. High inheritance value have high MIF value and vice versa. Each method which is inherited contributes in an equivalent way to MIF. Different programs can have the same MIF value.

Coupling: coupling factor (CF) i.e., coupling between the classes. Considering the pair sets of classes, the CF to be a direct and indirect of an attribute to which it is related [70]

Complexity lack of encapsulation, lack of reuse potential, lack of understandability, lack of maintainability.

CF as a direct measure meets all the properties of it same as the MIF properties and in indirect measure high coupling does not mean high complexity.

Polymorphism: Polymorphism factor (PF) measures polymorphism potential. PF is the number of methods that redefine inherited methods, divided by the maximum number of possible distinct polymorphic situations [69]. Here PF is considered as an indirect metric via division by using descendent classes and multiplication so if there is no inheritance it will be undefined.

Here data analysis is done by using three releases of electronic rail system (ERS). By using this, the six metrics are measured . Here the problem is measuring imprecise attributes definition. Empirical results shows that metrics operate at system level. Different assessment of system is done by comparing Chidamber and Kemerer [73+], here the evaluation of system is done at class level.

*5) SUMMMARY*

Basing on the discussion within our team mates, a priority list is made mentioning metrics that are to be implemented. QMOOD is placed first as it deals with features including both quality and design attributes. Quality has gained importance with our increasing dependence on software so QMOOD is one property that gives Object oriented design quality assessment with lower level metrics which are well defined in terms of design characteristics. It helps in identifying six important design quality attributes namely Functionality, Effectiveness, Understandability, Extendibility, Reusability and Flexibility. Helps in describing the internal and external structure of classes in term of Abstraction, Encapsulation, coupling, cohesion, complexity, design size, messaging, composition, inheritance, and polymorphism and class hierarchies. Additionally it identifies whether properties have positive and negative influence on attributes. The main attribute of this model is that it can be easily modified according to their relations and weight .This model has ability to estimate overall design quality by using various functional equivalent projects where model has significant correlation between quality characteristics that determine by independent evaluators. It gives facility of practical quality assessment tool, which can be used in variety of demands. This gives an indication that model can effectively use in order to monitor the quality of software. Secondly comes the Quality validation for few factors such as LCOM, WMC, NOC DIT CBO, and RFC which is a part of the QMOOD metrics. Thirdly the cohesion and coupling measure. In this case the cohesion is measured in object oriented analysis model based on crosscutting concerns. Which means the metric is measured based on the use cases and use case model. Fourthly, MOOD metrics measures quality properties namely encapsulation, inheritance, coupling polymorphism based on direct and indirect metrics. Then follows framework for coupling this is performed in 3 steps. The measures generated with this framework are counts of connections between classes. Namely defining terminology and formalism reviewing the framework, applying changes to the reviewed existing framework. This generally is done during the development phase and doesn't not give much required effective results. And lastly Simplified implementation of OO design metrics this is time consuming and space consuming uses hierarchy based approach. Alternatively there is query based language but problem with query based language is complexity will rise with respect to depth of the structure.

## III. Methodology

### *A. Metrics with Tools: McCabe and Logiscope*

*1)* Logiscope *[76]*

This tool was used to measure metrics of the code for MARF and GIPSY. The main objectives can be shortlisted as below:

1. Improve quality via efficient bug prevention
2. Reduce development & maintenance costs
3. Improve reuse
4. Outsourced development validation
5. Customer and regulation acceptance [76]

After getting a hands on experience with this tool. The following results were obtained. After loading the project for (MARF) following results were obtained

Application Factor Description

Factor: STATISTICS
This factor applies to Application for JAVA language.
The formula to compute the factor is:

STATISTICS = SIZE

It is already known that LOGISCOPE decomposes the

Maintainability factor into the following four criteria:

1) Analyzability
2) Changeability
3) Stability
4) Testability

Definition of Maintainability obtained from Logiscope is as below

MAINTAINABILITY: the capability of the software product to be modified.

Modifications may include corrections, improvements or adaptation of the software to changes in environment, and in requirements. and functional specifications [ISO/IEC 9126-1:2001].

Factor: MAINTAINABILITY (applicable to Packages) This factor applies to Packages for JAVA language.

The formula to compute the factor is:
MAINTAINABILITY = Analyzabilityp + CHANGEABILITYp + STABILITYp + TESTABILITYp
p=packages

Factor : MAINTAINABILITY(applicable to modules)

This factor applies to Modules for JAVA language.
The formula to compute the factor is:

MAINTAINABILITY = ANALYZABILITYm + CHANGEABILITYm + STABILITYm + TESTABILITYm
m=modules

MAINTAINABILITY (applicable for classes)
This factor applies to Classes for JAVA language.

The formula to compute the factor is :
MAINTAINABILITY = ANALYZABILITYc + CHANGEABILITYc + STABILITYc + TESTABILITYc
c=classes

Factor : MAINTAINABILITY applicable to functions
This factor applies to Functions for JAVA language.

The formula to compute the factor is :
MAINTAINABILITY = ANALYZABILITY + CHANGEABILITY + STABILITY + TESTABILITY

Criteria : SIZE

This criteria applies to Application for JAVA language. The formula to compute the criteria is :

SIZE = ap_stat + ap_func + ap_sline + ap_vg + ap_wmc + ap_eloc + ap_comf + ap_inhg_levl

| Name of the operand | What it measures |
|---|---|
| ap_stat (metric applicable to application) | Number of statements (total number of statements in the application) |
| ap_func | Contains total number of application functions |
| ap_sline | Number of physical lines in the application |
| ap_vg | Sum of cyclomatic numbers (VG) for all functions in the application |
| ap_wmc=ap_vg / ap_func | Average cyclomatic number (VG) of the functions defined in the project: |
| ap_eloc=ap_sloc - ap_ssbra | Total number of effective lines of code in the project source files. Empty lines, lines of comments or lines containing only lone braces<br><br>are not counted |
| ap_comf(ap_scomm) / (ap_sline) | Percentage of comments in the project source files |
| ap_inhg_levl | The depth of the inheritance tree is the number of classes in the longest inheritance link. |
| Ap_func | ap_func is the number of functions in the application |
| Ap_sloc | ap_sloc is the number of lines of code in the project source files |
| Ap_ssbra | ap_ssbra is the number of lines containing only lone braces in the project source files. |
| Ap_scomm | ap_scomm is the number of lines of comments in the project source files |

Table 1

Application Size is based on:

1) Number of executable statements
2) Number of functions
3) Total number of lines
4) Sum of cyclomatic numbers
5) Weighted methods per application
6) Number of effective lines of code
7) Comment rate
8) Depth of inheritance

Definition of ANALYZABILITY can be given as: the capability of the software product to be diagnosed for deficiencies or causes of failures in the software, or for the parts to be modified to be identified [ISO/IEC 9126-1:2001].

Criteria : ANALYZABILITYp(applicable to packages in java) This criteria applies to Packages for JAVA language.

The formula to compute the criteria is :
ANALYZABILITYp = pk_vg + pk_comf + pk_inh_levl_max + pk_pkused

Criteria : ANALYZABILITYm (applicable to modules in java) This criteria applies to Modules for JAVA language.

The formula to compute the criteria is :
ANALYZABILITYm = md_comf + md_line + md_import_pack

Criteria : ANALYZABILITYc(applicable to classes in java) This criteria applies to Classes for JAVA language.

The formula to compute the criteria is :
ANALYZABILITYc = cl_wmc + cl_comf + in_bases + cu_cdused

| Name of the operand | What it measures |
|---|---|
| cl_wmc | Weighted methods per class |
| cl_comf | Class comment rate |
| in_bases | Number of base classes |
| cu_cdused | Number of direct used classses |

Table 2

Criteria : ANALYZABILITY (applicable to functions in java) This criteria applies to Functions for JAVA language.

The formula to compute the criteria is :
ANALYZABILITY = ct_vg + avg_size + com_freq + lc_stat

Definition of CHANGEABILITYcan be given : the capability of the software product to enable a specified modification to be implemented [ISO/IEC 9126-1:2001].

Criteria : CHANGEABILITYp(applicable to packages in java) This criteria applies to Packages for JAVA language.

The formula to compute the criteria is :
CHANGEABILITYp = pk_stmt + pk_file + pk_func + pk_data

Criteria : CHANGEABILITYm(applicable to modules in java) This criteria applies to Modules for JAVA language.

The formula to compute the criteria is :
CHANGEABILITYm = md_stat + md_dclstat + md_class

Criteria : CHANGEABILITYc(applicable to classes in java) This criteria applies to Classes for JAVA language.

The formula to compute the criteria is :
CHANGEABILITYc = cl_stat + cl_func + cl_data

| Name of the metric | What it measures |
|---|---|
| cl_stat | Number of statements |
| cl_func | Total number of methods |
| cl_data | Total number of attributes |

Table 3

Criteria : CHANGEABILITY(applicable to functions in java) This criteria applies to Functions for JAVA language.
The formula to compute the criteria is :
CHANGEABILITY = voc_freq + n2 + ct_nest

Definition of STABILITY can be given as “the capability of the software product to avoid unexpected effects from modifications of the software” [ISO/IEC 9126-1:2001].

Criteria : STABILITYp(applicable to packages in java) This criteria applies to Packages for JAVA language.

The formula to compute the criteria is :
STABILITYp = pk_data_publ + pk_func_publ + pk_class + pk_inh

Criteria : STABILITYm(applicable to modules in java) This criteria applies to Modules for JAVA language.

The formula to compute the criteria is :
STABILITYm = md_dclstat + md_class + md_interf

Criteria : STABILITYc(applicable to classes in java. This criteria applies to Classes for JAVA language.

The formula to compute the criteria is :

STABILITYc = cl_data_publ + cu_cdusers + in_noc + cl_func_publ

| Name of the metric | What it measures |
|---|---|
| cl_data_publ | Number of Public attributes |
| cu_cdusers | Number of direct user calsses |
| in_noc | Number of children |
| cl_func_publ | Number of public methods |

Table 4

Criteria : STABILITY(applicable to functions in JAVA) This criteria applies to Functions for JAVA language. The formula to compute the criteria is :

STABILITY = struc_pg + ct_nest + ct_npath

Definition for TESTABILITY can be given as" the capability of the software product to enable modified software to be validated" [ISO/IEC 9126-1:2001].

Criteria : TESTABILITYp(applicable for packages in JAVA)
This criteria applies to Packages for JAVA language.
The formula to compute the criteria is :

TESTABILITYp = pk_vg + pk_func + pk_inh + pk_class

Criteria : TESTABILITYm(applicable to Modules in JAVA)
This criteria applies to Modules for JAVA language.
The formula to compute the criteria is :

TESTABILITYm = md_stat + md_class + md_import
Criteria : TESTABILITYc(applicable to classes in JAVA)
This criteria applies to Classes for JAVA language. The formula to compute the criteria is :

TESTABILITYc = cl_wmc + cl_func + cu_cdused

| Name of the metric | What it measures |
|---|---|
| cl_wmc | Weighted methods per class |
| cl_func | Total number of methods |
| cu_cdused | Number of direct used classes |

Table 5

Criteria : TESTABILITY(applicable to functions in JAVA)
This criteria applies to Functions for JAVA language.
The formula to compute the criteria is :
TESTABILITY = ct_vg + ic_param + ct_npath
According to LOGISCOPE these are few class metric levels obtained for MARF and GIPSY

| Mnemonic | Metric Name |
|---|---|
| cl_comf | Class comment rate |
| cl_comm | Number of lines of comment |
| cl_data | Total number of attributes |
| cl_data_publ | Number of public attributes |
| cl_func | Total number of methods |
| cl_func_publ | Number of public methods |
| cl_line | Number of lines |
| cl_stat | Number of statements |
| cl_wmc | Weighted Methods per Class |
| cu_cdused | Number of direct used classes |
| cu_cdusers | Number of direct users classes |
| in_bases | Number of base classes |
| in_noc | Number of children |

Table 6

Comparison

Let us consider minimum and maximum values of both class metrics and assume that two values M1 M2 lie between the min and max values. The table below shows if M1 is or M2 is better for each class metric level.

| Name of the Class metric | Assumption | M2 is better | M1 is better |
|---|---|---|---|
| cl_comf | M2>M1 | ✓ | |
| cl_comm | M2>M1 | ✓ | |
| cl_data | M2>M1 | | ✓ |
| cl_data_publ | M2>M1 | | ✓ |
| cl_func | M2>M1 | ✓ | |
| cl_func_publ | M2>M1 | | ✓ |
| cl_line | M2>M1 | | ✓ |
| cl_stat | M2>M1 | | ✓ |
| cl_wmc | M2>M1 | | ✓ |
| cu_cdused | M2>M1 | | ✓ |
| cu_cdusers | M2>M1 | | ✓ |
| in_bases | M2>M1 | ✓ | |
| in_noc | M2>M1 | ✓ | |

Table 7

*Class comment rate*: Given two values M1 and M2 within the range of minimum and maximum values M2 is greater than M1 therefore higher comment rate increases the understandability of the application. In this case.M2 being greater proves to be good.

*Number of lines of comments*: Given two values M1 and M2 within the range of minimum and maximum values assuming M2 is greater than M1.More number of comments results in better understanding of code. Therefore M2 being greater than M1 proves to be a good case.

*Total number of attributes:* Given two values M1 and M2 within the range of minimum and maximum values assuming M2 is greater than M1.More number of attributes results in increasing the complexity of the code. Therefore M2 being greater than M1 might not be a good scenario in his case.

*Total number of public attributes*: Given two values M1 and M2 within the range of minimum and maximum values assuming M2 is greater than M1.More number of public attributes results decreasing the security level of the a class. Hence M1 is better in this case.

*Number of public methods*. : Given two values M1 and M2 within the range of minimum and maximum values assuming M2 is greater than M1.More number of public methods might result in decreasing the security levels of code as the methods can be used anywhere. These methods are visible to everyone therefore diminishing the security levels of the application.M2 being greater than M1 might not be a good scenario in this case. Having less number of public methods in a class is good.

*Total number of methods*: Given two values M1 and M2 within the range of minimum and maximum values assuming M2 is greater than M1.Having more number of methods in a class can result in higher complexity of the code. In few cases having more number of methods might increase method calling which implies re usability.

*Number of lines*: Given two values M1 and M2 within the range of minimum and maximum values assuming M2 is greater than M1.More number of lines of code result in having a high possibility of dead code. Therefore M2 being greater than M1 in this case might be a bad scenario.

*Total number of statements*: Given two values M1 and M2 within the range of minimum and maximum values assuming M2 is greater than M1.Having more number of statements increases the complexity of a class. Therefore M1 is better in this case.

*Weighted methods per class:* Given two values M1 and M2 within the range of minimum and maximum values assuming M2 is greater than M1.Having more number of weighted methods per class results in increasing the complexity of the application. Therefore M2 being greater than M1 is not a good case here.

*Number of direct used classes*: Given two values M1 and M2 within the range of minimum and maximum values assuming M2 is greater than M1.More number of direct used classes result in effective function of the class.M1 is better in this case.

*Number of direct user classes*: Given two values M1 and M2 within the range of minimum and maximum values assuming M2 is greater than M1.More number of direct user classes result in effective function of the class.M2 is better in this case.

*Number of base classes*: Given two values M1 and M2 within the range of minimum and maximum values assuming M2 is greater than M1.Having more number of base classes might result in increasing the polymorphism therefore M2 being greater than M1 is a good scenario in this case.

*Number of children*: Given two values M1 and M2 within the range of minimum and maximum values assuming M2 is greater than M1.More number of children classes means higher the level of inheritance. Therefore M2 being greater than M1 is a good scenario in this case.

Comparative study of MARF and GIPSY with the minimum, maximum and out of bound values

| MARF | | | | | GIPSY | | |
|---|---|---|---|---|---|---|---|
| Mnemonic | Metric Name | Min | Max | Out | Min | Max | Out |
| cl_comf | Class comment rate | 0.20 | +oo | 2.78% | 0.20 | +oo | 40.24% |
| cl_comm | Number of lines of comment | -oo | +oo | 0.00% | -oo | +oo | 0.00% |
| cl_data | Total number of attributes | 0 | 7 | 13.89% | 0 | 7 | 17.84% |
| cl_data_publ | Number of public attributes | 0 | 0 | 32.87% | 0 | 0 | 28.66% |
| cl_func | Total number of methods | 0 | 25 | 5.56% | 0 | 25 | 5.64% |
| cl_func_publ | Number of public methods | 0 | 15 | 16.20% | 0 | 15 | 7.01% |
| cl_line | Number of lines | -oo | +oo | 0.00% | -oo | +oo | 0.00% |
| cl_stat | Number of statements | 0 | 100 | 10.65% | 0 | 100 | 11.43% |
| cl_wmc | Weighted Methods per Class | 0 | 60 | 4.63% | 0 | 60 | 6.10% |
| cu_cdused | Number of direct used classes | 0 | 10 | 22.22% | 0 | 10 | 21.65% |
| cu_cdusers | Number of direct users classes | 0 | 5 | 20.37% | 0 | 5 | 14.18% |
| in_bases | Number of base classes | 0 | 3 | 33.80% | 0 | 3 | 13.87% |
| in_noc | Number of children | 0 | 3 | 4.17% | 0 | 3 | 5.18% |

Table 8

Comparison of out of bound rates for MARF and GIPSY

Cl_comf MARF < cl_comf_GIPSY
Cl_comm_MARf==cl_comf_GIPSY
Cl_data_MARF<cl_data_GIPSY
Cl_data_publ>cl_data_publ_GIPSY
Cl_func_publ>cl_func_publ
Cl_line_MARF=cl_line_GIPSY
Cl_stat_MARF<Cl_stat_GIPSY
Cl_wmc<cl_wmc_GIPSY
Cu_cdused>cu_cdused_GIPSY
Cu_cdusers_MARF>cu_cdusers_GIPSY
In_bases_MARF>in_bases_GIPSY
In_noc_MARF<in_noc_GIPSY

LOGICSCOPE

Extraction: in terms of LOGISCOPE(tool used) Class Factor Level (MARF AND GIPSY)

At Class Factor Level- the percentage of poor and fair is 16.21 in MARF where as in GIPSY it is 15.39 so maintainability of GIPSY is quite easy as compared to MARF. Therefore, taking into considerations all possible statistics MARF has worst quality code when compared to GIPSY as an application.

In analyzability – the percentage of poor and fair is 13.43 in MARF where as in GIPSY it is 17.07, it means the classes that are categorized under analyzability of gipsy are less maintainable as compared to MARF and the same applies to stability. It is noticeable that the code needs improvement in classes that come under stability criteria level of MARF(19.44 %) is more as compared to GIPSY(11.43 %).
In the table below the percentage of GIPSY and MARF in terms of poor and fair percentage is calculated.

| | GIPSY (fair +poor)% | MARF (fair +poor)% |
|---|---|---|
| Class Factor Level | 15.39 | 16.21 |
| Class Criteria Level | | |
| Analyzability | 17.07 | 13.43 |
| Changeability | 9.45 | 6.48 |
| **Stability** | **11.43** | **19.44** |
| Testability | 6.56 | 7.01 |

Table 9

Class Metric Level: Kiviat Diagram for poor class obtained from Logiscope:

**Kiviat Diagram for marf.MARF**

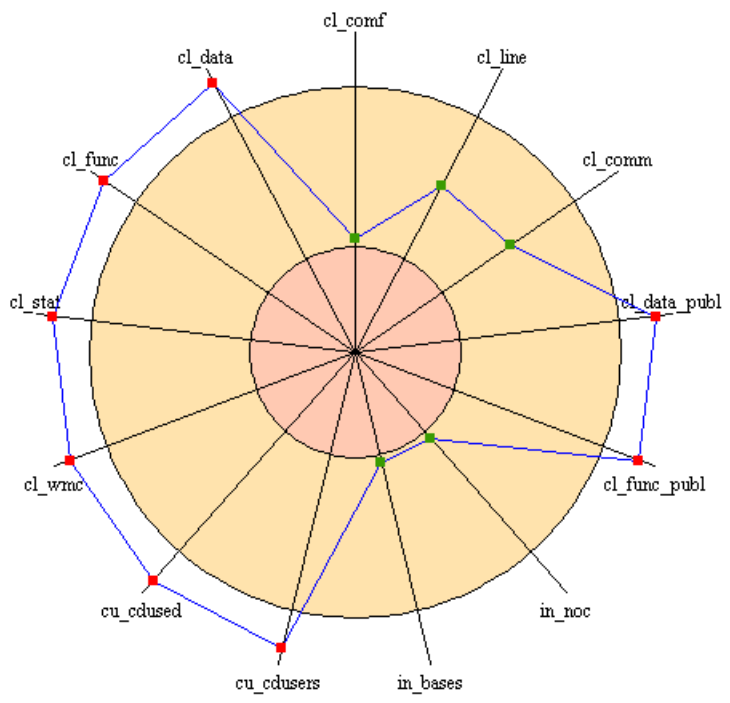

*Figure 14 Kiviat graph for class marf.MARF.java is taken from Kalimetrix Logiscope tool*

| | MARF | GIPSY |
|---|---|---|
| Class Factor level | 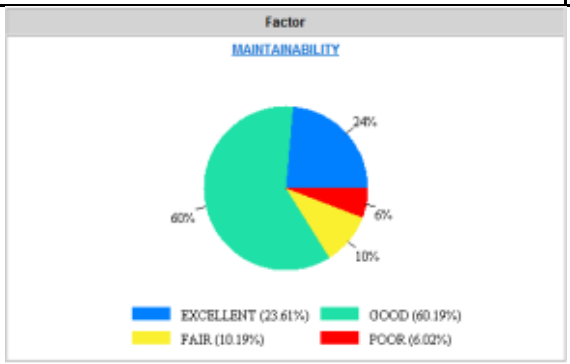 | 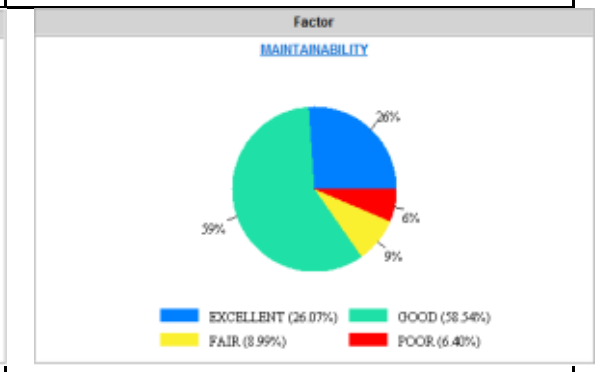 |
| Class Criteria Level | 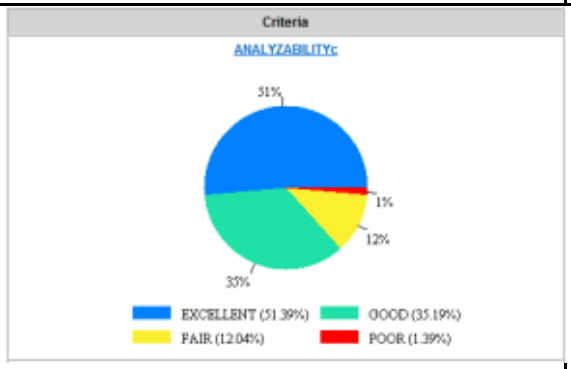 | 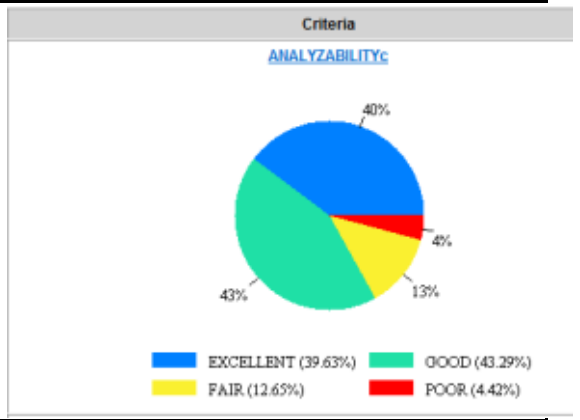 |
| | 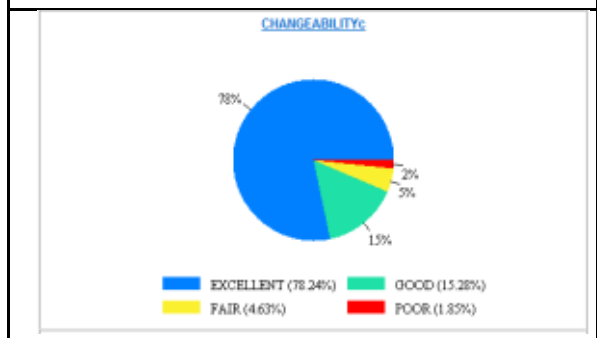 | 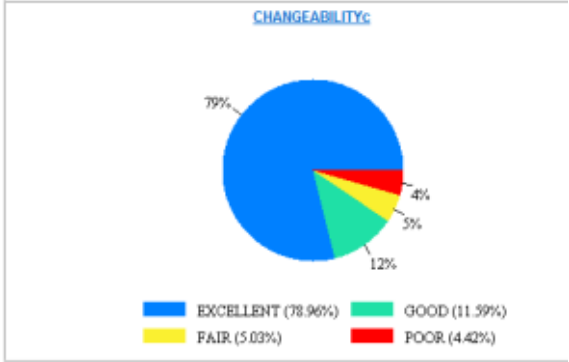 |
| | 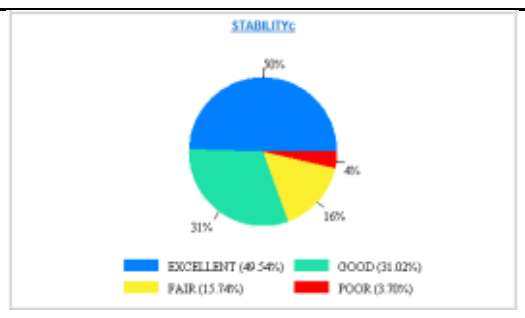 | 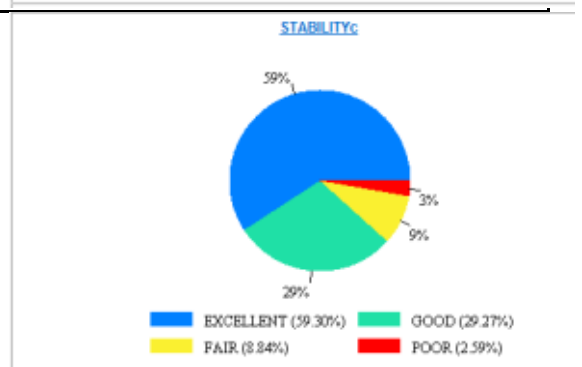 |
| | 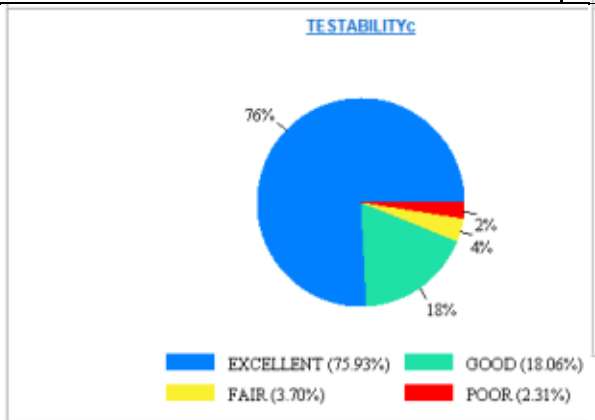 | 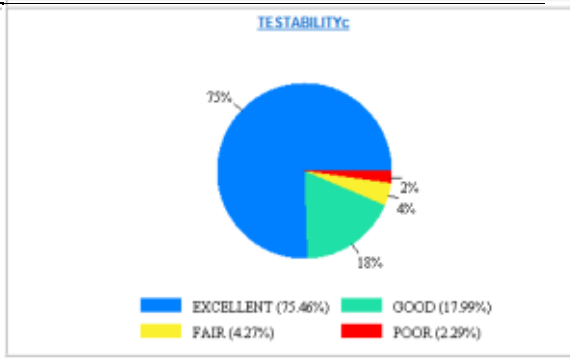 |

*Figure 15 Figures for Class factor level and Class Criteria level is taken from Kalimetrix Logiscope tool*

**Kiviat Diagram for marf.Classification.NeuralNetwork.NeuralNetwork**

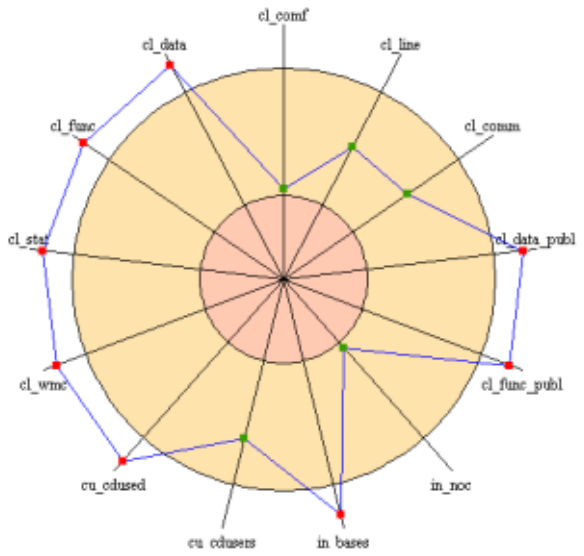

*Figure 16 Kiviat graph for class marf.Classification.NeuralNetwork.NeuralNetwork.java is taken from Kalimetrix Logiscope tool*

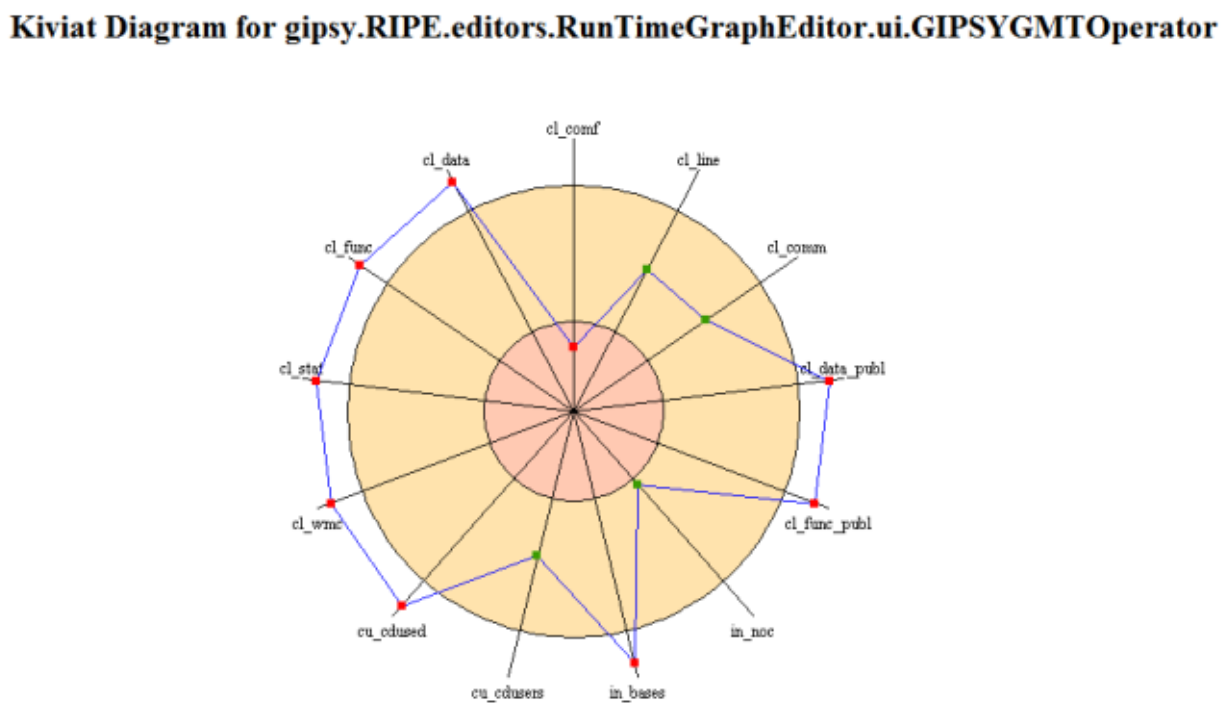

*Figure 17 KiviatGraph for class gipsy.RIPE.editors.RunTimeGraphEditor.ui.GIPSYGMTOperator is taken from Kalimetrix Logiscope tool*

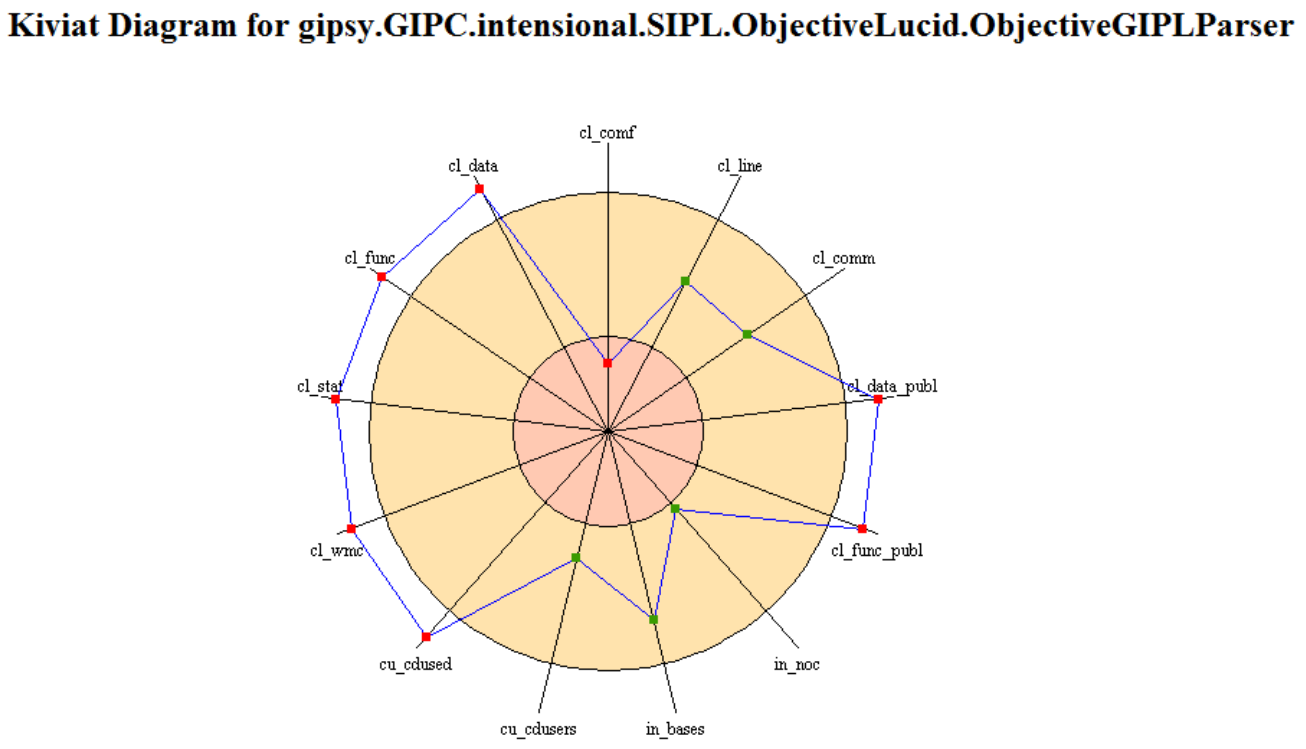

*Figure 18 Kiviat Graph for class gipsy.GIPC.intensional.SIPL.ObjectiveLucid.ObjectiveGIPLParser is taken form Kalimetrix Logiscope tool*

1) List of classes

*a)* MARF

Ranking of the list of fair and poor classes according to the status of metric which is equal to -1 which is obtained from Logiscope tool. According to the ranking the list of worst classes among all fair and poor classes is found. A list of classes of Maintainability which is taken from factor level and the other classes of Analyzability, Changeability, Stability and Testability is taken from criteria level. The highest number in the ranking is treated as worst class.

Ranking = Count the number of status (-1)

MAINTAINABILTY:
Factor: MAINTAINABILITY : FAIR

| Classes name | Ranking |
|---|---|
| marf.Classification.Classification | 4 |
| marf.FeatureExtraction.FFT.FFT | 3 |
| marf.FeatureExtraction.FeatureExtractionAggregator | 3 |
| marf.FeatureExtraction.LPC.LPC | 3 |
| marf.Preprocessing.CFEFilters.CFEFilter | 4 |
| marf.Preprocessing.FFTFilter.FFTFilter | 5 |
| marf.Preprocessing.Preprocessing | 4 |
| marf.Stats.ProbabilityTable | 4 |
| marf.Storage.Loaders.AudioSampleLoader | 3 |
| marf.Storage.Loaders.TextLoader | 3 |
| marf.Storage.Sample | 4 |
| marf.Storage.TrainingSet | 5 |
| marf.nlp.Parsing.GenericLexicalAnalyzer | 4 |
| marf.nlp.Parsing.GrammarCompiler.GrammarAnalyzer | 3 |
| marf.nlp.Parsing.LexicalAnalyzer | 4 |
| marf.nlp.Parsing.LexicalError | 4 |
| marf.nlp.Parsing.Parser | 4 |
| marf.nlp.Parsing.ProbabilisticParser | 3 |
| marf.nlp.Parsing.SyntaxError | 5 |
| marf.nlp.Parsing.TransitionTable | 3 |
| marf.util.OptionProcessor | 4 |
| test | 4 |

Table 10

Factor: MAINTAINABILITY : POOR

| Classes name | Ranking |
|---|---|
| marf.Classification.NeuralNetwork.NeuralNetwork | 8 |
| marf.Classification.Stochastic.ZipfLaw | 6 |
| marf.Configuration | 5 |
| marf.MARF | 8 |
| marf.Stats.StatisticalEstimators.StatisticalEstimator | 6 |
| marf.Storage.ResultSet | 5 |
| marf.Storage.StorageManager | 7 |
| marf.math.ComplexMatrix | 5 |
| marf.math.Matrix | 7 |
| marf.nlp.Parsing.GrammarCompiler.Grammar | 6 |
| marf.nlp.Parsing.GrammarCompiler.GrammarCompiler | 6 |
| marf.nlp.Storage.Corpus | 6 |
| marf.util.Arrays | 6 |

Table 11

ANALYZABILITY
Criteria: ANALYZABILITYc : FAIR

| Classes name | Ranking |
|---|---|
| marf.Classification.Classification | 4 |
| marf.Classification.RandomClassification.RandomClassification | 2 |
| marf.Classification.Stochastic.MaxProbabilityClassifier | 2 |
| marf.Classification.Stochastic.ZipfLaw | 6 |
| marf.FeatureExtraction.FFT.FFT | 3 |
| marf.FeatureExtraction.FeatureExtractionAggregator | 3 |
| marf.FeatureExtraction.LPC.LPC | 3 |
| marf.MARF | 8 |
| marf.Preprocessing.CFEFilters.BandStopFilter | 2 |
| marf.Preprocessing.CFEFilters.CFEFilter | 4 |
| marf.Preprocessing.CFEFilters.LowPassFilter | 2 |
| marf.Preprocessing.FFTFilter.FFTFilter | 5 |
| marf.Preprocessing.Preprocessing | 4 |
| marf.Preprocessing.WaveletFilters.WaveletFilter | 2 |
| marf.Stats.ProbabilityTable | 4 |
| marf.Stats.StatisticalEstimators.StatisticalEstimator | 6 |
| marf.Storage.Cluster | 2 |
| marf.Storage.SampleRecorder | 2 |
| marf.Storage.StorageManager | 7 |
| marf.Storage.TrainingSet | 5 |
| marf.math.Matrix | 7 |
| marf.nlp.Parsing.GrammarCompiler.Grammar | 6 |
| marf.nlp.Parsing.GrammarCompiler.ProbabilisticGrammarCompiler | 2 |
| marf.nlp.Parsing.ProbabilisticParser | 3 |
| marf.util.Arrays | 6 |
| test | 4 |

Table 12

Criteria: ANALYZABILITYc: POOR

| Classes name | Ranking |
|---|---|
| marf.Classification.NeuralNetwork.NeuralNetwork | 8 |
| marf.nlp.Parsing.GrammarCompiler.GrammarCompiler | 6 |
| marf.nlp.Parsing.LexicalAnalyzer | 4 |

Table 13

CHANGEABILITY
Criteria: CHANGEABILITYc : FAIR

| Classes name | Ranking |
|---|---|
| marf.Classification.Stochastic.ZipfLaw | 6 |
| marf.Storage.ResultSet | 5 |
| marf.Storage.StorageManager | 7 |
| marf.math.ComplexMatrix | 5 |
| marf.math.Matrix | 7 |
| marf.nlp.Parsing.GrammarCompiler.Grammar | 6 |
| marf.nlp.Parsing.GrammarCompiler.GrammarCompiler | 6 |
| marf.nlp.Parsing.Parser | 4 |
| marf.util.Arrays | 6 |
| marf.util.OptionProcessor | 4 |

Table 14

Criteria: CHANGEABILITYc : POOR

| Classes name | Ranking |
|---|---|
| marf.Classification.NeuralNetwork.NeuralNetwork | 8 |
| marf.Configuration | 5 |
| marf.MARF | 8 |
| marf.nlp.Storage.Corpus | 6 |

Table 15

STABILITY
Criteria: STABILITYc : FAIR

| Classes name | Ranking |
|---|---|
| marf.Classification.Classification | 4 |
| marf.Classification.Distance.Distance | 3 |
| marf.Classification.NeuralNetwork.NeuralNetwork | 8 |
| marf.Classification.Stochastic.ZipfLaw | 6 |
| marf.FeatureExtraction.FeatureExtraction | 3 |
| marf.FeatureExtraction.IFeatureExtraction | 2 |
| marf.MARF.ENgramModels | 2 |
| marf.MARF.NLP | 3 |
| marf.Preprocessing.IPreprocessing | 2 |
| marf.Storage.Loaders.AudioSampleLoader | 3 |
| marf.Storage.MARFAudioFileFormat | 3 |
| marf.Storage.ModuleParams | 2 |
| marf.Storage.Result | 2 |
| marf.Storage.ResultSet | 5 |
| marf.Storage.TrainingSet | 5 |
| marf.math.ComplexMatrix | 5 |
| marf.math.ComplexVector | 3 |
| marf.nlp.Parsing.CodeGenerator | 2 |
| marf.nlp.Parsing.CompilerError | 3 |
| marf.nlp.Parsing.GenericLexicalAnalyzer | 4 |
| marf.nlp.Parsing.GrammarCompiler.Grammar | 6 |
| marf.nlp.Parsing.GrammarCompiler.GrammarTokenType | 3 |
| marf.nlp.Parsing.LexicalError | 4 |
| marf.nlp.Parsing.SymTabEntry | 3 |
| marf.nlp.Parsing.SymbolTable | 2 |
| marf.nlp.Parsing.TokenSubType | 3 |
| marf.nlp.Parsing.TokenType | 3 |
| marf.nlp.Storage.Corpus | 6 |
| marf.util.Arrays | 6 |
| marf.util.BaseThread | 2 |
| marf.util.Debug | 2 |
| marf.util.MARFException | 2 |
| marf.util.OptionProcessor | 4 |
| marf.util.SortComparator | 2 |

Table 16

Criteria: STABILITYc : POOR

| Classes name | Ranking |
|---|---|
| marf.MARF | 8 |
| marf.Preprocessing.FFTFilter.FFTFilter | 5 |
| marf.Stats.StatisticalEstimators.StatisticalEstimator | 6 |
| marf.Storage.Sample | 4 |
| marf.Storage.StorageManager | 7 |
| marf.math.Matrix | 7 |
| marf.math.Vector | 3 |
| marf.nlp.Parsing.SyntaxError | 5 |

Table 17

TESTABILITY, Criteria: TESTABILITYc : FAIR

| Classes name | Ranking |
|---|---|
| marf.Configuration | 5 |
| marf.Stats.ProbabilityTable | 4 |
| marf.Storage.ResultSet | 5 |
| marf.math.ComplexMatrix | 5 |
| marf.nlp.Parsing.GrammarCompiler.Grammar | 6 |
| marf.nlp.Parsing.GrammarCompiler.GrammarCompiler | 6 |
| marf.nlp.Parsing.LexicalAnalyzer | 4 |
| marf.nlp.Storage.Corpus | 6 |

Table 18

Criteria: TESTABILITYc : POOR

| Classes name | Ranking |
|---|---|
| marf.Classification.NeuralNetwork.NeuralNetwork | 8 |
| marf.MARF | 8 |
| marf.Storage.StorageManager | 7 |
| marf.math.Matrix | 7 |
| marf.util.Arrays | 6 |

Table 19

GIPSY List of fair and poor classes
MAINTAINABILITY, Factor : MAINTAINABILITY : FAIR

| Classes name | Ranking |
|---|---|
| gipsy.GEE.GEE | 4 |
| gipsy.GEE.IDP.DemandGenerator.DemandGenerator | 3 |
| gipsy.GEE.IDP.DemandGenerator.LegacyEductiveInterpreter | 4 |
| gipsy.GEE.IDP.DemandGenerator.LegacyInterpreter | 3 |
| gipsy.GEE.IDP.DemandGenerator.jini.rmi.JINITransportAgent | 4 |
| gipsy.GEE.IDP.DemandGenerator.jms.DemandController | 3 |
| gipsy.GEE.IVW.Warehouse.NetCDFFileManager | 3 |
| gipsy.GEE.multitier.DGT.DGTWrapper | 5 |
| gipsy.GEE.multitier.DST.jini.JiniDSTWrapper | 5 |
| gipsy.GEE.multitier.DST.jini.JiniERIDSTWrapper | 5 |
| gipsy.GEE.multitier.DST.jms.JMSDSTWrapper | 3 |
| gipsy.GEE.multitier.DWT.DWTWrapper | 6 |
| gipsy.GEE.multitier.TAExceptionHandler | 4 |
| gipsy.GIPC.DFG.DFGAnalyzer.LucidCodeGenerator | 5 |
| gipsy.GIPC.imperative.ImperativeCompiler | 5 |
| gipsy.GIPC.imperative.Java.JavaCompiler | 3 |
| gipsy.GIPC.intensional.GIPL.GIPLParserTreeConstants | 5 |
| gipsy.GIPC.intensional.GenericTranslator.TranslationLexer | 4 |
| gipsy.GIPC.intensional.IntensionalCompiler | 4 |
| gipsy.GIPC.intensional.SIPL.IndexicalLucid.IndexicalLucidParserTreeConstants | 4 |
| gipsy.GIPC.intensional.SIPL.JOOIP.JOOIPCompiler | 3 |

Table 20

| | |
|---|---|
| gipsy.GIPC.intensional.SIPL.JOOIP.ast.Node | 4 |
| gipsy.GIPC.intensional.SIPL.JOOIP.ast.body.TypeDeclaration | 4 |
| gipsy.GIPC.intensional.SIPL.JOOIP.ast.expr.StringLiteralExpr | 4 |
| gipsy.GIPC.intensional.SIPL.JOOIP.ast.visitor.GenericVisitor | 4 |
| gipsy.GIPC.intensional.SIPL.JOOIP.ast.visitor.VoidVisitor | 4 |
| gipsy.RIPE.RIPE | 4 |
| gipsy.RIPE.editors.RunTimeGraphEditor.core.GIPSYTier | 4 |
| gipsy.RIPE.editors.RunTimeGraphEditor.core.GraphDataManager | 4 |
| gipsy.RIPE.editors.RunTimeGraphEditor.ui.ActionsLog | 3 |
| gipsy.RIPE.editors.RunTimeGraphEditor.ui.InstancesNodesPanel | 3 |
| gipsy.RIPE.editors.RunTimeGraphEditor.ui.MapEditor | 4 |
| gipsy.RIPE.editors.RunTimeGraphEditor.ui.dialogs.GIPSYNodeDialog | 4 |
| gipsy.RIPE.editors.RunTimeGraphEditor.ui.dialogs.TierPropertyDialog | 3 |
| gipsy.RIPE.editors.WebEditor.WebEditor | 4 |
| gipsy.apps.marfcat.MARFCATDWT | 3 |
| gipsy.apps.marfcat.MARFPCATDWT | 3 |
| gipsy.apps.memocode.genome.AlignDGT | 3 |
| gipsy.apps.memocode.genome.AlignDWT | 2 |
| gipsy.interfaces.GIPSYProgram | 4 |
| gipsy.lang.GIPSYFunction | 3 |
| gipsy.lang.GIPSYInteger | 4 |
| gipsy.lang.GIPSYType | 5 |
| gipsy.lang.context.TagSet | 4 |
| gipsy.tests.GEE.IDP.demands.DemandTest | 3 |
| gipsy.tests.GEE.multitier.GMT.GMTTestConsole | 4 |
| gipsy.tests.GEE.simulator.DGTDialog | 4 |
| gipsy.tests.GEE.simulator.DGTSimulator | 3 |
| gipsy.tests.GEE.simulator.DSTSpaceScalabilityTester | 4 |
| gipsy.tests.GEE.simulator.GlobalDef | 4 |
| gipsy.tests.GEE.simulator.ProfileDialog | 4 |
| gipsy.tests.GEE.simulator.demands.WorkResultPi | 3 |
| gipsy.tests.GEE.simulator.jini.WorkerJTA | 3 |
| gipsy.tests.Regression | 4 |

Table 21

Factor : MAINTAINABILITY : POOR

| Classes name | Ranking |
|---|---|
| gipsy.Configuration | 5 |
| gipsy.GEE.IDP.DemandGenerator.jini.rmi.JINITA | 6 |
| gipsy.GEE.IDP.DemandGenerator.jini.rmi.JiniDemandDispatcher | 6 |
| gipsy.GEE.IDP.DemandGenerator.jms.JMSTransportAgent | 7 |
| gipsy.GEE.IDP.demands.Demand | 8 |
| gipsy.GEE.multitier.GIPSYNode | 6 |
| gipsy.GEE.multitier.GMT.GMTWrapper | 7 |
| gipsy.GIPC.DFG.DFGAnalyzer.DFGParser | 8 |
| gipsy.GIPC.DFG.DFGAnalyzer.DFGParserTokenManager | 6 |
| gipsy.GIPC.DFG.DFGGenerator.DFGCodeGenerator | 6 |
| gipsy.GIPC.DFG.DFGGenerator.DFGTranCodeGenerator | 6 |
| gipsy.GIPC.GIPC | 7 |
| gipsy.GIPC.Preprocessing.PreprocessorParser | 8 |
| gipsy.GIPC.Preprocessing.PreprocessorParserTokenManager | 6 |
| gipsy.GIPC.SemanticAnalyzer | 6 |
| gipsy.GIPC.intensional.GIPL.GIPLParser | 8 |
| gipsy.GIPC.intensional.GIPL.GIPLParserTokenManager | 6 |
| gipsy.GIPC.intensional.GenericTranslator.TranslationParser | 6 |
| gipsy.GIPC.intensional.SIPL.ForensicLucid.ForensicLucidParser | 8 |
| gipsy.GIPC.intensional.SIPL.ForensicLucid.ForensicLucidParserTokenManager | 6 |
| gipsy.GIPC.intensional.SIPL.ForensicLucid.ForensicLucidSemanticAnalyzer | 7 |
| gipsy.GIPC.intensional.SIPL.IndexicalLucid.IndexicalLucidParser | 8 |
| gipsy.GIPC.intensional.SIPL.IndexicalLucid.IndexicalLucidParserTokenManager | 6 |
| gipsy.GIPC.intensional.SIPL.JLucid.JGIPLParser | 8 |
| gipsy.GIPC.intensional.SIPL.JLucid.JGIPLParserTokenManager | 6 |
| gipsy.GIPC.intensional.SIPL.JLucid.JIndexicalLucidParser | 8 |
| gipsy.GIPC.intensional.SIPL.JLucid.JIndexicalLucidParserTokenManager | 6 |
| gipsy.GIPC.intensional.SIPL.JOOIP.JavaCharStream | 7 |
| gipsy.GIPC.intensional.SIPL.JOOIP.JavaParser | 8 |
| gipsy.GIPC.intensional.SIPL.JOOIP.JavaParserTokenManager | 7 |
| gipsy.GIPC.intensional.SIPL.JOOIP.ast.visitor.DumpVisitor | 6 |
| gipsy.GIPC.intensional.SIPL.Lucx.LucxParser | 8 |
| gipsy.GIPC.intensional.SIPL.Lucx.LucxParserTokenManager | 6 |
| *gipsy.GIPC.intensional.SIPL.ObjectiveLucid.ObjectiveGIPLParser* | 8 |
| gipsy.GIPC.intensional.SIPL.ObjectiveLucid.ObjectiveGIPLParserTokenManager | 6 |
| gipsy.GIPC.intensional.SIPL.ObjectiveLucid.ObjectiveIndexicalLucidParser | 8 |
| gipsy.GIPC.intensional.SIPL.ObjectiveLucid.ObjectiveIndexicalLucidParserTokenManager | 6 |
| gipsy.GIPC.util.SimpleCharStream | 8 |
| gipsy.RIPE.editors.RunTimeGraphEditor.core.GlobalInstance | 6 |
| *gipsy.RIPE.editors.RunTimeGraphEditor.ui.GIPSYGMTOperator* | 9 |
| gipsy.lang.GIPSYContext | 7 |
| gipsy.tests.GIPC.intensional.SIPL.Lucx.SemanticTest.LucxSemanticAnalyzer | 7 |

Table 22

ANALYZABILITY

Criteria : ANALYZABILITYc : FAIR

| Classes name | Ranking |
|---|---|
| gipsy.GEE.IDP.DemandGenerator.Interpreter | 2 |
| gipsy.GEE.IDP.DemandGenerator.LegacyInterpreter | 3 |
| gipsy.GEE.IDP.DemandGenerator.jini.rmi.JINITA | 6 |
| gipsy.GEE.IDP.DemandGenerator.jini.rmi.JINITransportAgent.JTABackend | 2 |
| gipsy.GEE.IDP.DemandGenerator.jini.rmi.JINITransportAgent.JINITransportAgentProxy | 2 |
| gipsy.GEE.IDP.DemandGenerator.jini.rmi.JiniDemandDispatcher | 2 |
| gipsy.GEE.IDP.DemandGenerator.jini.rmi.MulticastJiniServiceDiscoverer | 2 |
| gipsy.GEE.IDP.DemandGenerator.jms.DemandController | 3 |
| gipsy.GEE.IDP.DemandGenerator.jms.JMSDemandDispatcher | 2 |
| gipsy.GEE.IDP.DemandGenerator.rmi.IdentifierContextServer | 2 |
| gipsy.GEE.IDP.demands.Demand | 8 |
| gipsy.GEE.IVW.Warehouse.NetCDFFileManager | 3 |
| gipsy.GEE.multitier.DST.jms.JMSDSTWrapper | 3 |
| gipsy.GEE.multitier.DWT.DWTFactory | 2 |
| gipsy.GEE.multitier.GIPSYNode | 6 |
| gipsy.GEE.multitier.GMT.demands.DSTRegistration | 2 |
| gipsy.GEE.multitier.GMT.demands.NodeRegistration | 2 |
| gipsy.GEE.multitier.GMT.demands.TierAllocationResult | 2 |
| gipsy.GEE.multitier.TAExceptionHandler | 4 |
| gipsy.GIPC.DFG.DFGAnalyzer.DFGParserTokenManager | 6 |
| gipsy.GIPC.DFG.DFGAnalyzer.LucidCodeGenerator | 5 |
| gipsy.GIPC.GIPC | 7 |
| gipsy.GIPC.Preprocessing.PreprocessorParserTokenManager | 6 |
| gipsy.GIPC.imperative.ImperativeCompiler | 5 |
| gipsy.GIPC.imperative.Java.JavaCommunicationProcedureGenerator | 2 |
| gipsy.GIPC.intensional.GIPL.GIPLParserTokenManager | 6 |
| gipsy.GIPC.intensional.GenericTranslator.TranslationLexer | 4 |
| gipsy.GIPC.intensional.IntensionalCompiler | 4 |
| gipsy.GIPC.intensional.SIPL.ForensicLucid.ForensicLucidParserTokenManager | 6 |
| gipsy.GIPC.intensional.SIPL.IndexicalLucid.IndexicalLucidParserTokenManager | 6 |
| gipsy.GIPC.intensional.SIPL.JLucid.JGIPLParserTokenManager | 6 |
| gipsy.GIPC.intensional.SIPL.JLucid.JIndexicalLucidParserTokenManager | 6 |
| gipsy.GIPC.intensional.SIPL.JLucid.JLucidCompiler | 2 |
| gipsy.GIPC.intensional.SIPL.JOOIP.JOOIPCompiler | 3 |
| gipsy.GIPC.intensional.SIPL.JOOIP.JavaCharStream | 7 |
| gipsy.GIPC.intensional.SIPL.JOOIP.ast.expr.CharLiteralExpr | 2 |
| gipsy.GIPC.intensional.SIPL.JOOIP.ast.expr.DoubleLiteralExpr | 2 |
| gipsy.GIPC.intensional.SIPL.JOOIP.ast.expr.IntegerLiteralMinValueExpr | 2 |
| gipsy.GIPC.intensional.SIPL.JOOIP.ast.expr.LongLiteralMinValueExpr | 2 |
| gipsy.GIPC.intensional.SIPL.JOOIP.ast.visitor.GenericVisitor | 4 |
| gipsy.GIPC.intensional.SIPL.JOOIP.ast.visitor.VoidVisitor | 4 |
| gipsy.GIPC.intensional.SIPL.Lucx.LucxParserTokenManager | 6 |
| gipsy.GIPC.intensional.SIPL.ObjectiveLucid.ObjectiveGIPLParserTokenManager | 6 |
| gipsy.GIPC.intensional.SIPL.ObjectiveLucid.ObjectiveIndexicalLucidParserTokenManager | 6 |
| gipsy.GIPC.util.SimpleCharStream | 8 |
| gipsy.RIPE.RIPE | 4 |
| gipsy.RIPE.editors.RunTimeGraphEditor.core.GraphDataManager | 4 |
| gipsy.RIPE.editors.RunTimeGraphEditor.operator.GIPSYGMTController | 2 |
| gipsy.RIPE.editors.RunTimeGraphEditor.operator.GIPSYTiersController | 2 |
| gipsy.RIPE.editors.RunTimeGraphEditor.ui.dialogs.GIPSYNodeDialog | 4 |
| gipsy.RIPE.editors.WebEditor.WebEditor | 4 |
| gipsy.apps.marfcat.MARFCATDGT | 2 |
| gipsy.apps.marfcat.MARFCATDWT | 3 |
| gipsy.apps.marfcat.MARFCATDWT.MARFCATDWTApp | 2 |
| gipsy.apps.marfcat.MARFPCATDGT | 2 |
| gipsy.apps.marfcat.MARFPCATDWT | 3 |
| gipsy.apps.marfcat.MARFPCATDWT.MARFCATDWTApp | 2 |
| gipsy.lang.GIPSYContext | 7 |
| gipsy.lang.context.OrderedFinitePeriodicTagSet | 2 |
| gipsy.lang.context.OrderedInfiniteNonPeriodicTagSet | 3 |
| gipsy.lang.context.OrderedInfinitePeriodicTagSet | 2 |
| gipsy.lang.context.UnorderedFiniteNonPeriodicTagSet | 2 |
| gipsy.lang.context.UnorderedFinitePeriodicTagSet | 2 |
| gipsy.lang.context.UnorderedInfinitePeriodicTagSet | 2 |
| gipsy.tests.GEE.IDP.demands.DemandTest | 3 |
| gipsy.tests.GEE.multitier.DGT.PseudoDGT | 2 |
| gipsy.tests.GEE.multitier.DST.PseudoJiniDSTWrapper | 2 |
| gipsy.tests.GEE.multitier.GIPSYNodeTestDriver | 2 |
| gipsy.tests.GEE.multitier.GMT.GMTTestConsole | 4 |
| gipsy.tests.GEE.multitier.GMT.GMTTestConsole.KeyInputProcessor | 2 |
| gipsy.tests.GEE.simulator.DGTDialog | 4 |
| gipsy.tests.GEE.simulator.DSTSpaceScalabilityTester | 4 |
| gipsy.tests.GEE.simulator.DemandResponseTimeTester | 2 |
| gipsy.tests.GEE.simulator.ResultAnalyst | 2 |
| gipsy.tests.GEE.simulator.demands.LightUniqueDemand | 2 |
| gipsy.tests.GEE.simulator.demands.SizeAdjustableDemand | 2 |
| gipsy.tests.GEE.simulator.demands.WorkResultHD | 2 |
| gipsy.tests.GEE.simulator.demands.WorkResultPi | 3 |
| gipsy.tests.jooip.CopyOfGIPLtestVerbose | 4 |
| gipsy.tests.junit.GEE.multitier.DGT.DGTWrapperTest | 3 |
| gipsy.tests.junit.GEE.multitier.DWT.DWTWrapperTest | 3 |

Table 23

Criteria : ANALYZABILITYc : POOR

| Classes name | Ranking |
|---|---|
| gipsy.GEE.IDP.DemandGenerator.LegacyEductiveInterpreter | 4 |
| gipsy.GEE.IDP.DemandGenerator.jms.JMSTransportAgent | 7 |
| gipsy.GEE.multitier.DST.jini.JiniDSTWrapper | 5 |
| gipsy.GEE.multitier.DST.jini.JiniERIDSTWrapper | 5 |
| gipsy.GEE.multitier.GMT.GMTWrapper | 7 |
| gipsy.GIPC.DFG.DFGAnalyzer.DFGParser | 8 |
| gipsy.GIPC.DFG.DFGGenerator.DFGCodeGenerator | 6 |
| gipsy.GIPC.DFG.DFGGenerator.DFGTranCodeGenerator | 6 |
| gipsy.GIPC.Preprocessing.PreprocessorParser | 8 |
| gipsy.GIPC.SemanticAnalyzer | 6 |
| gipsy.GIPC.imperative.Java.JavaCompiler | 3 |
| gipsy.GIPC.intensional.GIPL.GIPLParser | 8 |
| gipsy.GIPC.intensional.GenericTranslator.TranslationParser | 6 |
| gipsy.GIPC.intensional.SIPL.ForensicLucid.ForensicLucidParser | 8 |
| gipsy.GIPC.intensional.SIPL.ForensicLucid.ForensicLucidSemanticAnalyzer | 7 |
| gipsy.GIPC.intensional.SIPL.IndexicalLucid.IndexicalLucidParser | 8 |
| gipsy.GIPC.intensional.SIPL.JLucid.JGIPLParser | 8 |
| gipsy.GIPC.intensional.SIPL.JLucid.JIndexicalLucidParser | 8 |
| gipsy.GIPC.intensional.SIPL.JOOIP.JavaParser | 8 |
| gipsy.GIPC.intensional.SIPL.JOOIP.JavaParserTokenManager | 7 |
| gipsy.GIPC.intensional.SIPL.JOOIP.ast.visitor.DumpVisitor | 6 |
| gipsy.GIPC.intensional.SIPL.Lucx.LucxParser | 8 |
| *gipsy.GIPC.intensional.SIPL.ObjectiveLucid.ObjectiveGIPLParser* | 8 |
| gipsy.GIPC.intensional.SIPL.ObjectiveLucid.ObjectiveIndexicalLucidParser | 8 |
| *gipsy.RIPE.editors.RunTimeGraphEditor.ui.GIPSYGMTOperator* | 9 |
| gipsy.apps.memocode.genome.AlignDGT | 3 |
| gipsy.apps.memocode.genome.AlignDWT | 2 |
| gipsy.tests.GEE.simulator.DGTSimulator | 3 |
| gipsy.tests.GIPC.intensional.SIPL.Lucx.SemanticTest.LucxSemanticAnalyzer | 7 |

Table 24

CHANGEABILITY

Criteria : CHANGEABILITYc : FAIR

| Classes name | Ranking |
|---|---|
| gipsy.GEE.GEE | 4 |
| gipsy.GEE.IDP.DemandGenerator.jini.rmi.JINITA | 6 |
| gipsy.GEE.IDP.DemandGenerator.jini.rmi.JINITransportAgent | 4 |
| gipsy.GEE.IDP.DemandGenerator.jini.rmi.JiniDemandDispatcher | 6 |
| gipsy.GEE.IDP.DemandGenerator.jms.JMSTransportAgent | 4 |
| gipsy.GEE.IDP.demands.Demand | 8 |
| gipsy.GEE.IVW.Warehouse.Cache | 3 |
| gipsy.GEE.multitier.DST.jini.JiniDSTWrapper | 5 |
| gipsy.GEE.multitier.DST.jini.JiniERIDSTWrapper | 5 |
| gipsy.GEE.multitier.GIPSYNode | 6 |
| gipsy.GEE.multitier.GMT.GMTWrapper | 7 |
| gipsy.GIPC.DFG.DFGAnalyzer.LucidCodeGenerator | 5 |
| gipsy.GIPC.DFG.DFGGenerator.DFGCodeGenerator | 6 |
| gipsy.GIPC.DFG.DFGGenerator.DFGTranCodeGenerator | 6 |
| gipsy.GIPC.GIPC | 7 |
| gipsy.GIPC.SemanticAnalyzer | 6 |
| gipsy.GIPC.intensional.GenericTranslator.TranslationLexer | 4 |
| gipsy.GIPC.intensional.SIPL.ForensicLucid.ForensicLucidSemanticAnalyzer | 7 |
| gipsy.GIPC.intensional.SIPL.JOOIP.ast.visitor.DumpVisitor | 6 |
| gipsy.RIPE.editors.RunTimeGraphEditor.core.GIPSYTier | 4 |
| gipsy.RIPE.editors.RunTimeGraphEditor.core.GraphDataManager | 4 |
| gipsy.RIPE.editors.RunTimeGraphEditor.ui.InstancesNodesPanel | 3 |
| gipsy.RIPE.editors.RunTimeGraphEditor.ui.MapEditor | 4 |
| gipsy.RIPE.editors.RunTimeGraphEditor.ui.dialogs.GIPSYNodeDialog | 4 |
| gipsy.RIPE.editors.RunTimeGraphEditor.ui.dialogs.TierPropertyDialog | 3 |
| gipsy.RIPE.editors.WebEditor.WebEditor | 4 |
| gipsy.lang.GIPSYContext | 7 |
| gipsy.tests.GEE.multitier.GMT.GMTTestConsole | 4 |
| gipsy.tests.GEE.simulator.DGTDialog | 4 |
| gipsy.tests.GEE.simulator.ProfileDialog | 4 |
| gipsy.tests.GEE.simulator.jini.WorkerJTA | 3 |
| gipsy.tests.GIPC.intensional.SIPL.Lucx.SemanticTest.LucxSemanticAnalyzer | 7 |
| gipsy.tests.Regression | 4 |

Table 25

Criteria : CHANGEABILITYc : POOR

| Classes name | Ranking |
|---|---|
| gipsy.GIPC.DFG.DFGAnalyzer.DFGParser | 8 |
| gipsy.GIPC.DFG.DFGAnalyzer.DFGParserTokenManager | 6 |
| gipsy.GIPC.Preprocessing.PreprocessorParser | 8 |
| gipsy.GIPC.Preprocessing.PreprocessorParserTokenManager | 6 |
| gipsy.GIPC.intensional.GIPL.GIPLParser | 8 |
| gipsy.GIPC.intensional.GIPL.GIPLParserTokenManager | 6 |
| gipsy.GIPC.intensional.GenericTranslator.TranslationParser | 6 |
| gipsy.GIPC.intensional.SIPL.ForensicLucid.ForensicLucidParser | 8 |
| gipsy.GIPC.intensional.SIPL.ForensicLucid.ForensicLucidParserTokenManager | 6 |
| gipsy.GIPC.intensional.SIPL.IndexicalLucid.IndexicalLucidParser | 8 |
| gipsy.GIPC.intensional.SIPL.IndexicalLucid.IndexicalLucidParserTokenManager | 6 |
| gipsy.GIPC.intensional.SIPL.JLucid.JGIPLParser | 8 |
| gipsy.GIPC.intensional.SIPL.JLucid.JGIPLParserTokenManager | 6 |
| gipsy.GIPC.intensional.SIPL.JLucid.JIndexicalLucidParser | 8 |
| gipsy.GIPC.intensional.SIPL.JLucid.JIndexicalLucidParserTokenManager | 6 |
| gipsy.GIPC.intensional.SIPL.JOOIP.JavaCharStream | 7 |
| gipsy.GIPC.intensional.SIPL.JOOIP.JavaParser | 8 |
| gipsy.GIPC.intensional.SIPL.JOOIP.JavaParserTokenManager | 7 |
| gipsy.GIPC.intensional.SIPL.Lucx.LucxParser | 8 |
| gipsy.GIPC.intensional.SIPL.Lucx.LucxParserTokenManager | 6 |
| *gipsy.GIPC.intensional.SIPL.ObjectiveLucid.ObjectiveGIPLParser* | 8 |
| gipsy.GIPC.intensional.SIPL.ObjectiveLucid.ObjectiveGIPLParserTokenManager | 6 |
| gipsy.GIPC.intensional.SIPL.ObjectiveLucid.ObjectiveIndexicalLucidParser | 8 |
| gipsy.GIPC.intensional.SIPL.ObjectiveLucid.ObjectiveIndexicalLucidParserTokenManager | 6 |
| gipsy.GIPC.util.SimpleCharStream | 8 |
| gipsy.RIPE.editors.RunTimeGraphEditor.core.GlobalInstance | 6 |
| *gipsy.RIPE.editors.RunTimeGraphEditor.ui.GIPSYGMTOperator* | 9 |
| gipsy.tests.junit.lang.GIPSYContextTest | 4 |
| gipsy.tests.junit.lang.context.GIPSYContextTest | 4 |

Table 26

## STABILITY

Criteria : STABILITYc : FAIR

| Classes name | Ranking |
|---|---|
| gipsy.GEE.IDP.DMSException | 2 |
| gipsy.GEE.IDP.DemandDispatcher.DemandDispatcherException | 3 |
| gipsy.GEE.IDP.DemandGenerator.DemandGenerator | 3 |
| gipsy.GEE.IDP.DemandGenerator.jini.rmi.JINITA | 6 |
| gipsy.GEE.IDP.DemandGenerator.jini.rmi.JiniDemandDispatcher | 6 |
| gipsy.GEE.IDP.DemandGenerator.jms.JMSTransportAgent | 7 |
| gipsy.GEE.IDP.DemandGenerator.threaded.IdentifierContext | 2 |
| gipsy.GEE.IDP.demands.DemandSignature | 2 |
| gipsy.GEE.IDP.demands.DemandState | 3 |
| gipsy.GEE.IDP.demands.DemandType | 3 |
| gipsy.GEE.IDP.demands.IDemand | 3 |
| gipsy.GEE.IDP.demands.ProceduralDemand | 3 |
| gipsy.GEE.IDP.demands.SystemDemand | 3 |
| gipsy.GEE.IDP.demands.TimeLine | 2 |
| gipsy.GEE.multitier.GIPSYNode | 6 |
| gipsy.GEE.multitier.GMT.GMTWrapper | 7 |
| gipsy.GEE.multitier.GenericTierWrapper | 2 |
| gipsy.GEE.multitier.IMultiTierWrapper | 2 |
| gipsy.GIPC.DFG.DFGAnalyzer.DFGParser | 8 |
| gipsy.GIPC.GIPCException | 2 |
| gipsy.GIPC.Preprocessing.PreprocessorParser | 8 |
| gipsy.GIPC.imperative.CommunicationProcedureGenerator.CommunicationProcedure | 2 |
| gipsy.GIPC.imperative.ImperativeCompiler | 5 |
| gipsy.GIPC.intensional.GIPL.GIPLParser | 8 |
| gipsy.GIPC.intensional.IntensionalCompiler | 4 |
| gipsy.GIPC.intensional.SIPL.ForensicLucid.ForensicLucidParser | 8 |
| gipsy.GIPC.intensional.SIPL.IndexicalLucid.IndexicalLucidParser | 8 |
| gipsy.GIPC.intensional.SIPL.IndexicalLucid.IndexicalLucidParserTreeConstants | 4 |
| gipsy.GIPC.intensional.SIPL.JLucid.JGIPLParser | 8 |
| gipsy.GIPC.intensional.SIPL.JLucid.JIndexicalLucidParser | 8 |
| gipsy.GIPC.intensional.SIPL.JOOIP.JavaCharStream | 7 |
| gipsy.GIPC.intensional.SIPL.JOOIP.JavaParser | 8 |
| gipsy.GIPC.intensional.SIPL.JOOIP.Token | 3 |
| gipsy.GIPC.intensional.SIPL.JOOIP.ast.body.BodyDeclaration | 2 |
| gipsy.GIPC.intensional.SIPL.JOOIP.ast.body.VariableDeclaratorId | 3 |
| gipsy.GIPC.intensional.SIPL.JOOIP.ast.expr.Expression | 2 |
| gipsy.GIPC.intensional.SIPL.JOOIP.ast.expr.NameExpr | 3 |
| gipsy.GIPC.intensional.SIPL.JOOIP.ast.stmt.BlockStmt | 3 |
| gipsy.GIPC.intensional.SIPL.JOOIP.ast.stmt.Statement | 2 |
| gipsy.GIPC.intensional.SIPL.JOOIP.ast.type.Type | 2 |
| gipsy.GIPC.intensional.SIPL.Lucx.LucxParser | 8 |
| *gipsy.GIPC.intensional.SIPL.ObjectiveLucid.ObjectiveGIPLParser* | 8 |
| gipsy.GIPC.intensional.SIPL.ObjectiveLucid.ObjectiveIndexicalLucidParser | 8 |
| gipsy.GIPC.util.ParseException | 2 |
| gipsy.GIPC.util.Token | 3 |
| gipsy.GIPC.util.TokenMgrError | 2 |
| gipsy.RIPE.editors.RunTimeGraphEditor.core.AppConstants | 3 |
| gipsy.RIPE.editors.RunTimeGraphEditor.core.GIPSYPhysicalNode | 3 |
| gipsy.RIPE.editors.RunTimeGraphEditor.core.GIPSYTier | 4 |
| gipsy.RIPE.editors.RunTimeGraphEditor.core.GlobalInstance | 6 |
| gipsy.RIPE.editors.RunTimeGraphEditor.ui.AppLogger | 2 |
| *gipsy.RIPE.editors.RunTimeGraphEditor.ui.GIPSYGMTOperator* | 9 |
| gipsy.interfaces.GIPSYProgram | 4 |
| gipsy.lang.GIPSYInteger | 4 |
| gipsy.storage.DictionaryItem | 3 |
| gipsy.tests.GEE.simulator.GlobalDef | 4 |
| gipsy.util.GIPSYException | 2 |
| gipsy.util.NetUtils | 2 |

Table 27

Criteria : STABILITYc : POOR

| Classes name | Ranking |
|---|---|
| gipsy.Configuration | 5 |
| gipsy.GEE.CONFIG | 3 |
| gipsy.GEE.IDP.ITransportAgent | 3 |
| gipsy.GEE.IDP.demands.Demand | 8 |
| gipsy.GEE.multitier.DGT.DGTWrapper | 5 |
| gipsy.GEE.multitier.DST.DSTWrapper | 3 |
| gipsy.GEE.multitier.DWT.DWTWrapper | 6 |
| gipsy.GIPC.GIPC | 7 |
| gipsy.GIPC.intensional.GIPL.GIPLParserTreeConstants | 5 |
| gipsy.GIPC.intensional.SIPL.JOOIP.ast.Node | 4 |
| gipsy.GIPC.intensional.SIPL.JOOIP.ast.body.TypeDeclaration | 4 |
| gipsy.GIPC.intensional.SIPL.JOOIP.ast.expr.StringLiteralExpr | 4 |
| gipsy.GIPC.intensional.SimpleNode | 3 |
| gipsy.GIPC.util.SimpleCharStream | 8 |
| gipsy.lang.GIPSYContext | 7 |
| gipsy.lang.GIPSYType | 5 |
| gipsy.lang.context.TagSet | 4 |

TESTABILITY

Criteria : TESTABILITYc : FAIR

| Classes name | Rank |
|---|---|
| gipsy.Configuration | 5 |
| gipsy.GEE.IDP.DemandGenerator.jini.rmi.JINITA | 6 |
| gipsy.GEE.IDP.DemandGenerator.jini.rmi.JiniDemandDispatcher | 6 |
| gipsy.GEE.IDP.DemandGenerator.jms.JMSTransportAgent | 7 |
| gipsy.GEE.IDP.demands.Demand | 8 |
| gipsy.GEE.multitier.GMT.GMTWrapper | 7 |
| gipsy.GIPC.DFG.DFGAnalyzer.DFGParserTokenManager | 6 |
| gipsy.GIPC.DFG.DFGGenerator.DFGCodeGenerator | 6 |
| gipsy.GIPC.DFG.DFGGenerator.DFGTranCodeGenerator | 6 |
| gipsy.GIPC.Preprocessing.PreprocessorParserTokenManager | 6 |
| gipsy.GIPC.SemanticAnalyzer | 6 |
| gipsy.GIPC.intensional.GIPL.GIPLParserTokenManager | 6 |
| gipsy.GIPC.intensional.SIPL.ForensicLucid.ForensicLucidParserTokenManager | 6 |
| gipsy.GIPC.intensional.SIPL.ForensicLucid.ForensicLucidSemanticAnalyzer | 7 |
| gipsy.GIPC.intensional.SIPL.IndexicalLucid.IndexicalLucidParserTokenManager | 6 |
| gipsy.GIPC.intensional.SIPL.JLucid.JGIPLParserTokenManager | 6 |
| gipsy.GIPC.intensional.SIPL.JLucid.JIndexicalLucidParserTokenManager | 6 |
| gipsy.GIPC.intensional.SIPL.JOOIP.JavaCharStream | 7 |
| gipsy.GIPC.intensional.SIPL.JOOIP.ast.visitor.GenericVisitor | 4 |
| gipsy.GIPC.intensional.SIPL.JOOIP.ast.visitor.VoidVisitor | 4 |
| gipsy.GIPC.intensional.SIPL.Lucx.LucxParserTokenManager | 6 |
| gipsy.GIPC.intensional.SIPL.ObjectiveLucid.ObjectiveGIPLParserTokenManager | 6 |
| gipsy.GIPC.intensional.SIPL.ObjectiveLucid.ObjectiveIndexicalLucidParserTokenManager | 6 |
| gipsy.GIPC.util.SimpleCharStream | 8 |
| gipsy.RIPE.editors.RunTimeGraphEditor.core.GlobalInstance | 6 |
| gipsy.lang.GIPSYContext | 7 |
| gipsy.lang.GIPSYInteger | 4 |
| gipsy.tests.GIPC.intensional.SIPL.Lucx.SemanticTest.LucxSemanticAnalyzer | 7 |

Criteria : TESTABILITYc : POOR

| Classes name | Ranking |
|---|---|
| gipsy.GIPC.DFG.DFGAnalyzer.DFGParser | 8 |
| gipsy.GIPC.Preprocessing.PreprocessorParser | 8 |
| gipsy.GIPC.intensional.GIPL.GIPLParser | 8 |
| gipsy.GIPC.intensional.GenericTranslator.TranslationParser | 6 |
| gipsy.GIPC.intensional.SIPL.ForensicLucid.ForensicLucidParser | 8 |
| gipsy.GIPC.intensional.SIPL.IndexicalLucid.IndexicalLucidParser | 8 |
| gipsy.GIPC.intensional.SIPL.JLucid.JGIPLParser | 8 |
| gipsy.GIPC.intensional.SIPL.JLucid.JIndexicalLucidParser | 8 |
| gipsy.GIPC.intensional.SIPL.JOOIP.JavaParser | 8 |
| gipsy.GIPC.intensional.SIPL.JOOIP.JavaParserTokenManager | 7 |
| gipsy.GIPC.intensional.SIPL.JOOIP.ast.visitor.DumpVisitor | 6 |
| gipsy.GIPC.intensional.SIPL.Lucx.LucxParser | 8 |
| *gipsy.GIPC.intensional.SIPL.ObjectiveLucid.ObjectiveGIPLParser* | 8 |
| gipsy.GIPC.intensional.SIPL.ObjectiveLucid.ObjectiveIndexicalLucidParser | 8 |
| *gipsy.RIPE.editors.RunTimeGraphEditor.ui.GIPSYGMTOperator* | 9 |

Table 28

Prioritized list of worst classes for Marf and Gipsy

Here the classes are prioritized based on the highest ranking of worst classes in both factor and criteria level classes which are mentioned above. The top two classes in the priority list to compare the quality of classes in both Marf and Gipsy are considered. Ranking being in the scale 1-10 where 10 being worse than consecutive ranking of lower numbers.

MARF

| Ranking of Classes based on worst quality (MARF) | |
|---|---|
| Name of Class | Priority |
| marf.Classification.NeuralNetwork.NeuralNetwork | 1 |
| marf.MARF | 2 |
| marf.Storage.StorageManager | 3 |
| marf.math.Matrix | 4 |
| marf.util.Arrays | 5 |
| marf.nlp.Parsing.GrammarCompiler.Grammar | 6 |
| marf.nlp.Parsing.GrammarCompiler.GrammarCompiler | 7 |
| marf.nlp.Storage.Corpus | 8 |
| marf.Classification.Stochastic.ZipfLaw | 9 |
| marf.Stats.StatisticalEstimators.StatisticalEstimator | 10 |
| marf.Configuration | 11 |
| marf.Storage.ResultSet | 12 |
| marf.math.ComplexMatrix | 13 |
| marf.Preprocessing.FFTFilter.FFTFilter | 14 |
| marf.nlp.Parsing.SyntaxError | 15 |
| marf.nlp.Parsing.LexicalAnalyzer | 16 |
| marf.Storage.Sample | 17 |
| marf.math.Vector | 18 |

GIPSY

| Ranking of Classes based on worst quality (GIPSY) | |
|---|---|
| Name of Class | Priority |
| | |
| *gipsy.RIPE.editors.RunTimeGraphEditor.ui.GIPSYGMTOperator* | 1 |
| *gipsy.GIPC.intensional.SIPL.ObjectiveLucid.ObjectiveGIPLParser* | 2 |
| gipsy.GIPC.intensional.SIPL.Lucx.LucxParser | 3 |
| gipsy.GIPC.intensional.SIPL.JOOIP.JavaParser | 4 |
| gipsy.GIPC.intensional.SIPL.JOOIP.JavaParserTokenManager | 5 |
| gipsy.GIPC.intensional.SIPL.JLucid.JIndexicalLucidParser | 6 |
| gipsy.GIPC.intensional.SIPL.JLucid.JGIPLParser | 7 |
| gipsy.GIPC.intensional.SIPL.IndexicalLucid.IndexicalLucidParser | 8 |
| gipsy.GIPC.intensional.SIPL.ForensicLucid.ForensicLucidParser | 9 |

Table29

| | |
|---|---|
| gipsy.GIPC.intensional.SIPL.ObjectiveLucid.ObjectiveIndexicalLucidParser | 10 |
| gipsy.GIPC.intensional.GIPL.GIPLParser | 11 |
| gipsy.GIPC.Preprocessing.PreprocessorParser | 12 |
| gipsy.GIPC.DFG.DFGAnalyzer.DFGParser | 13 |
| gipsy.GIPC.util.SimpleCharStream | 14 |
| gipsy.GIPC.intensional.GenericTranslator.TranslationParser | 15 |
| gipsy.GEE.IDP.demands.Demand | 16 |
| gipsy.GEE.IDP.DemandGenerator.jms.JMSTransportAgent | 17 |
| gipsy.GEE.multitier.GMT.GMTWrapper | 18 |
| gipsy.GIPC.intensional.SIPL.ObjectiveLucid.ObjectiveIndexicalLucidParserTokenManager | 19 |
| gipsy.GIPC.GIPC | 20 |
| gipsy.GIPC.intensional.SIPL.ForensicLucid.ForensicLucidSemanticAnalyzer | 21 |
| gipsy.GIPC.intensional.SIPL.JOOIP.JavaCharStream | 22 |
| gipsy.lang.GIPSYContext | 23 |
| gipsy.tests.GIPC.intensional.SIPL.Lucx.SemanticTest.LucxSemanticAnalyzer | 24 |
| gipsy.GIPC.DFG.DFGAnalyzer.DFGParserTokenManager | 25 |
| gipsy.GIPC.DFG.DFGGenerator.DFGCodeGenerator | 26 |
| gipsy.GIPC.DFG.DFGGenerator.DFGTranCodeGenerator | 27 |
| gipsy.GIPC.Preprocessing.PreprocessorParserTokenManager | 28 |
| gipsy.GIPC.SemanticAnalyzer | 29 |
| gipsy.GIPC.intensional.GIPL.GIPLParserTokenManager | 30 |
| gipsy.GIPC.intensional.SIPL.ForensicLucid.ForensicLucidParserTokenManager | 31 |
| gipsy.GIPC.intensional.SIPL.IndexicalLucid.IndexicalLucidParserTokenManager | 32 |
| gipsy.GIPC.intensional.SIPL.JLucid.JGIPLParserTokenManager | 33 |
| gipsy.GIPC.intensional.SIPL.JLucid.JIndexicalLucidParserTokenManager | 34 |
| gipsy.GIPC.intensional.SIPL.JOOIP.ast.visitor.DumpVisitor | 35 |
| gipsy.GIPC.intensional.SIPL.Lucx.LucxParserTokenManager | 36 |
| gipsy.GIPC.intensional.SIPL.ObjectiveLucid.ObjectiveGIPLParserTokenManager | 37 |
| gipsy.RIPE.editors.RunTimeGraphEditor.core.GlobalInstance | 38 |
| gipsy.Configuration | 39 |
| gipsy.GEE.IDP.DemandGenerator.jini.rmi.JINITA | 40 |
| gipsy.GEE.IDP.DemandGenerator.jini.rmi.JiniDemandDispatcher | 41 |
| gipsy.GEE.multitier.GIPSYNode | 42 |
| gipsy.GEE.IDP.DemandGenerator.LegacyEductiveInterpreter | 43 |
| gipsy.GEE.multitier.DST.jini.JiniDSTWrapper | 44 |
| gipsy.GEE.multitier.DST.jini.JiniERIDSTWrapper | 45 |
| gipsy.GIPC.imperative.Java.JavaCompiler | 46 |
| gipsy.apps.memocode.genome.AlignDGT | 47 |
| gipsy.apps.memocode.genome.AlignDWT | 48 |
| gipsy.tests.GEE.simulator.DGTSimulator | 49 |
| gipsy.tests.junit.lang.GIPSYContextTest | 50 |
| gipsy.tests.junit.lang.context.GIPSYContextTest | 51 |
| gipsy.GEE.CONFIG | 52 |
| gipsy.GEE.IDP.ITransportAgent | 53 |
| gipsy.GEE.multitier.DGT.DGTWrapper | 54 |
| gipsy.GEE.multitier.DST.DSTWrapper | 55 |
| gipsy.GEE.multitier.DWT.DWTWrapper | 56 |
| gipsy.GIPC.intensional.GIPL.GIPLParserTreeConstants | 57 |
| gipsy.GIPC.intensional.SIPL.JOOIP.ast.Node | 58 |
| gipsy.GIPC.intensional.SIPL.JOOIP.ast.body.TypeDeclaration | 59 |
| gipsy.GIPC.intensional.SIPL.JOOIP.ast.expr.StringLiteralExpr | 60 |
| gipsy.GIPC.intensional.SimpleNode | 61 |
| gipsy.lang.GIPSYType | 62 |
| gipsy.lang.context.TagSet | 63 |

Table 30

Metric data and kiviat graph for the worst classes for Marf and Gipsy

MARF
1. marf.Classification.NeuralNetwork.NeuralNetwork
Class Metric level measurement data

| Metric : marf.Classification.NeuralNetwork.NeuralNetwork | Value | Min | Max | Status |
|---|---|---|---|---|
| cl_comf: Class comment rate | 0.26 | 0.20 | +oo | 0 |
| cl_comm: Number of lines of comment | 354 | -oo | +oo | 0 |
| cl_data: Total number of attributes | 17 | 0 | 7 | -1 |
| cl_data_publ: Number of public attributes | 8 | 0 | 0 | -1 |
| cl_func: Total number of methods | 27 | 0 | 25 | -1 |
| cl_func_publ: Number of public methods | 21 | 0 | 15 | -1 |
| cl_line: Number of lines | 1348 | -oo | +oo | 0 |
| cl_stat: Number of statements | 372 | 0 | 100 | -1 |
| cl_wmc: Weighted Methods per Class | 115 | 0 | 60 | -1 |
| cu_cdused: Number of direct used classes | 33 | 0 | 10 | -1 |
| cu_cdusers: Number of direct users classes | 3 | 0 | 5 | 0 |
| in_bases: Number of base classes | 6 | 0 | 3 | -1 |
| in_noc: Number of children | 0 | 0 | 3 | 0 |

**Kiviat Diagram for marf.Classification.NeuralNetwork.NeuralNetwork**

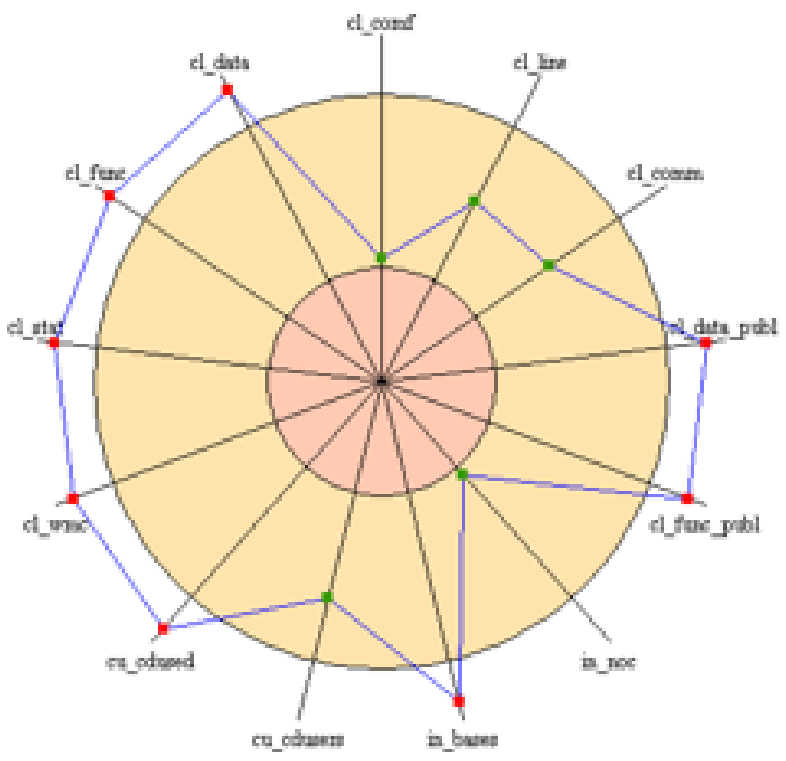

*Figure 19 Metrics and Kiviat graph for class marf.Classification.NeuralNetwork.NeuralNetwork.java is taken from Kalimetrix Logiscope tool*

2. marf.MARF

Class Metric level measurement data:

| Metric : marf.MARF | Value | Min | Max | Status |
|---|---|---|---|---|
| cl_comf: Class comment rate | 0.49 | 0.20 | +oo | 0 |
| cl_comm: Number of lines of comment | 932 | -oo | +oo | 0 |
| cl_data: Total number of attributes | 84 | 0 | 7 | -1 |
| cl_data_publ: Number of public attributes | 64 | 0 | 0 | -1 |
| cl_func: Total number of methods | 58 | 0 | 25 | -1 |
| cl_func_publ: Number of public methods | 55 | 0 | 15 | -1 |
| cl_line: Number of lines | 1892 | -oo | +oo | 0 |
| cl_stat: Number of statements | 219 | 0 | 100 | -1 |
| cl_wmc: Weighted Methods per Class | 82 | 0 | 60 | -1 |
| cu_cdused: Number of direct used classes | 26 | 0 | 10 | -1 |
| cu_cdusers: Number of direct users classes | 20 | 0 | 5 | -1 |
| in_bases: Number of base classes | 0 | 0 | 3 | 0 |
| in_noc: Number of children | 0 | 0 | 3 | 0 |

**Kiviat Diagram for marf.MARF**

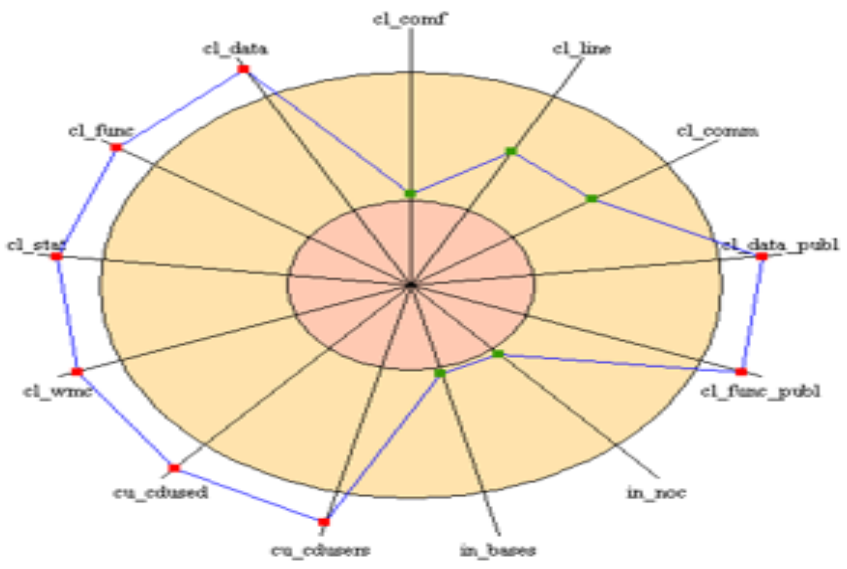

*Figure 20 Metrics and Kiviat graph for class marf.MARF .java is taken from Kalimetrix Logiscope tool*

GIPSY:

gipsy.RIPE.editors.RunTimeGraphEditor.ui.GIPSYGMTOperator

**Class Metric Level measurement data:**

| Metric : gipsy.RIPE.editors.RunTimeGraphEditor.ui.GIPSYGMTOperator | Value | Min | Max | Status |
|---|---|---|---|---|
| cl_comf: Class comment rate | 0.19 | 0.20 | +oo | -1 |
| cl_comm: Number of lines of comment | 160 | -oo | +oo | 0 |
| cl_data: Total number of attributes | 44 | 0 | 7 | -1 |
| cl_data_publ: Number of public attributes | 8 | 0 | 0 | -1 |
| cl_func: Total number of methods | 37 | 0 | 25 | -1 |
| cl_func_publ: Number of public methods | 19 | 0 | 15 | -1 |
| cl_line: Number of lines | 859 | -oo | +oo | 0 |
| cl_stat: Number of statements | 311 | 0 | 100 | -1 |
| cl_wmc: Weighted Methods per Class | 69 | 0 | 60 | -1 |
| cu_cdused: Number of direct used classes | 51 | 0 | 10 | -1 |
| cu_cdusers: Number of direct users classes | 2 | 0 | 5 | 0 |
| in_bases: Number of base classes | 6 | 0 | 3 | -1 |
| in_noc: Number of children | 0 | 0 | 3 | 0 |

**Kiviat Diagram for gipsy.RIPE.editors.RunTimeGraphEditor.ui.GIPSYGMTOperator**

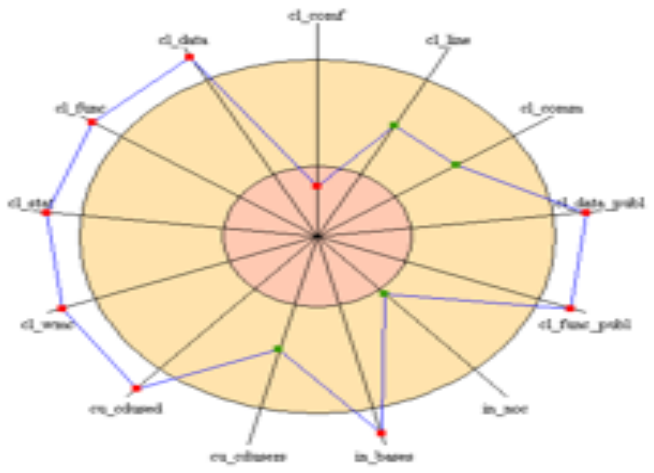

*Figure 21 Metrics and Kiviat graph for class gipsy.RIPE.editors.RunTimeGraphEditor.ui.GIPSYGMTOperator is taken from Kalimetrix Logiscope tool*

## 2 gipsy.GIPC.intensional.SIPL.ObjectiveLucid.ObjectiveGIPLParser

| Metric : gipsy.GIPC.intensional.SIPL.ObjectiveLucid.ObjectiveGIPLP... | Value | Min | Max | Status |
|---|---|---|---|---|
| cl_comf: Class comment rate | 0.04 | 0.20 | +oo | -1 |
| cl_comm: Number of lines of comment | 82 | -oo | +oo | 0 |
| cl_data: Total number of attributes | 25 | 0 | 7 | -1 |
| cl_data_publ: Number of public attributes | 4 | 0 | 0 | -1 |
| cl_func: Total number of methods | 143 | 0 | 25 | -1 |
| cl_func_publ: Number of public methods | 27 | 0 | 15 | -1 |
| cl_line: Number of lines | 2016 | -oo | +oo | 0 |
| cl_stat: Number of statements | 1595 | 0 | 100 | -1 |
| cl_wmc: Weighted Methods per Class | 605 | 0 | 60 | -1 |
| cu_cdused: Number of direct used classes | 20 | 0 | 10 | -1 |
| cu_cdusers: Number of direct users classes | 1 | 0 | 5 | 0 |
| in_bases: Number of base classes | 2 | 0 | 3 | 0 |
| in_noc: Number of children | 0 | 0 | 3 | 0 |

**Kiviat Diagram for gipsy.GIPC.intensional.SIPL.ObjectiveLucid.ObjectiveGIPLParser**

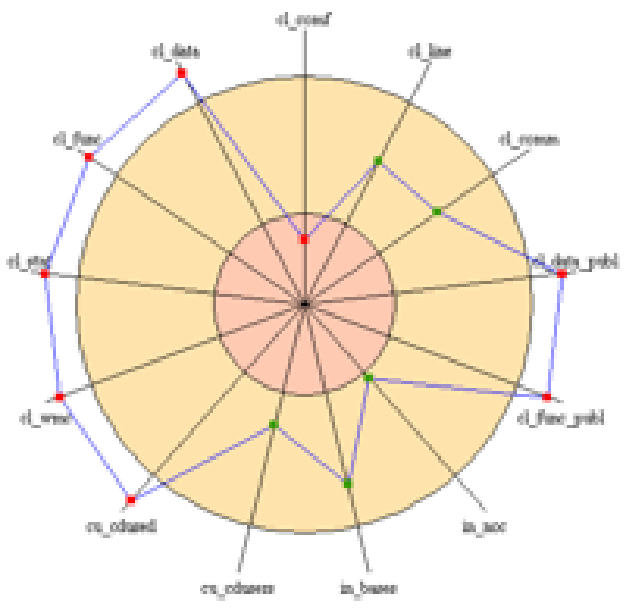

*Figure 22 Metrics and Kiviat graph for class gipsy.GIPC.intensional.SIPL.ObjectiveLucid.ObjectiveGIPLParser is taken from Kalimetrix Logiscope tool*

Comparing the quality of worst classes:

| | Class name | Ranks as per metric status |
|---|---|---|
| MC1. | marf.Classification.NeuralNetwork.NeuralNetwork | 8 |
| MC2. | marf.MARF | 8 |
| GC1. | gipsy.RIPE.editors.RunTimeGraphEditor.ui.GIPSYGMTOperator | 9 |
| GC2. | gipsy.GIPC.intensional.SIPL.ObjectiveLucid.ObjectiveGIPLParser | 8 |

Table 31

The above classes are used to compare the quality of the classes. The status of each metric is decided on its value provided by each metric based on maximum and minimum values of the metrics. If the value is not in the range of minimum and maximum value then the status will be -1 which is problematic. These problematic metrics is represented as red mark in the Kiviat graph. The big circle is considered for maximum values and small circle for minimum values.

Here for marf classes MC1 and MC2 the class comment rate value is in the range of minimum and maximum values but for gipsy classes GC1 and GC2 the class comment rate value is not in the range. The values for Number of lines of comments, Number of lines and Number of children for all the classes is in the range. The value for Total number of attributes, Number of Public attributes, Total number of methods, Total number of public methods, Number of statements, Weighted methods per class, Number of direct used classes for all the classes is not in the range which is to be improved. The value for MC2 for Number of direct user classes is not in the range and for MC1, GC1 and GC2. The value for MC1 and GC1 for Number of base classes is not in the range.

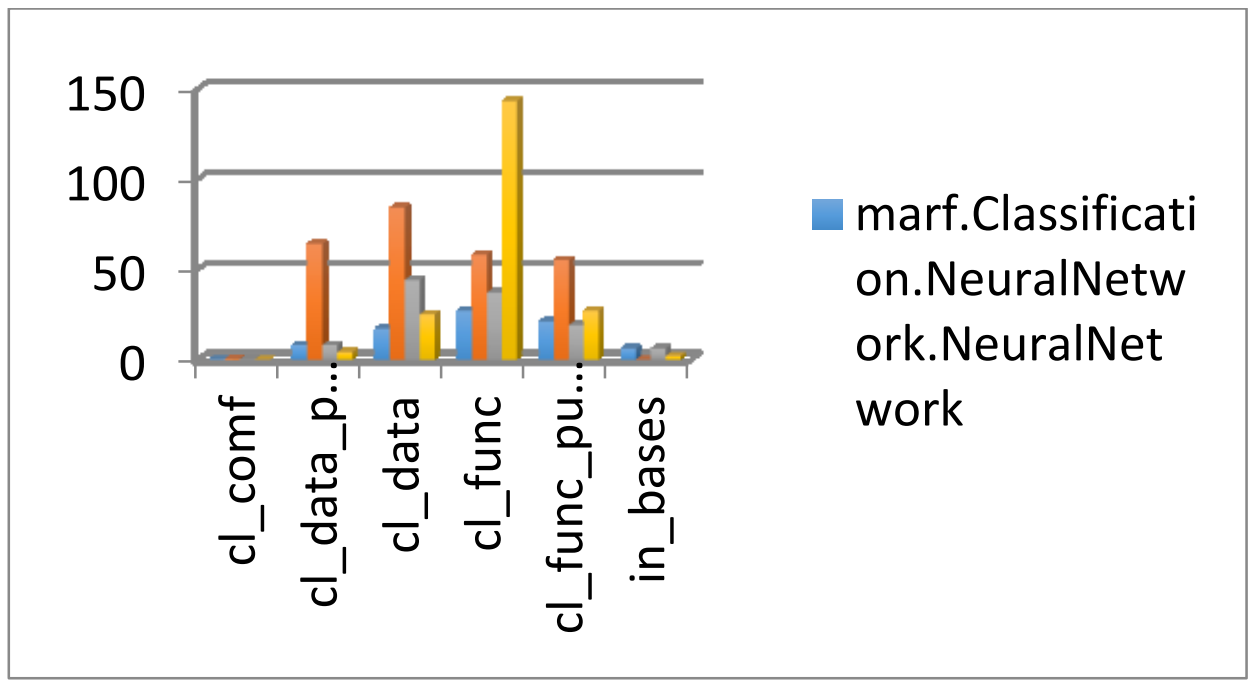

**Comparison of the chosen worst classes**

*Figure 23 Graph for comparision of chosen worst classes*

Recommendations

After careful assessment of lines of source code there are few recommendations noticed where changes can be made in different classes.

For instance let us consider a poor class for GIPSY

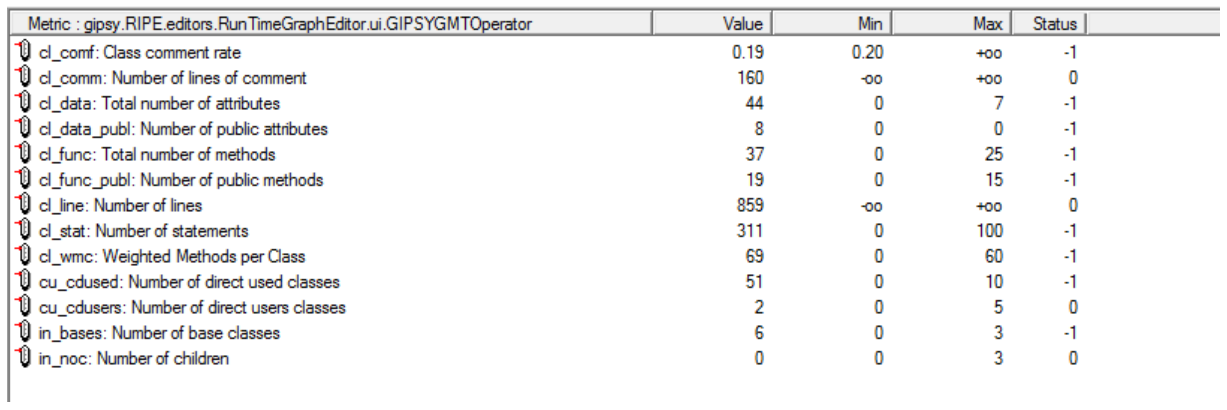

| Metric : gipsy.RIPE.editors.RunTimeGraphEditor.ui.GIPSYGMTOperator | Value | Min | Max | Status |
|---|---|---|---|---|
| cl_comf: Class comment rate | 0.19 | 0.20 | +oo | -1 |
| cl_comm: Number of lines of comment | 160 | -oo | +oo | 0 |
| cl_data: Total number of attributes | 44 | 0 | 7 | -1 |
| cl_data_publ: Number of public attributes | 8 | 0 | 0 | -1 |
| cl_func: Total number of methods | 37 | 0 | 25 | -1 |
| cl_func_publ: Number of public methods | 19 | 0 | 15 | -1 |
| cl_line: Number of lines | 859 | -oo | +oo | 0 |
| cl_stat: Number of statements | 311 | 0 | 100 | -1 |
| cl_wmc: Weighted Methods per Class | 69 | 0 | 60 | -1 |
| cu_cdused: Number of direct used classes | 51 | 0 | 10 | -1 |
| cu_cdusers: Number of direct users classes | 2 | 0 | 5 | 0 |
| in_bases: Number of base classes | 6 | 0 | 3 | -1 |
| in_noc: Number of children | 0 | 0 | 3 | 0 |

*Figure 24 Metrics and Kiviat graph for class gipsy.RIPE.editors.RunTimeGraphEditor.ui.GIPSYGMTOperator is taken from Kalimetrix Logiscope tool*

The figure above shows different metrics and corresponding values for each metric for this particular class.

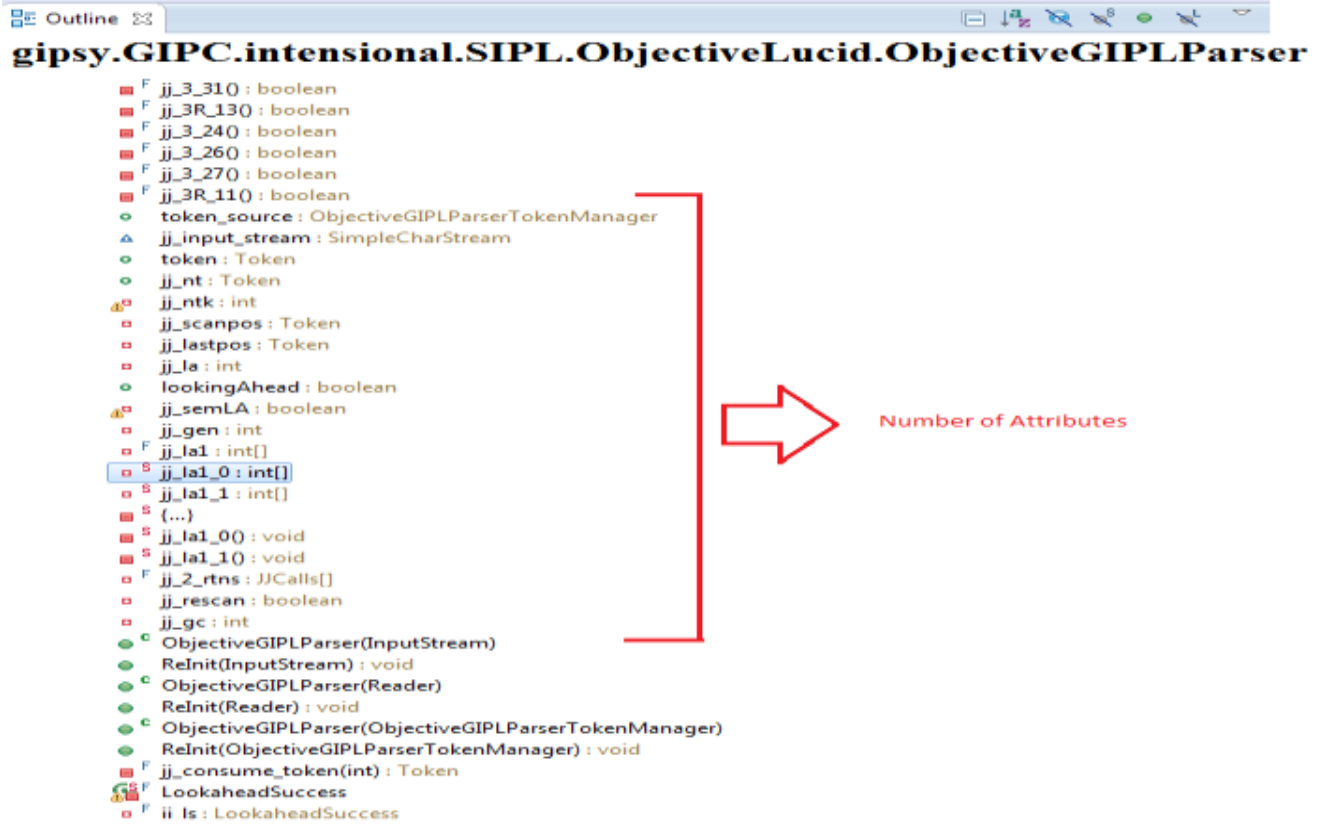

*Figure 25 Outline for class gipsy.GIPC.intensional.SIPL.ObjectiveLucid.ObjectiveGIPLParser is taken from Kalimetrix Logiscope tool*

For this particular class the number of public attributes are more. Therefore reducing the number of public attributes can result in higher security level of the class. It also increases the encapsulation. It can also be noted that the inheritance level for this is class is less as there are no children classes. The number of comments can be increased to increase the understandability. The public methods can be made private or protected so that the security issues of the class are addressed. The lines of code can be made less to reduce the dead code. So that the efficiency of the classes is maintained.

For MARF

Let us consider an instance class for MARF to analyze the code.

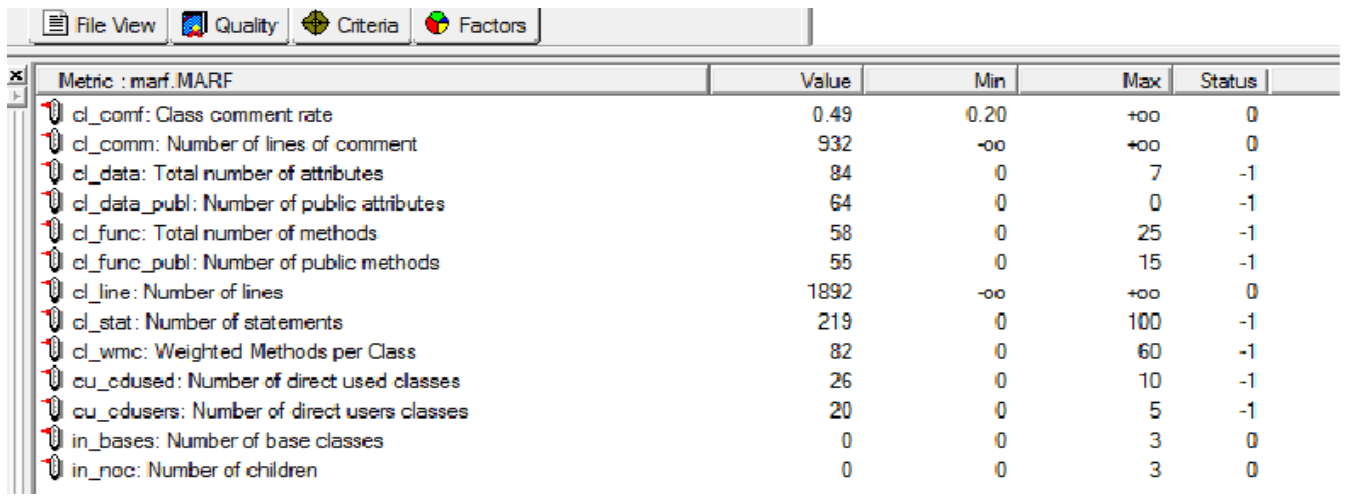

| Metric : marf.MARF | Value | Min | Max | Status |
|---|---|---|---|---|
| cl_comf: Class comment rate | 0.49 | 0.20 | +oo | 0 |
| cl_comm: Number of lines of comment | 932 | -oo | +oo | 0 |
| cl_data: Total number of attributes | 84 | 0 | 7 | -1 |
| cl_data_publ: Number of public attributes | 64 | 0 | 0 | -1 |
| cl_func: Total number of methods | 58 | 0 | 25 | -1 |
| cl_func_publ: Number of public methods | 55 | 0 | 15 | -1 |
| cl_line: Number of lines | 1892 | -oo | +oo | 0 |
| cl_stat: Number of statements | 219 | 0 | 100 | -1 |
| cl_wmc: Weighted Methods per Class | 82 | 0 | 60 | -1 |
| cu_cdused: Number of direct used classes | 26 | 0 | 10 | -1 |
| cu_cdusers: Number of direct users classes | 20 | 0 | 5 | -1 |
| in_bases: Number of base classes | 0 | 0 | 3 | 0 |
| in_noc: Number of children | 0 | 0 | 3 | 0 |

*Figure 26 Metrics for Class marf.MARF.java is taken from Kalimetrix Logiscope tool*

These are few public attributes that were present in a class of the MARF system. The number of lines of MARF is high when compared to GIPSY

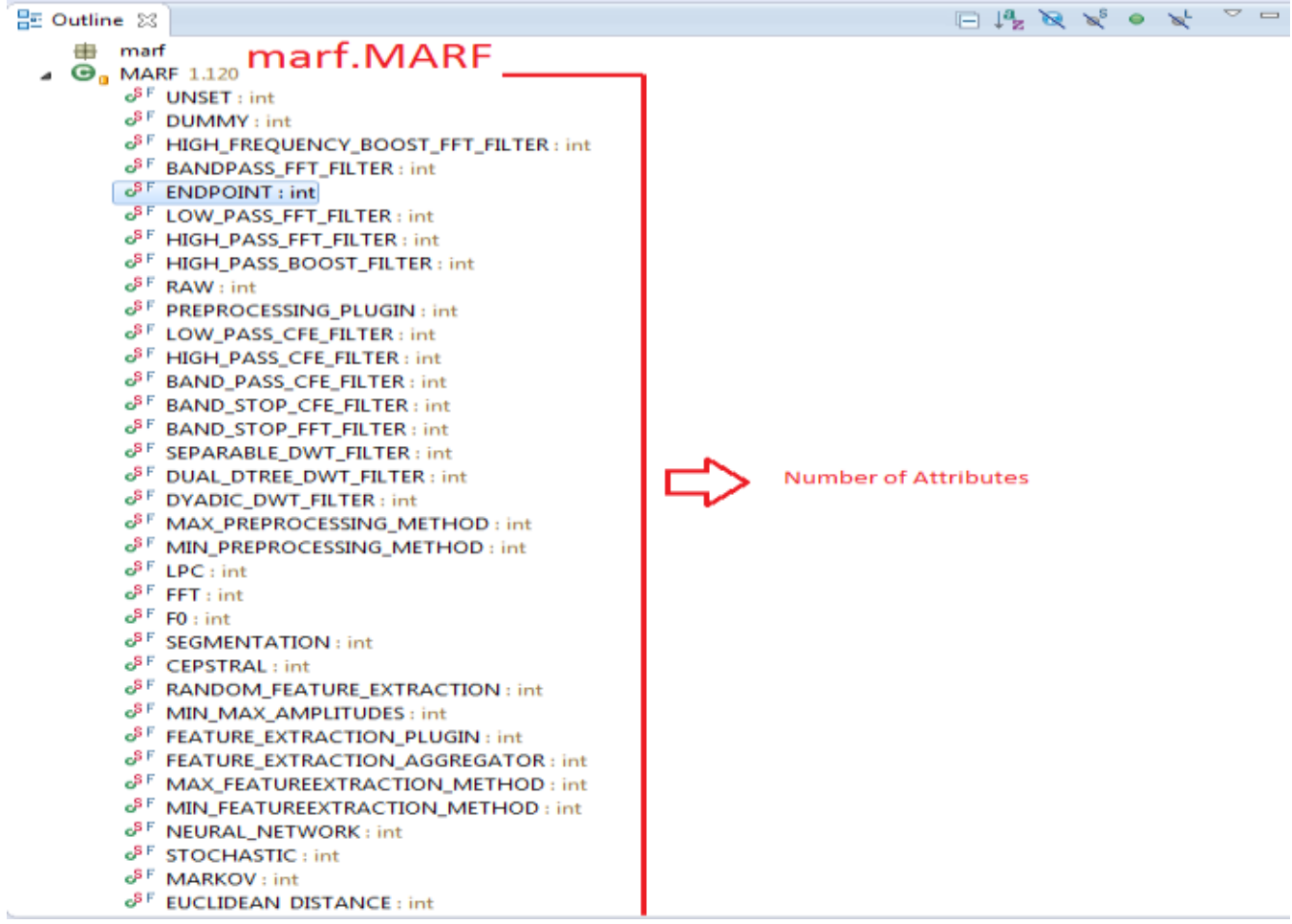

*Figure 27 Outline for Class marf.MARF.java is taken from Kalimetrix Logiscope tool*

Number of lines can be reduced. This is because more number of lines result in dead code which decreases the efficiency of the classes. Number of attended states during the compilation can be increased by reducing the lines of code. Classes can be inherited which possess the same functionality so that inheritance level of each class is maintained.

Class factor level

Class factor level importantly deals with maintainability. This is done to increase the rate of efficiency of the all the classes. The poor classes can be identified in GIPSY and MARF and can be re factored by increasing the re usability. The classes can be inherited wherever the methods of classes needed to be used. Maintainability is decomposed into analyzability, testability, stability, changeability.

MARF and GIPSY can be modified in subject to “changeability “to make it more reliable. The lines of code can be reduced to a certain extent to maintain the efficiency levels.

| | Excellent | Good | Fair | Poor |
|---|---|---|---|---|
| MARF | 23% | 60% | 10% | 6% |
| | | | | |
| GIPSY | 26% | 58% | 8% | 6% |

Table 32

2) *McCabe [75]*

To perform analysis on two case studies MARF and GIPSY McCabe IQ is used. McCabe IQ is the integrated set of products and solutions designed to help you throughout the software development cycle. The following products and solutions make up the McCabe IQ suite:

- McCabe QA helps you assess your system's design and quality.
- McCabe Test helps you plan, implement, and assess functional and code-based testing.
- McCabe TestCompress can be used as a stand-alone tool McCabe ReTest helps you create regression test suites for command line programs.
- McCabe 2000 is a solution that combines the features of McCabe Test and McCabe Data to help you throughout a Year 2000 conversion process. “ [77]

There are two section that were the part of project outline

- Quality trends of methods
- Quality trends of classes

a) *Quality rend for Methods for MARF and GIPSY*

Following table represent the threshold value for metric provided by McCabe Tool [74], establishing a meaning full threshold will help in analysis.

*Threshold table*

| | | Threshold | Min | Max |
|---|---|---|---|---|
| Cyclomatic Complexity [v(G)] | Cyclomatic Complexity (v(G)) is a measure of the complexity of a module's decision structure. It is the number of linearly independent paths and therefore, the minimum number of | 10 | 1 | Infinity |
| | paths that should be tested. [78] | | | |
| Essential Complexity [ev(G)] | Essential complexity (abbreviated as ev(G)) is a measure of the degree to which a module contains unstructured constructs. [78] | 4 | 1 | [v(G)] |
| Module Design Complexity | Module Design Complexity (iv(G)) is the complexity of the design-reduced module and reflects the complexity of the module's calling patterns to its immediate subordinate modules. [78] | 7 | 1 | [v(G)] |

Table 33

**Cyclomatic Complexity:**

Low Cyclomatic complexity is desired and high Cyclomatic means more complexity data collected form McCabe Tool for both MARF and GIPSY is shown below.

MARF (Max 56, found in LexicalAnalyzer.getNextToken())

| MARF | Average Complexity | Total Complexity |
|---|---|---|
| | 1.75 | 3722 |

Table 34

GIPSY (Max 294, found in JGIPLParserTokenManager.jjMoveNfa_0 (int,int) , ObjectiveGIPLParserTokenManager.jjMoveNfa_0(int,int), GIPLParserTokenManager.jjMoveNfa_0(int,int) )

| GIPSY | Average Complexity | Total Complexity |
|---|---|---|
| | 1.75 | 3722 |

Table 35

**Essential Complexity**

High value of essential complexity implies programmer is not using more structured programming.

For MARF (Max 37, found in LexicalAnalyzer.getNextToken())

| MARF | Average Complexity | Total Complexity |
|---|---|---|
| | 1.20 | 2556 |

Table 36

GIPSY (Max 96, found in LucxSemanticAnalyzer.check(gipsy.GIPC.intensional.SimpleNode,gipsy.GIPC.intensional.SimpleNode) , ForensicLucidSemanticAnalyzer.check(gipsy.GIPC.intensional.SimpleNode,gipsy.GIPC.intensional.SimpleNode) )

| GIPSY | Average Complexity | Total Complexity |
|---|---|---|
| | 1.84 | 11172 |

Table 37

**Module Design Complexity** (calculate for each module in battlemap)

Calculate amount of logic involved within predicates, higher value represent more complex logic, hence lower value is desired.

MARF (Max 50, found in LexicalAnalyzer.getNextToken())

| MARF | Average Complexity | Total Complexity |
|---|---|---|
| | 1.75 | 3722 |

Table 38

GIPSY (Max 160, found in SemanticAnalyzer.check(gipsy.GIPC.intensional.SimpleNode, gipsy.GIPC.intensional.SimpleNode))

| | Average Complexity | Total Complexity |
|---|---|---|
| GIPSY | 3.01 | 18729 |

Table 39

**Conclusion**

Comparatively Analysis of both the Case Studies on the above mention modular complexities.

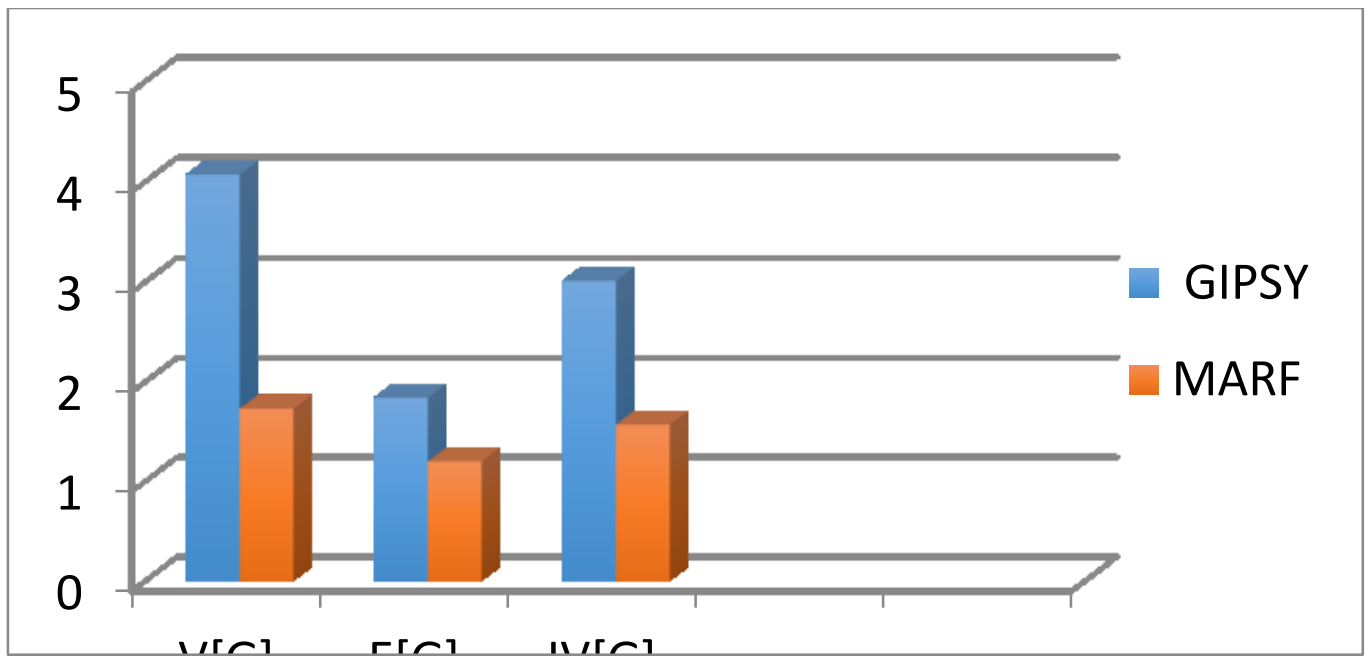

*Figure 28 Graph for comparing complexities for MARF and GIPSY*

It is clear fromt the above graph that GIPSY hold higher values in all the three complexity and hence GIPSY code is not well structured, holds poor sub routine call and have deep complexities as compared to the MARF.

**Scatter Plotter** (by default cyclomatic complexity (x-axis) vs essential metric (y-axis)) provide a mechanism to reliable / Unreliable / maintainable / Unmaintainable mechanism inside a quadrant based graph.

| | | |
|---|---|---|
| Low complexity | V[G] ≥ 10 | EV[G] ≥ 4 |
| Moderate complexity | V[G] > 10 | EV[G] ≥ 4 |
| High Complexity | (only dependant on EV[G]] | EV[G] ] > 4 |

Table 40

The following graph was obtained for MARF
Most of the code fall under III quadrant which implies code is reliable and maintainable)

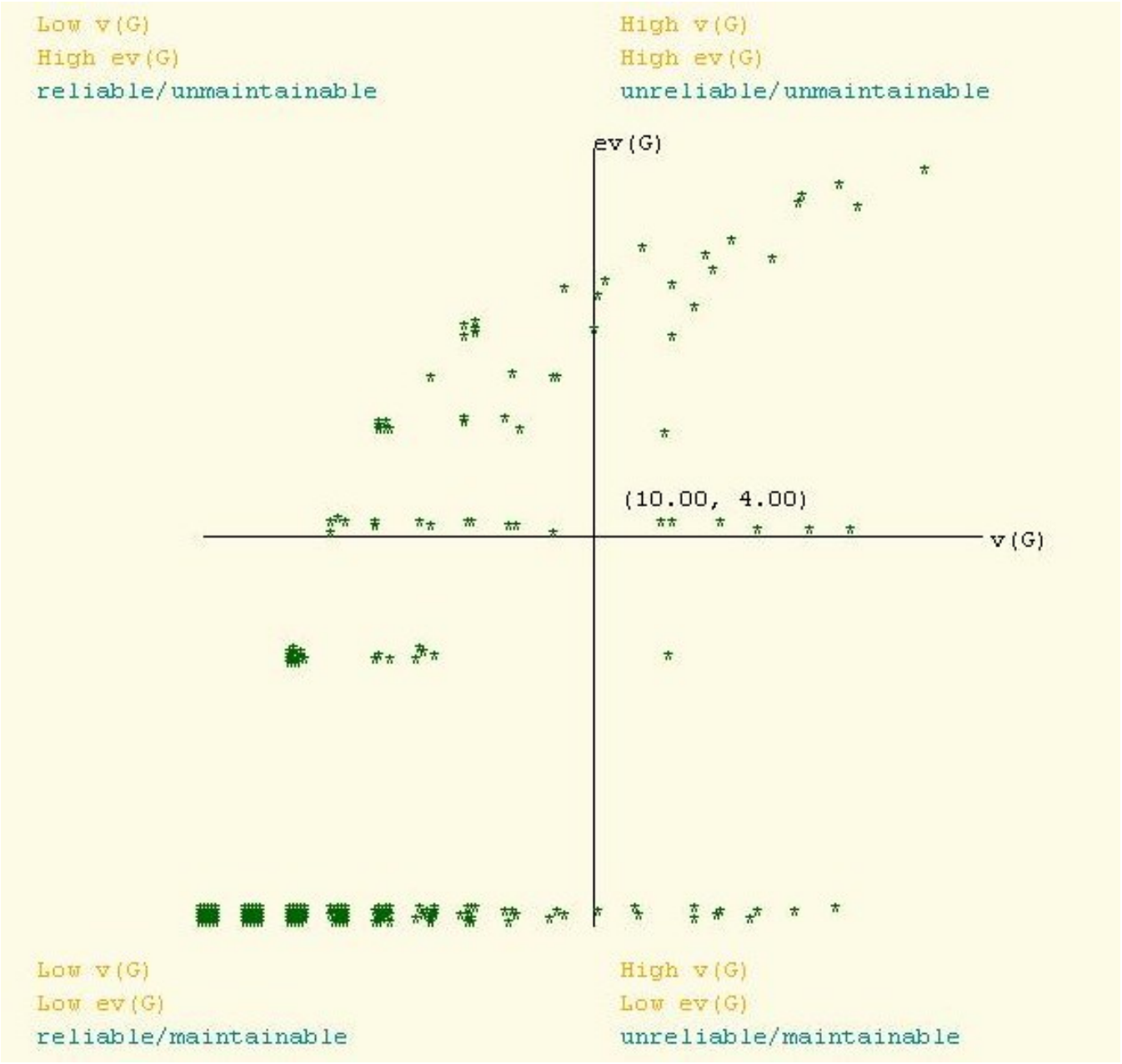

*Figure 29 Graph for Average Cyclomatic Complexity and Essential*

The following graph was obtained for GIPSY
Gipsy code is hard to maintain and more reliable as most of the code belongs to II and III quadrant in scatter chart.

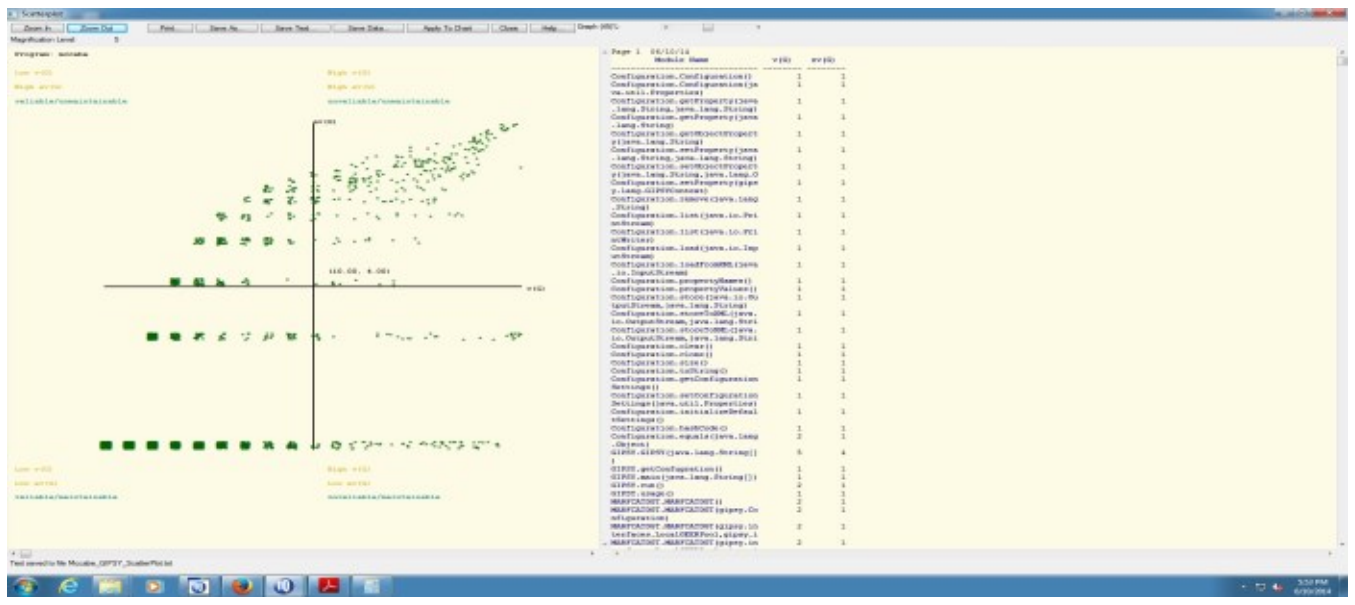

*Figure 30 Module Names and Graph for Average Cyclomatic complexity and Essential complexity for GIPSY using Mccabe tool*

*b) Quality Trend for Classess for MARF and GIPSY*

As per the project milestone 3 description data can be analyzed the following

Threshold table

| | Brief Description | Threshold |
|---|---|---|
| Coupling Between the Objects (CBO) | CBO is the number of classes to which a class is coupled. [79] | 2 |
| Number of Weighted | WMC is the number of locally implemented methods. Larger | 14 |
| Methods per class (WMC) | values (e.g., 14 or greater) indicate lesser polymorphism. [79] | |
| Response for Messages (RFC) | RFC is the number of methods in the set of all methods that can be invoked in response to a message sent to an object of a class. Larger values (e.g., 40 or greater) indicate lesser polymorphism. [79] | 100 |
| Depth in Inheritance Tree (DIT) | Discussed later | 7 |
| Number of Children (NOC) | Discussed later | 3 |

Table 41

Analysis results from McCabe

| Case Studies | [CBO] | | [WMC] | | [RFC] | |
|---|---|---|---|---|---|---|
| | Total | Avg | Total | Avg | Total | Avg |
| MARF | 1 | 0.01 | 2066 | 11.41 | 2973 | 16.43 |
| GIPSY | 38 | 0.07 | 6134 | 10.54 | 7347 | 12.62 |

Table 42

| [DIT] | | Number of Children[NOC] | |
|---|---|---|---|
| Total | Avg | Total | Avg |
| 387 | 2.14 | 45 | 0.25 |
| 1183 | 2.02 | 120 | 0.21 |

Table 42

Corresponding graph

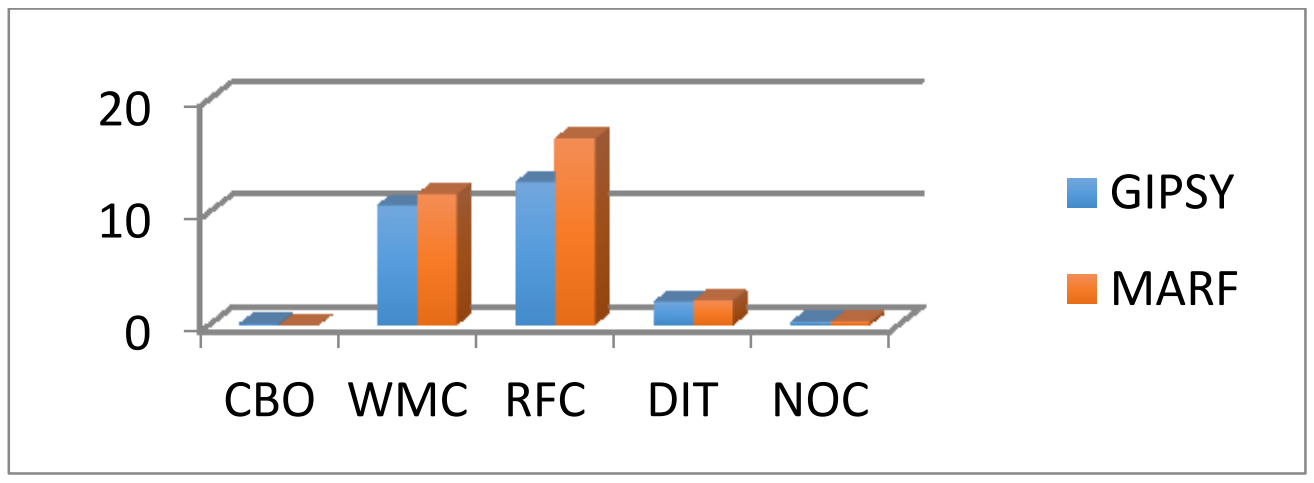

*Figure 31 Graph for comparing metrics for both MARF and GIPSY*

Observations (MARF as Compared to GIPSY)

1. CBO (The larger the number of couples, the higher the sensitivity to changes in other parts of the design, and therefore maintenance is more difficult [80] ). Hence GIPSY is difficult to maintain due to its High CBO value as compared with MARF
2. WMC (Classes with a large Weighted Methods Per Class value can often be refactored into two or more classes. [81]). Hence MARF is difficult to reuse and can be subject to refactoring due to its High WMC value as compared with GIPSY
3. RFC (The response set of a class is the set of all methods and constructors that can be invoked as a result of a message sent to an object of the class.)
4. GIPSY is easier to understand due its low RFC value as compare with MARF
5. DIT (The deeper the hierarchy the more difficult it might be to understand where particular methods and fields are defined or/and redefined. [82]). There is slight difference in DIT value of MARF and GISPY, MARF hold the higher value and hence its more difficult to understand due to its hierarchy structure.
6. NOC (Number of children), MARF holds the high value hence reuse value holds high
7. Scatter Plotter (by default cyclomatic complexity (x-axis) vs essential metric (y-axis)) provide a mechanism to reliable / Unreliable / maintainable / Unmaintainable mechanism inside a quadrant based graph.

For MARF

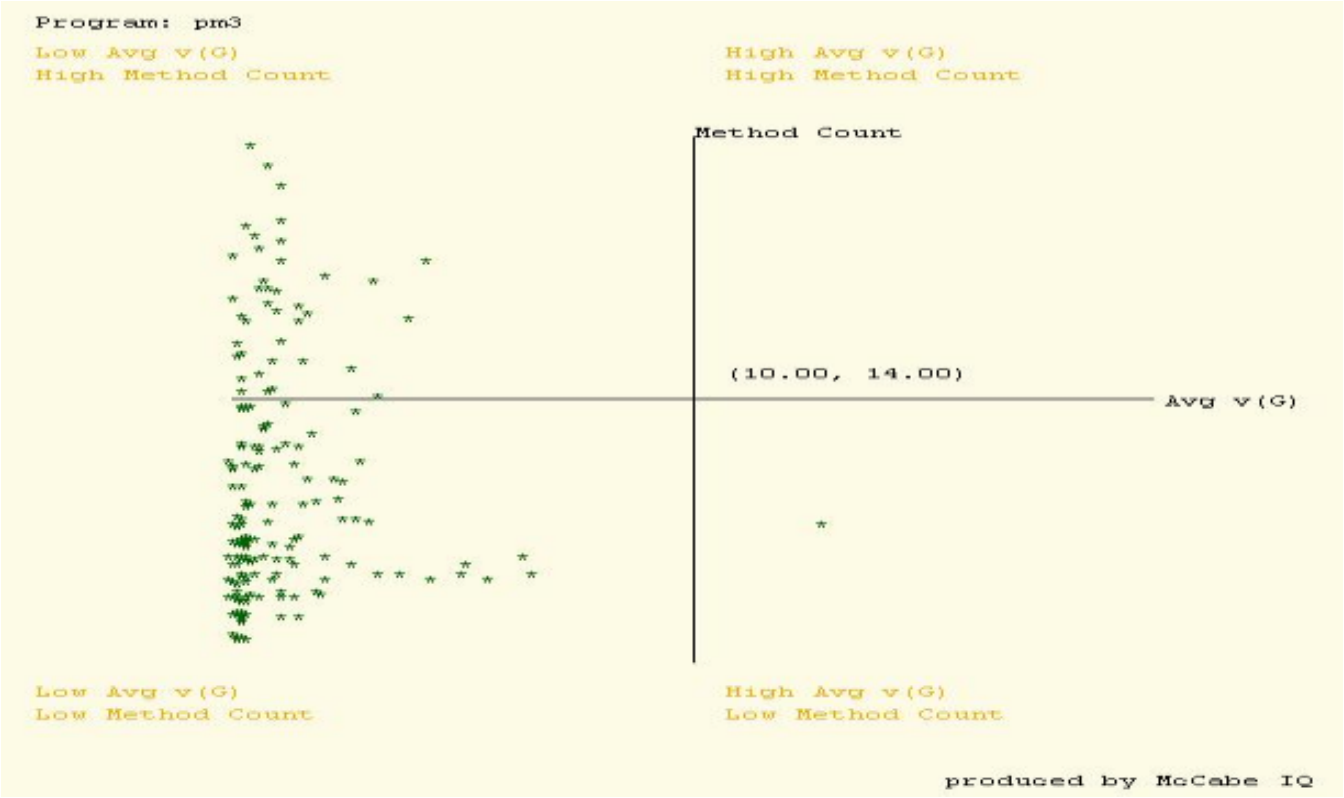

*Figure 32 Graph for checking code for Average Cyclomatic Complexity and method count for MARF using McCabe IQ tool*

For GIPSY

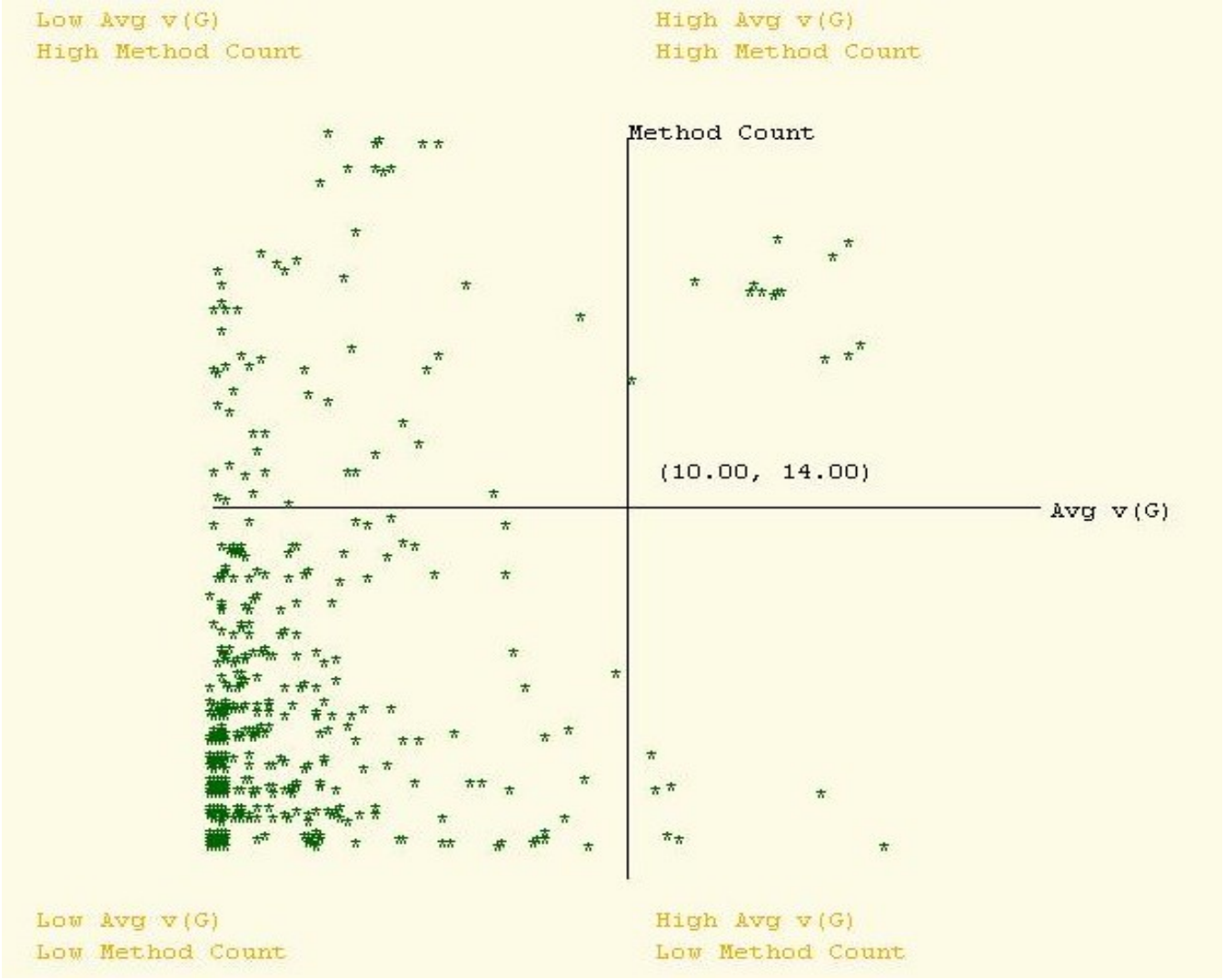

*Figure 33 Graph for Average Cyclomatic Complexity and method count for GIPSY using McCabe IQ tool*

Analysis from Scatter Plotter

Gipsy code is hard to maintain and more reliable as most of the code belongs to II and III quadrant in scatter chart as compared to MARF.

*3) Summary*

During the process of analyzing two systems, (GIPSY and MARF) files Logiscope and McCabe were used. McCabe is a tool used to compute Complexity factor for a program. The complexity is measured by taking consideration of independent paths in the program. It can be concluded that it counts the number of different test conditions in a program.

Some of the advantages of McCabe's Cyclomatic Complexity identified can be listed as below: Value is easily computed.

Other than other measurements, it is computed immediately in development lifecycle of a software system.

It makes easy code maintenance and focuses on testing effort. Easily find complex codes in a program for review.

On the other hand, there are some of the disadvantages of McCabe Cyclomatic Complexity as:

It is a measure of program's control complexity rather than data complexity.

As same weight is granted to both nested and non-nested loops, do deeply nested conditions are harder to understand than non-nested code.

Logiscope is a tool used to analyze code automatically for helping software analysis. It is used to drastically reduce both time and error while understanding the unfamiliar source code. Moreover it provides test path coverage analysis to reduce testing effort while improving its effectiveness.

Its advantages are:
Support different formats, e.g. code rule checking, code quality metric, code reducer and test checker for dynamic test coverage analysis.
□
Reduces testing effort
Reducing of maintenance effort on small section of source code.

Better code understandability.
Easily reuse of factorized code

Conclusions from above, both the tools take time to load source code files in case of large systems.McCabe and Logiscope only analyze what can be changed but cannot provide resources for any change to happen. In other words only analyzing and extraction can be performed execution is not possible.

## *B. Design and Implementation with JDeodorant and MARFCAT*

### *1) Overview*

The following section gives a generic explanation of both MARFCAT and JDeodrant followed by the experiences with both the tools.

#### *a) MARFCAT*

MARFCAT is a java based application which stands for MARF based Code Analysis tool. At SATE 2010 workshop, this application was presented on static code analysis tool exposition which is held at NIST 2010. Basically MARFCAT application is used to detect, classify and report the vulnerabilities and weaknesses or coding errors by using machine learning, data mining and classical NLP techniques. Static code Analysis tool is used to detect fingerprint security related weaknesses in code [84]. Normally machine learning is used to detect weak code very fastly compared to other tools. For classification and identification of vulnerabilities we use signal and NLP processing techniques. MARFCAT design was made independent of the programming language (source code, byte code or binary) being analyzed [84].

#### *b) JDeodorant*

Java 5.0 is the new version for JDeodorant plug-in and is compatible with eclipse. Based on latest version of eclipse classic JDeodorant plug-in is built. Basically this plug-in is used to spot the design issues in the software which are called as bad smells. JDeodorant applies different techniques to identify code smells and uses specific refactorings to solve them.

- Normally JDeodorant recognizes four different kinds of bad smells and can be resolved using different methods and techniques. They are
- Feature Envy issue can be solved by applying Move method refactorings[83] i.e.by creating a new method with the same body in the class.
- Type checking issue can be solved by applying Replace Conditional with Polymorphism and Replace Type code with state/strategy refactorings[83] i.e.by moving each part of conditional to an overriding method in a class and by replacing the type code with state object.
- Long Method issue can be solved by changing the fragments into methods so that the name of the method itself describes the purpose of the method which is called Extract Method refactorings[83].
- God Class issue can be solved by moving the required fields and methods of old class into new class which is called Extract Class refactoring [83].

*c) Experiences with JDeodorant and MARFCAT*

JDeodrant:

By JDeodorant can analyze the properties of cohesion, coupling, inheritance, and encapsulation. Moreover this plugin helps to refine data analysis of source code. One more advantage is this plugin provides refactoring option to refactor analyzed code so that the code can be improved and easily maintainable. One of the disadvantage is it takes time to load an heavy file and takes more time to parse that class.

MARFCAT:

It helps us to analyze for weak/vulnerable source code. This code scan gives result in terms of result threshold and warning notation.one of the disadvantage is it takes absolute path instead of relative path to analyze the source code

*2) Design and Implementation*

*a) Metrics definition*

We implemented all QMOOD [85] metrics and two of the CK Metrics Suite (1994) metric [86], definition and detail logic for the implementation is explained in subsequent sections.

1. QMOOD (Quality Model for Object-Oriented Design) model [85]

Design size in classes (DSC): "a count of the total number of classes in the design."[85]

Number of hierarchies (NOH): "a count of the number of class hierarchies in the design."[85]

Average number of ancestors (ANA): "the average number of classes from which each class inherits information."[85]

Number of polymorphic methods (NOP): "a count of the number of the methods that can exhibit polymorphic behaviour."[85]

Class interface size (CIS): "a count of the number of public methods in a class."[85]

Number of methods (NOM): "a count of all the methods defined in a class."[85]

Data access metric (DAM): "the ratio of the number of private (protected) attributes to the total number of attributes declared in the class."[85]

Direct class coupling (DCC): "a count of the number of different classes that a class is directly related to. The metric includes classes that are directly related by attribute declarations and message passing (parameters) in methods." [85]

Cohesion among methods of class (CAM): "the relatedness among methods of a class, computed using the summation of the intersection of parameters of a method with the maximum independent set of all parameter types in the class." [85]

Measure of aggregation (MOA): "a count of the number of data declarations whose types are user-defined classes." [85]

Measure of functional abstraction (MFA): "the ratio of the number of methods inherited by a class to the number of methods accessible by member methods of the class." [85]

2. The CK Metrics Suite (1994) metric suite [86]

DIT Depth of Inheritance Tree

"DIT: maximum inheritance path from the class to the root class The deeper a class is in the hierarchy, the more methods and variables it is likely to inherit, making it more complex. Deep trees as such indicate greater design complexity. Inheritance is a tool to manage complexity, really, not to not increase it. As a positive factor, deep trees promote reuse because of method inheritance." [86]

NOC Number of Children

"NOC = number of immediate sub-classes of a class
NOC equals the number of immediate child classes derived from a base class. In Visual Basic .NET one uses the Inherits statement to derive sub-classes. In classic Visual Basic inheritance is not available and thus NOC is always zero." [86]

*b) Metrics Implementation details*

QMOOD metrics are implemented out of which we will explain four

CACM:
1) Get the methods from the class
2) Obtain parameters method by method and add parameter data types in list
3) Sum distinct parameter in incremental way, Increment one in each method type, and incorporate .this pointier of the class
4) used formula

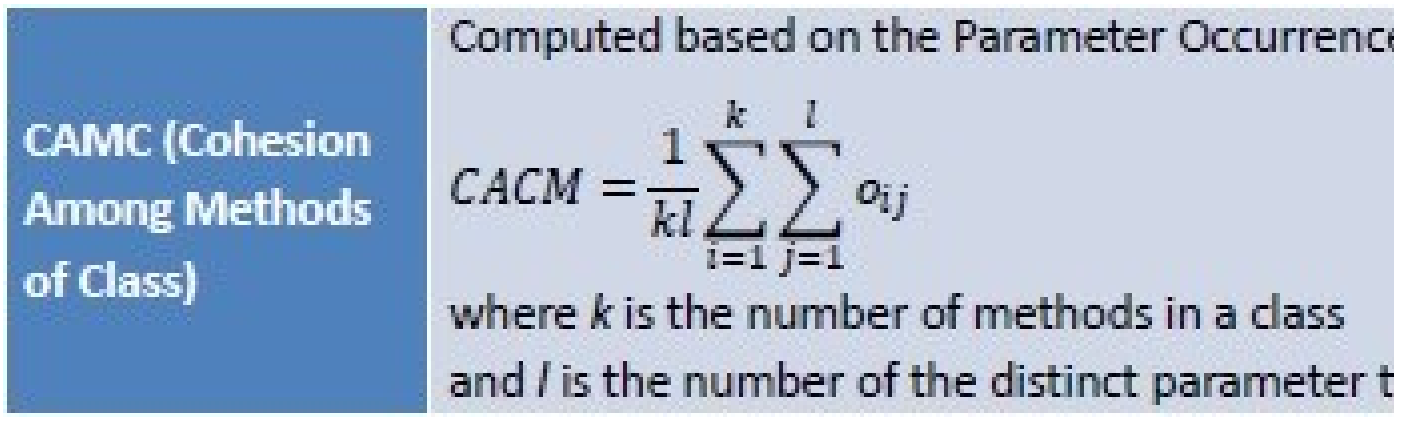

*Figure 34 CAMC*

2) ANA: The average number of ancestors of a praticular class is to be calculated.Instead of calculating using the top-bottom approach bottom top approach was used.According to this a counter is maintained initially starting at 0. when a super class is found the counter is incemented by one.This is done until there are no super classes.The count value of ANA is then printed.

3) NOM : For NOM need to calculate the number of methods in a class and for that JDeodrant function has in built function name, getNumberOfMethod() defined under ast.ClassObject.java file

4) CIS : for CIS, JDeodrant plugin take getMethodList() to get the list of methods from the past class defined under ast.ClassObject.java file. After that those method list is more refined with the help of getAccessMehod()defined under ast.ClassObject.java

CK metrics

DIT: if getSuperClass() && getClassType() if not null it mean they have child and count vai incremental logic.

Number of child : inheritenceDetectionTree () provide inheritance for the classs then we use get root node getChildcount and count vai incremental logic.

*c) Test Cases*

Three test cases has been written, Two test manual cases were written for QMODD and one for CK in order to validate all the metrics .

**Test CASE 1 for QMOOD**

Value of all metrics are calculated with help of uml diagram that is compared with test.suite2 value in order to check the result whether it is correct or not.

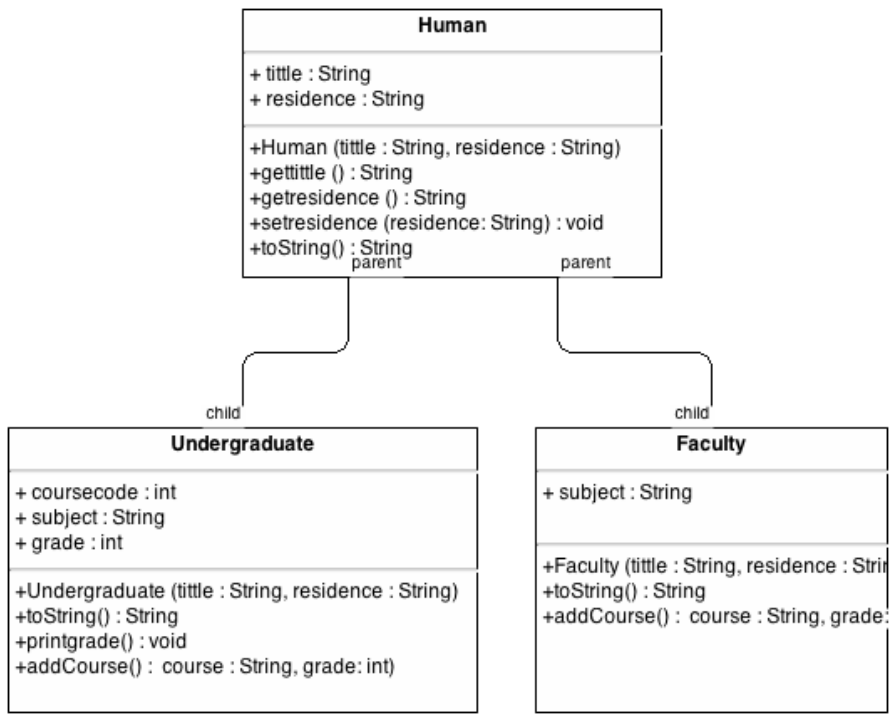

*Figure 35 UML Diagram*

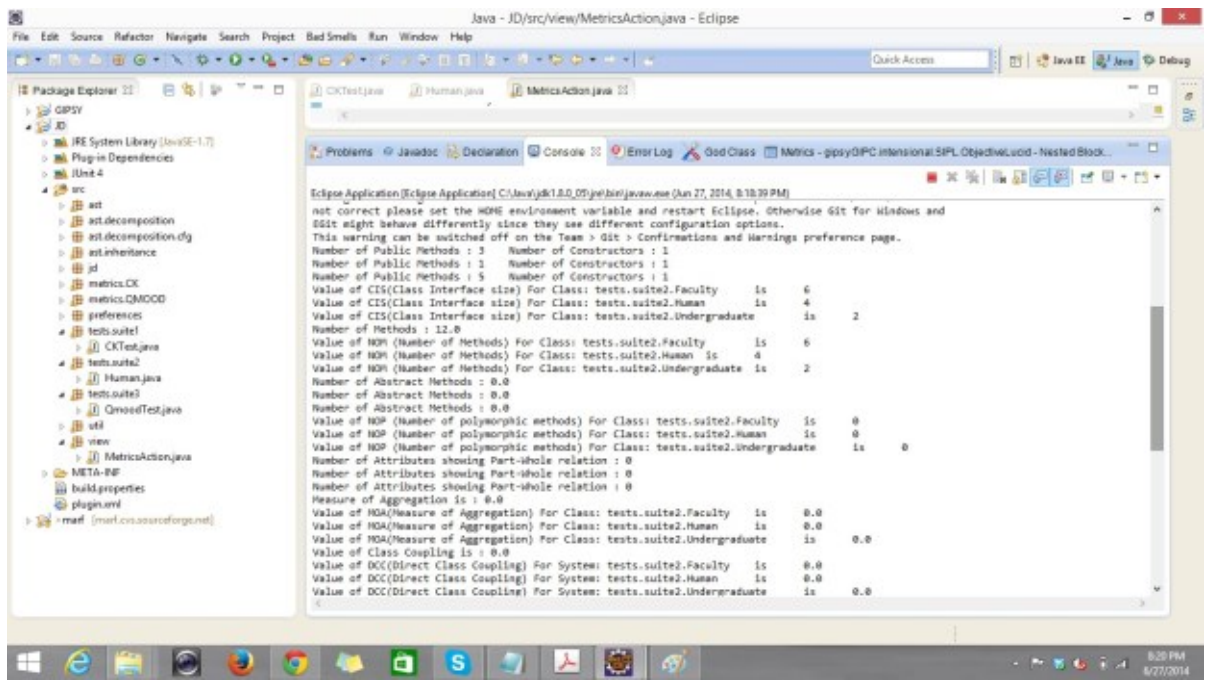
*Figure 36*

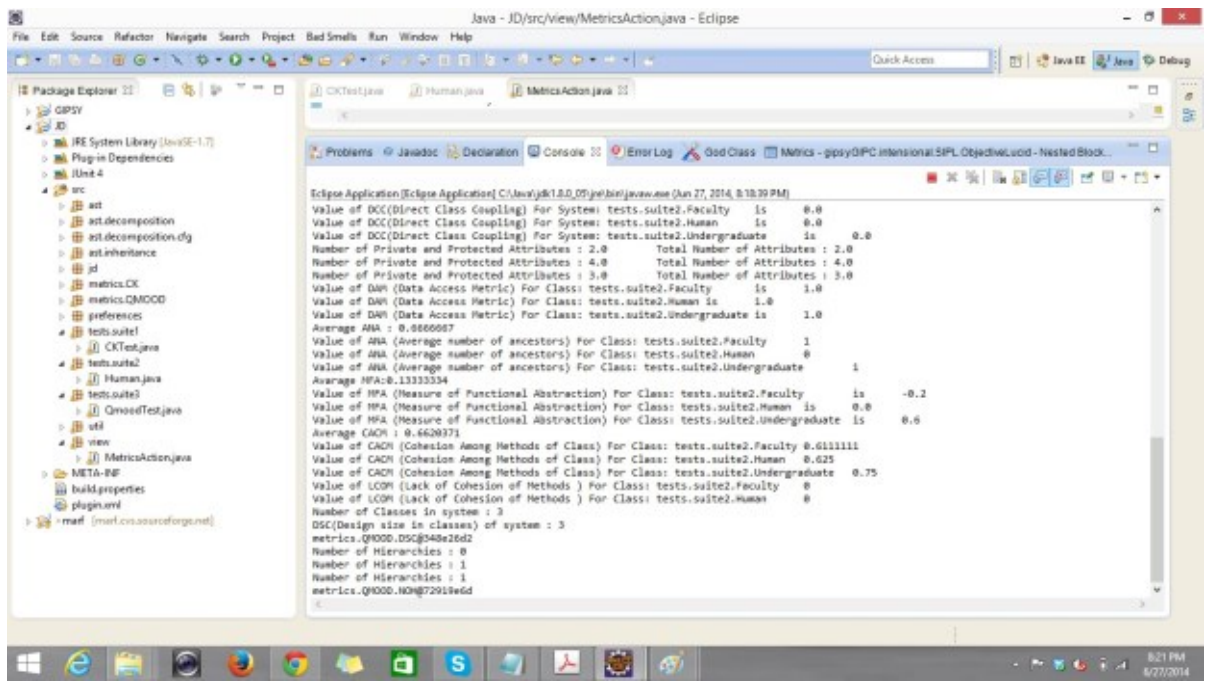
*Figure 37*

Test CASE 2 for QMOOD

All metric values are calculated with help of uml diagram that is compared with test.suite3 value.

## QMOOD TEST CASE

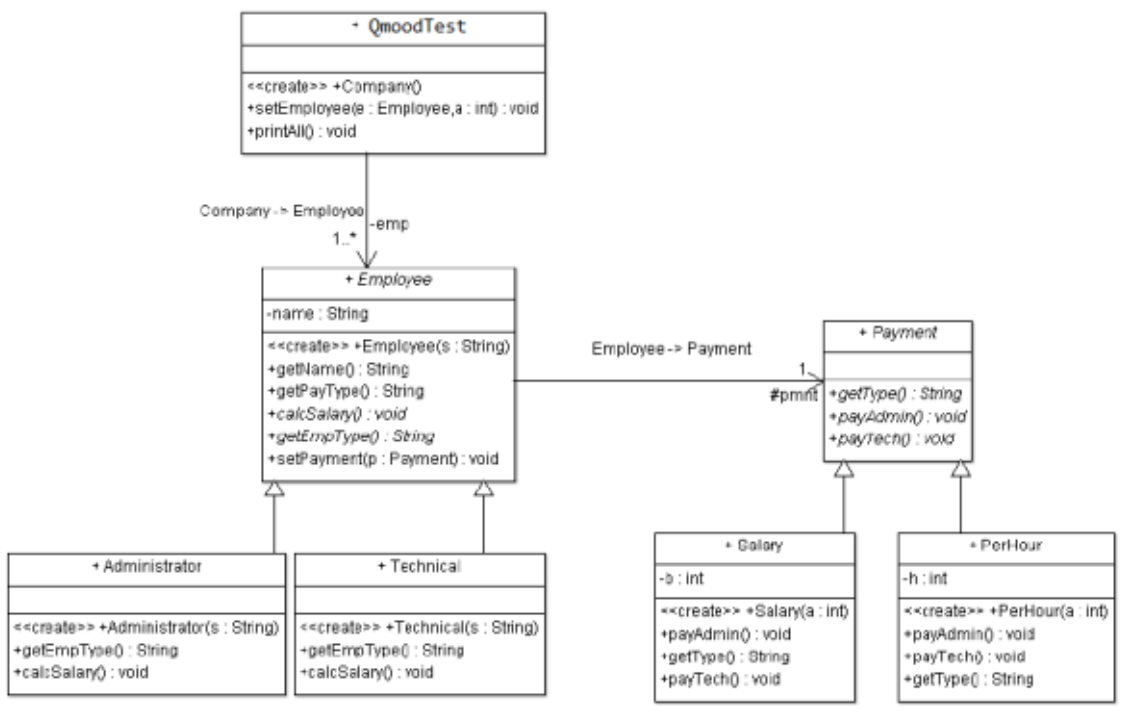

*Figure 38*

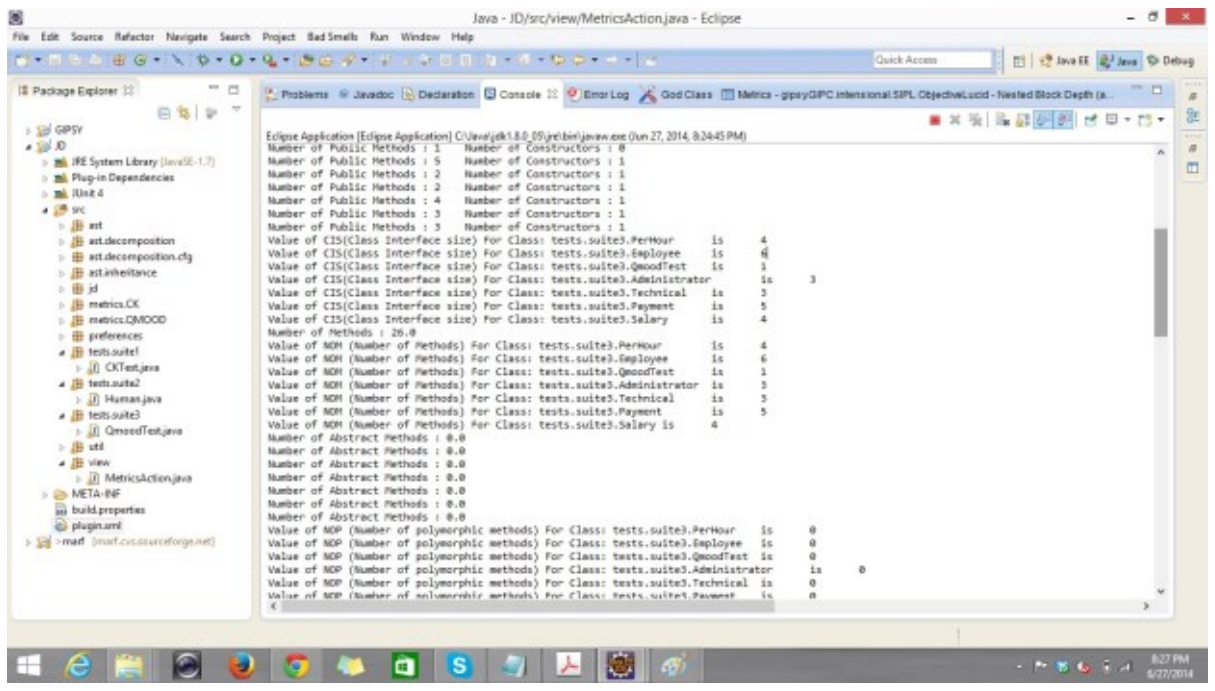

*Figure 39*

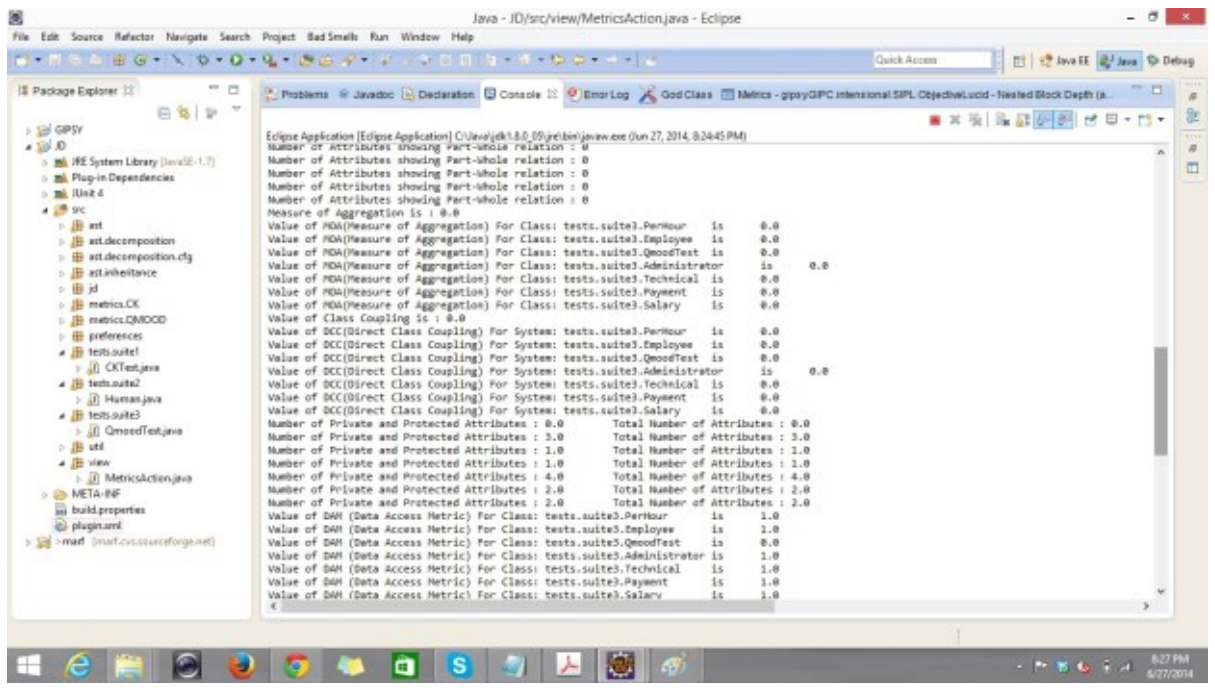

*Figure 40*

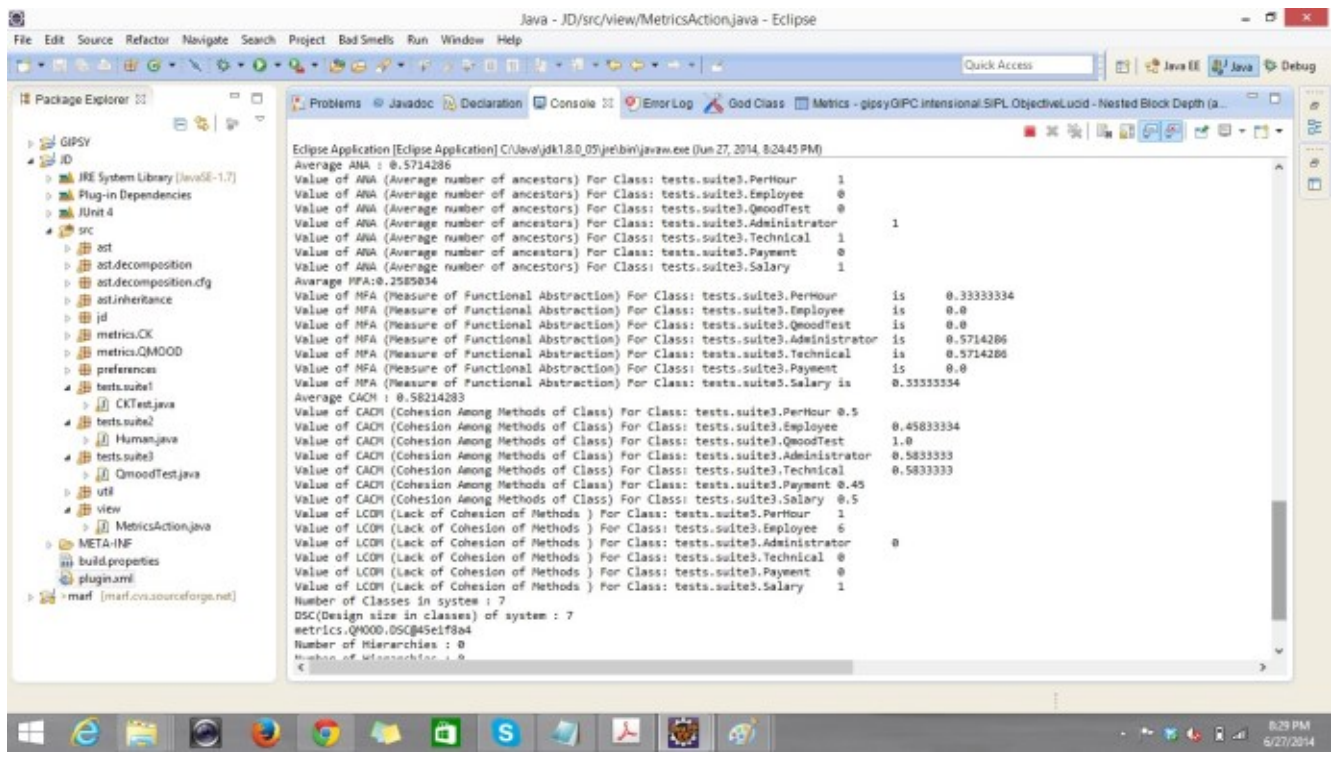

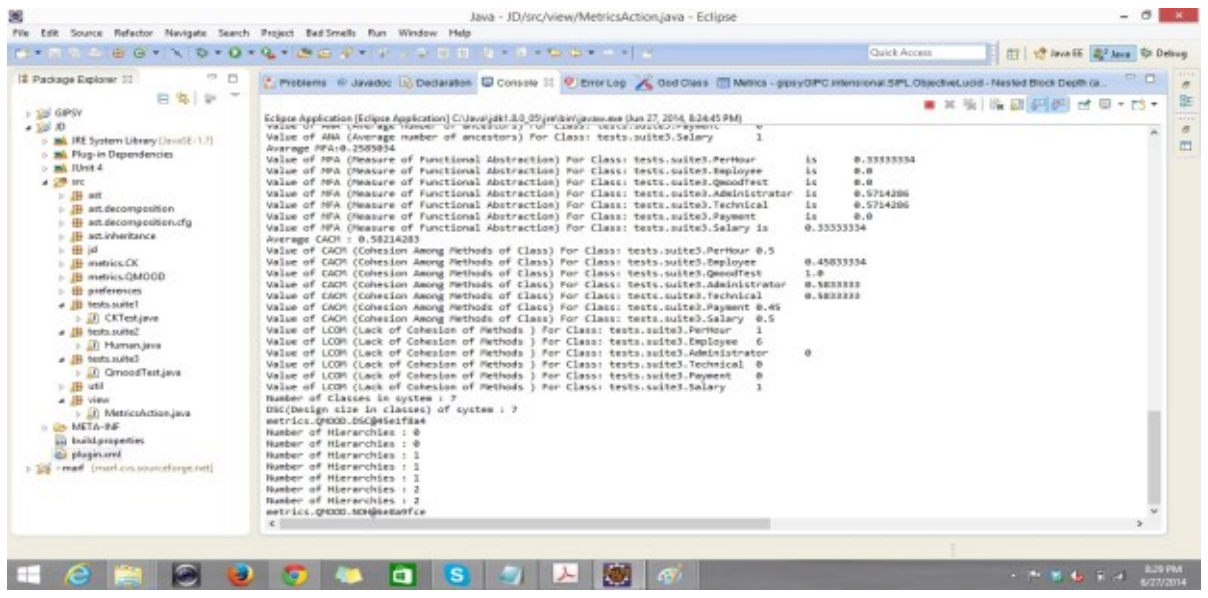

*Figure 41*

## Test CASE for CK

Value of Number of child (NOC) and Value of Depth of inheritance (DIT ) are calculated manually with help of UML diagram that is compared with test.suit1 value show in figure

CK –test case

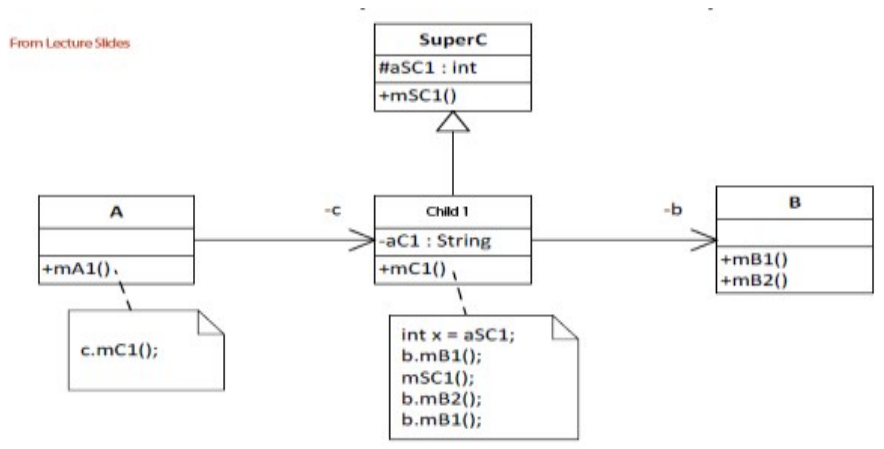

*Figure 42*

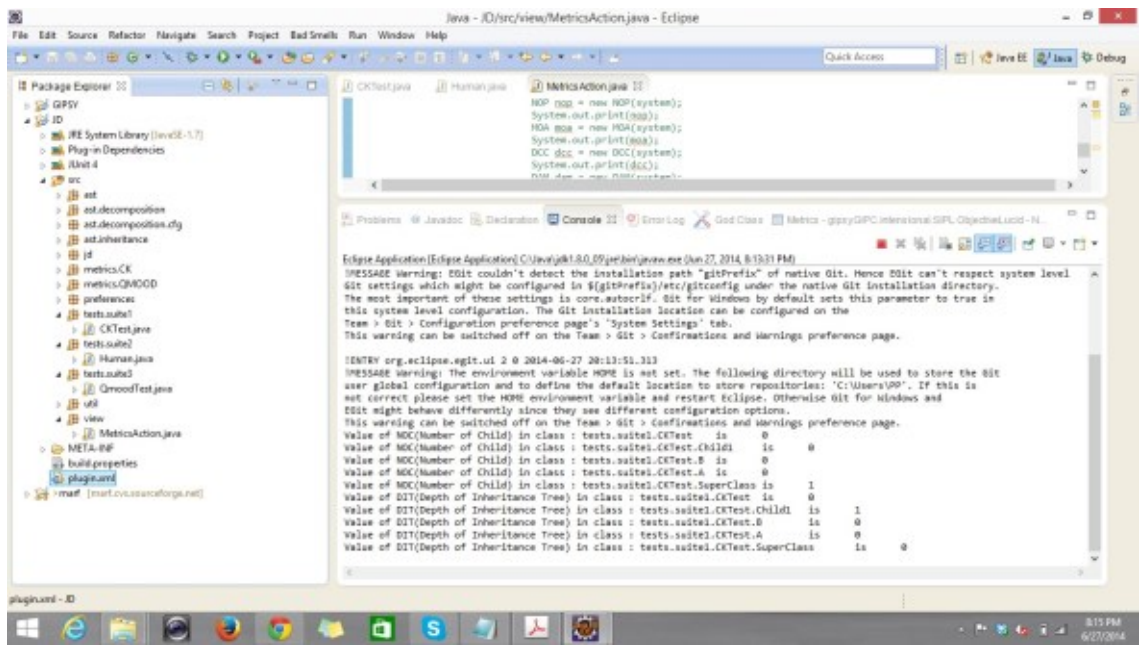

*Figure 43*

*3) Summery*

*a) Results and Analysis Interpretation*

JDeodrant as a plugin is installed on eclipse and further actions are performed to analyze both the systems. The classes on which the analysis is to be made are selected loading the project on eclipse. Option "metrics" is selected from a drop down menu. This will parse the java file sending it to AST (Abstract Syntax Tree). The results obtained can be seen on the console. The two problematic classes were identified using the JDeodrant tool. We have implemented top-ranked metrics that were prioritized earlier. Using the JDeodrant tool.

*a)* Metrics results & Interpretation (on our own metrics)

Here metrics are implemented on QMOOD and CK
The ANA and CIS metrics from QMOOD Metrics are taken for implementation

ANA ( Average no of ancestors):

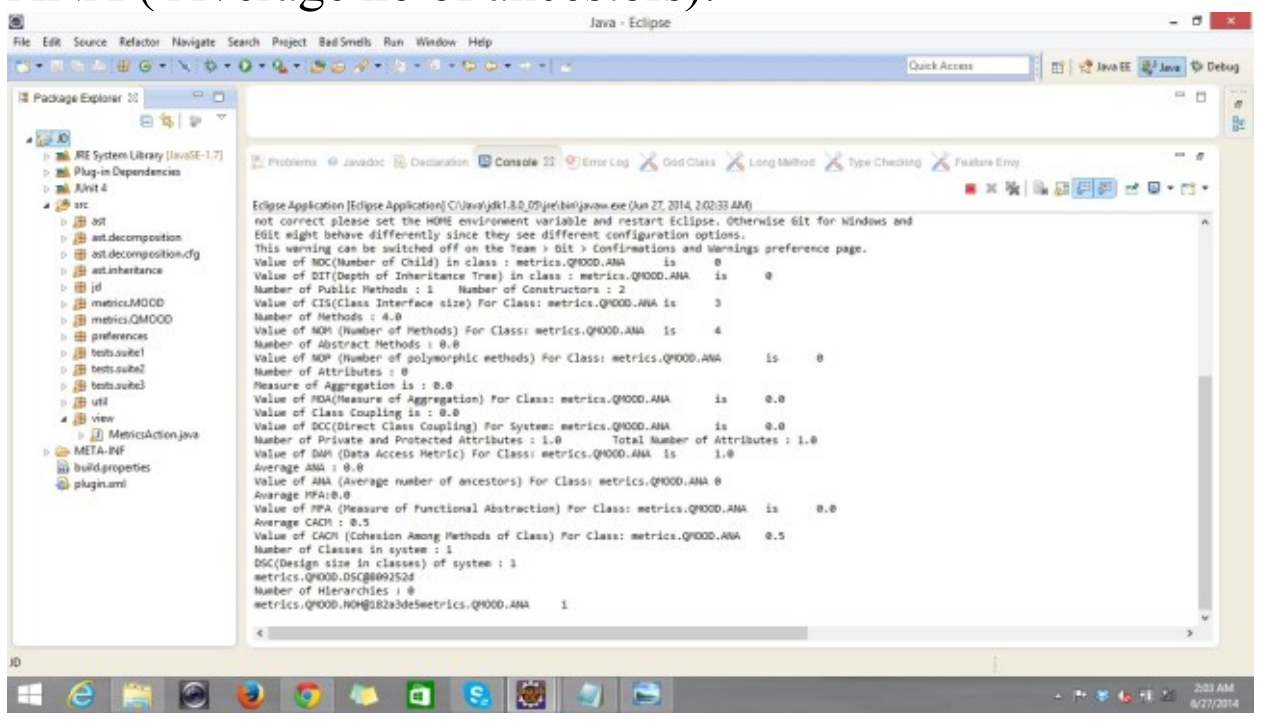

*Figure 44*

CIS (Class interface size):

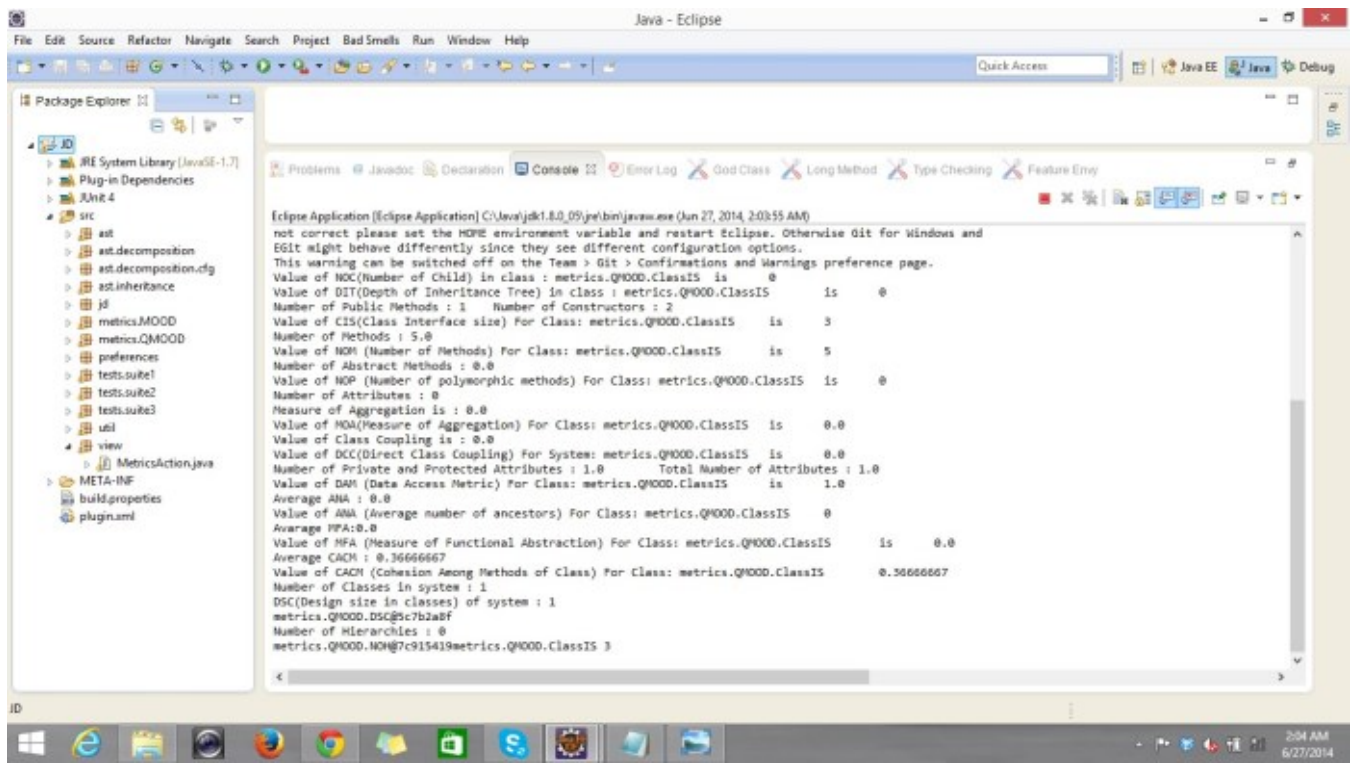

*Figure 45*

**DIT and NOC metrics from CK metrics**

DIT (Depth of inheritance) metrics:

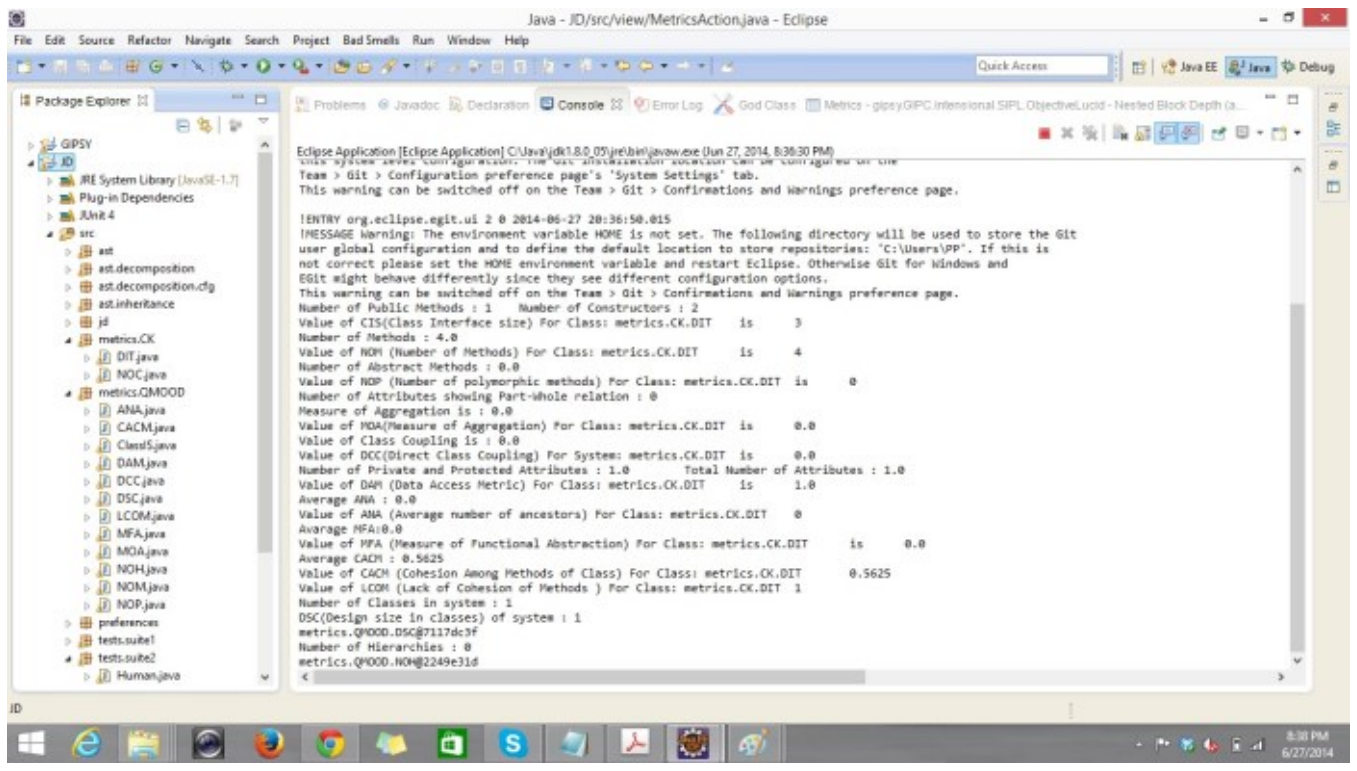

NOC (Number of Children) metrics:

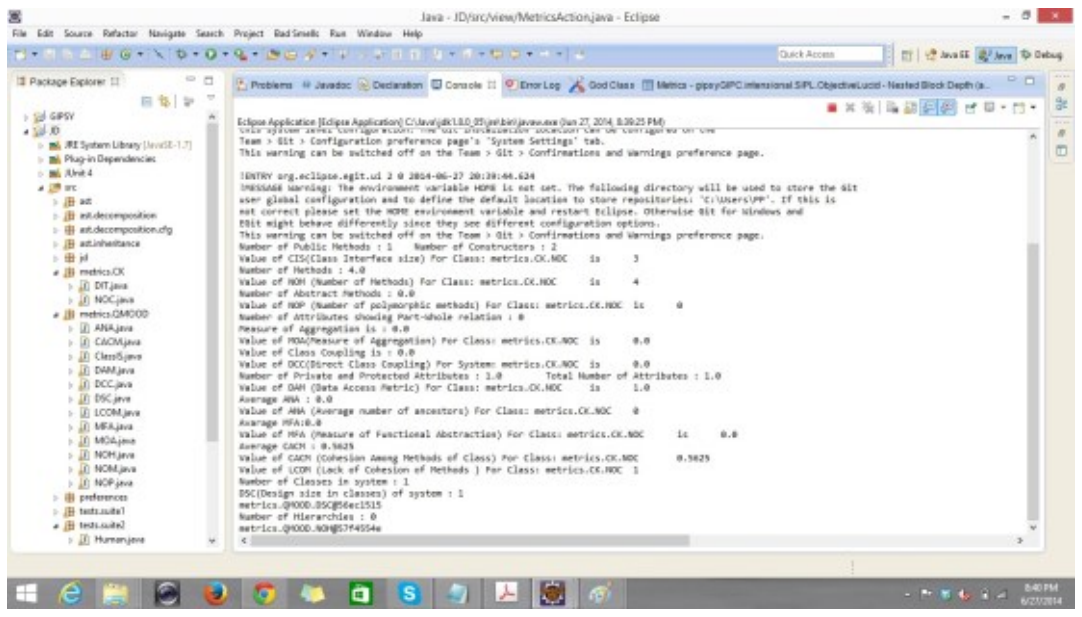

*Figure 46*

*b)* Metrics results & Interpretation (problematic classes)

QMOOD metrics are applied to the selected classes along with the LCOM values provided by the JDeodrant tool will be present in the console part of the tool.

For GIPSY class1:
gipsy/src/gipsy/RIPE/editors/RunTimeGraphEditor/ui/GIPSYGMTOperator

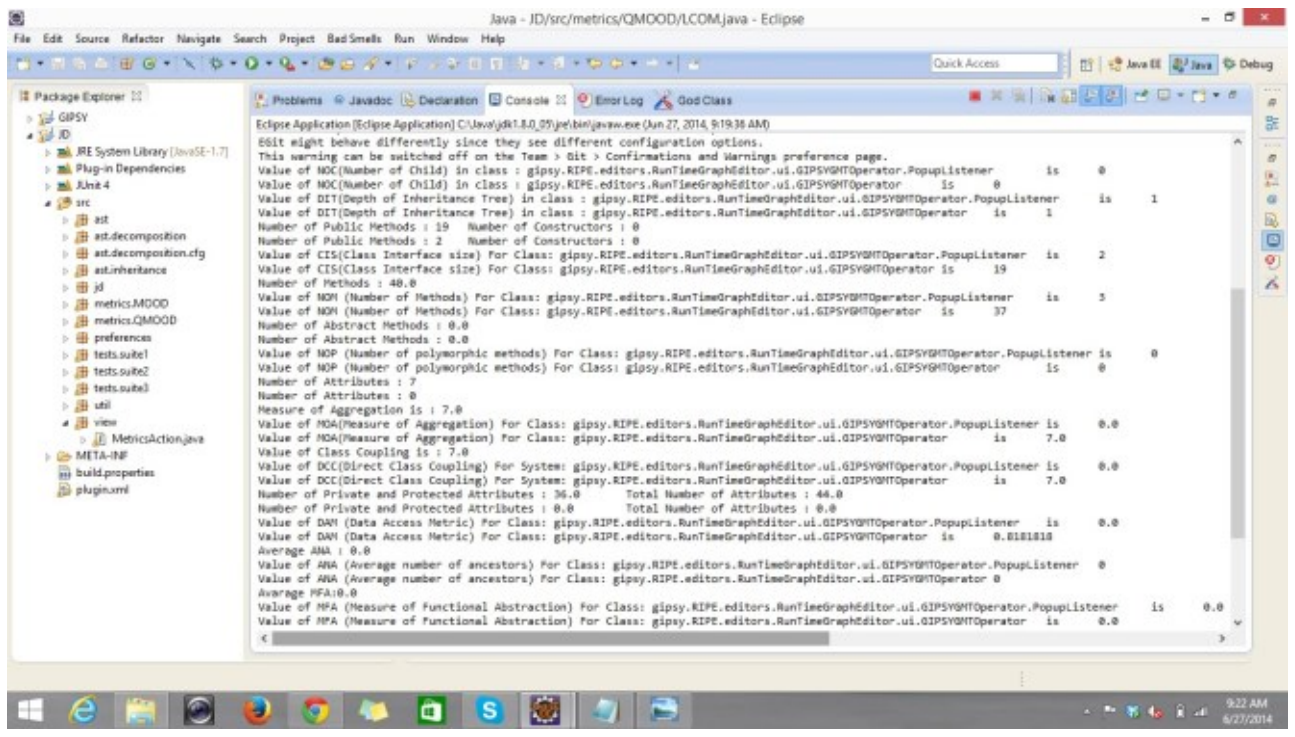

*Figure 47*

All QMOOD metrics including LCOM values are

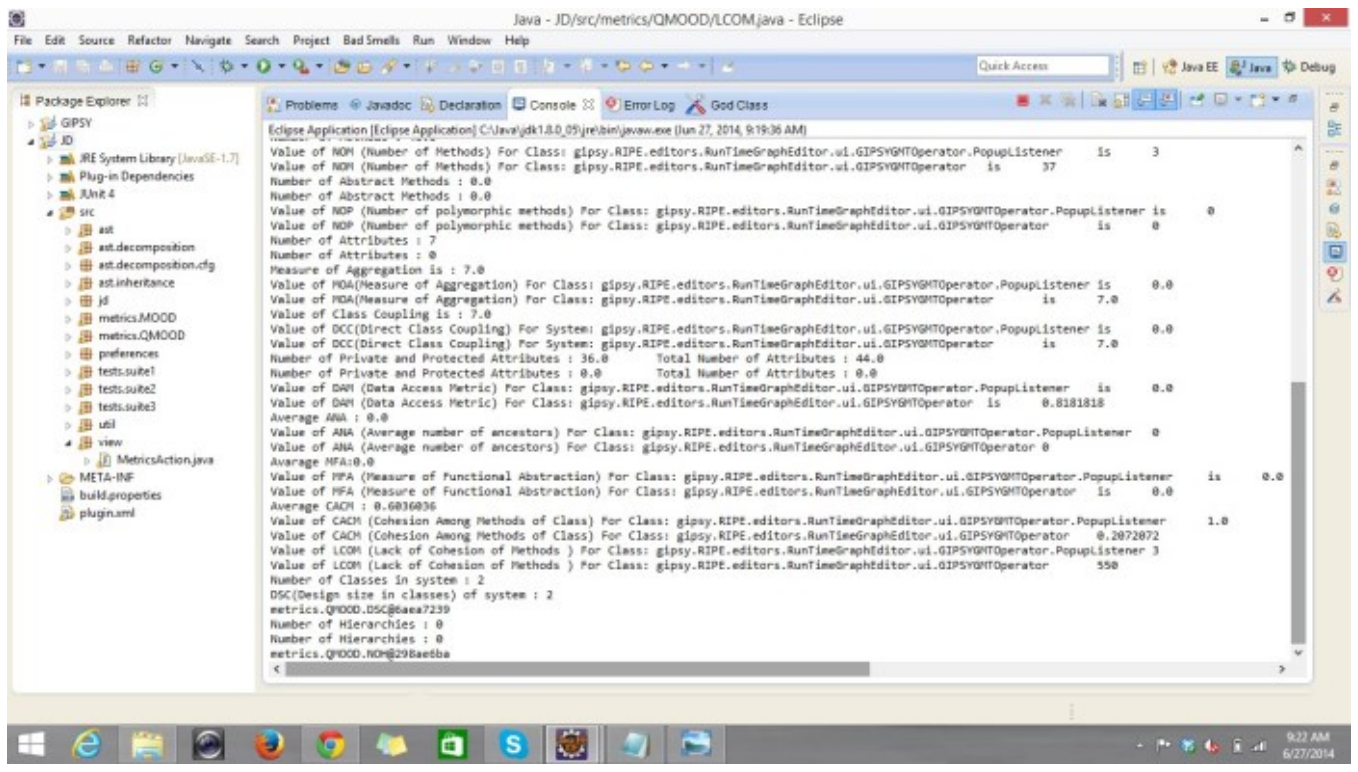

*Figure 48*

For GIPSY class 2:
gipsy/src/gipsy/GIPC/intensional/SIPL/ObjectiveLucid.ObjectiveGIPLParser

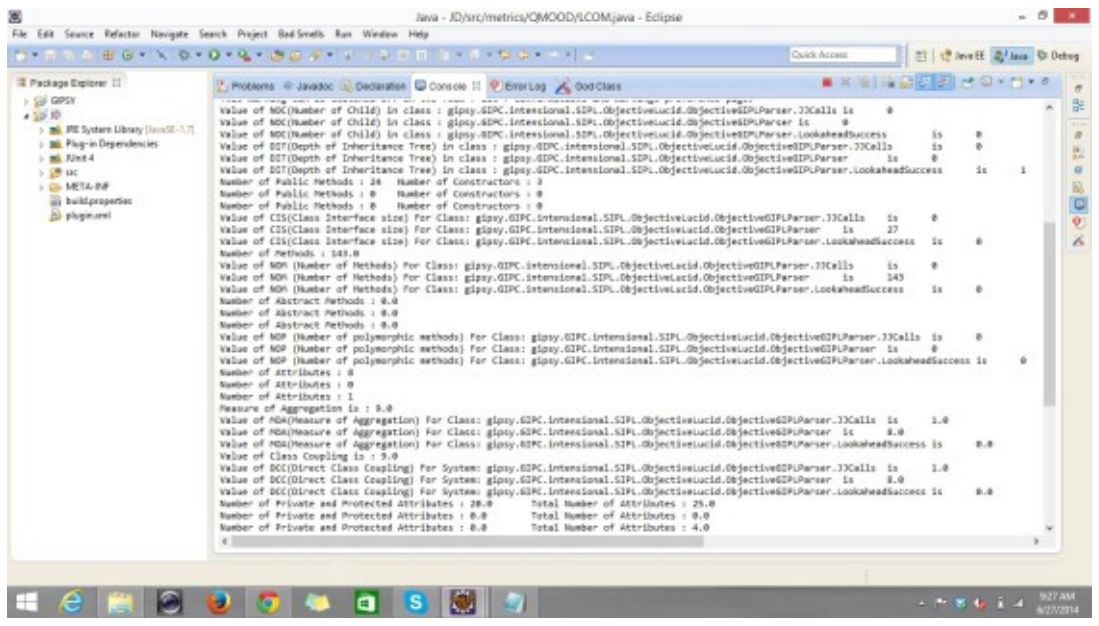

*Figure 49*

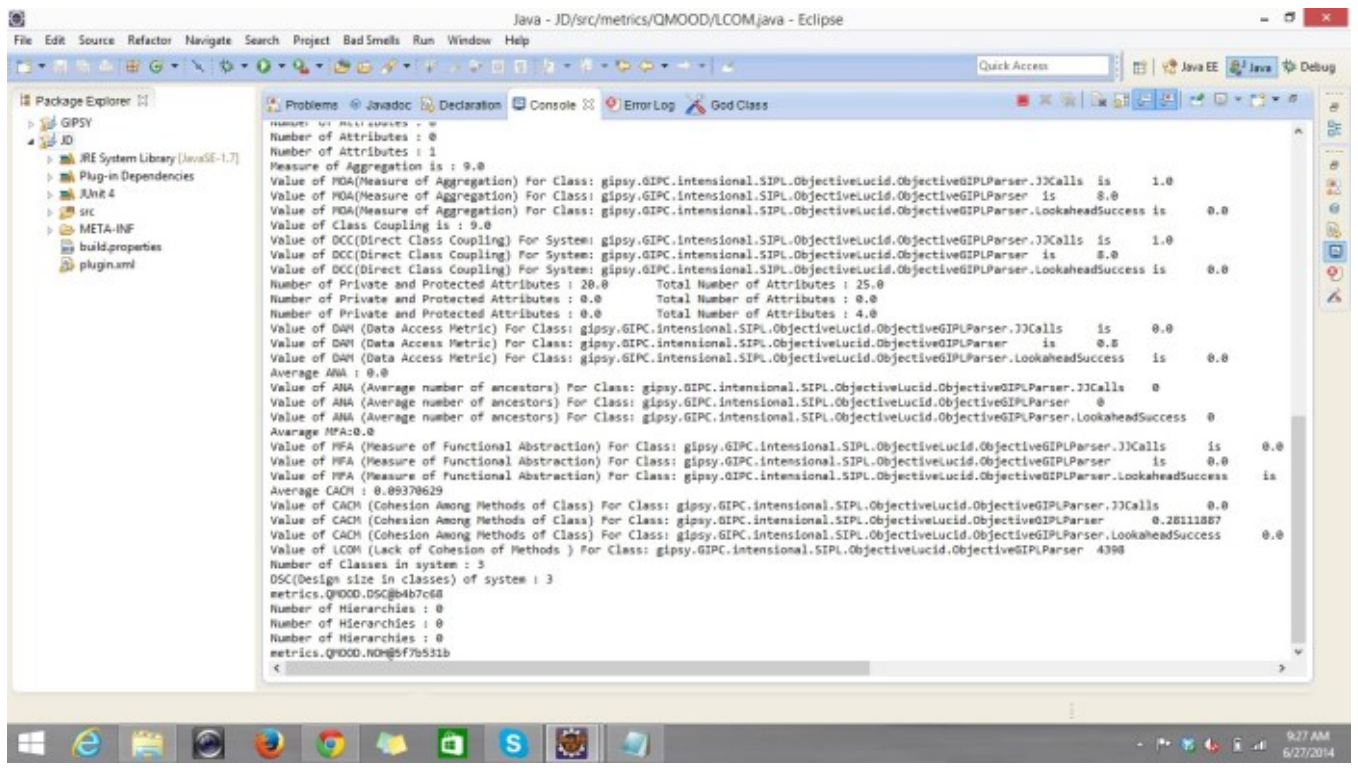

*Figure 50*

For MARF class 1:
marf/src/marf/MARF.java

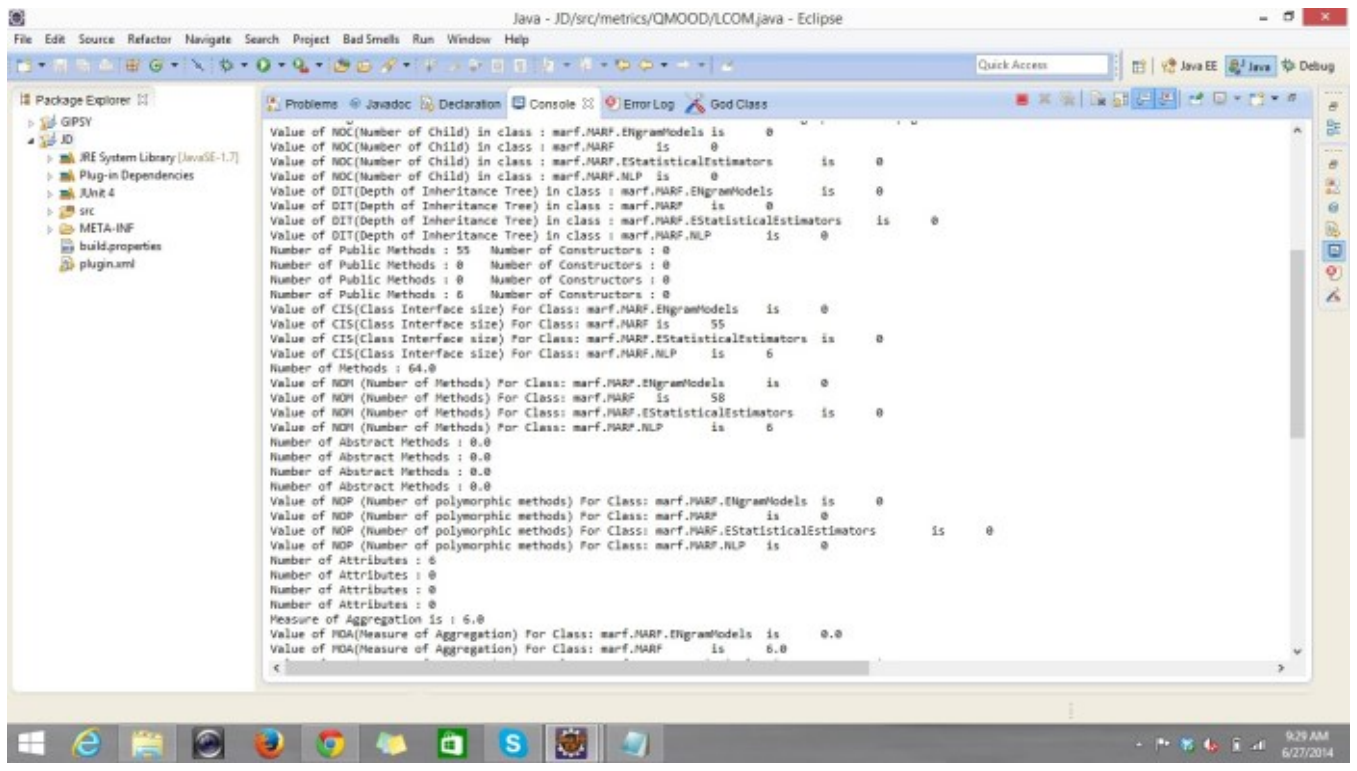

*Figure 51*

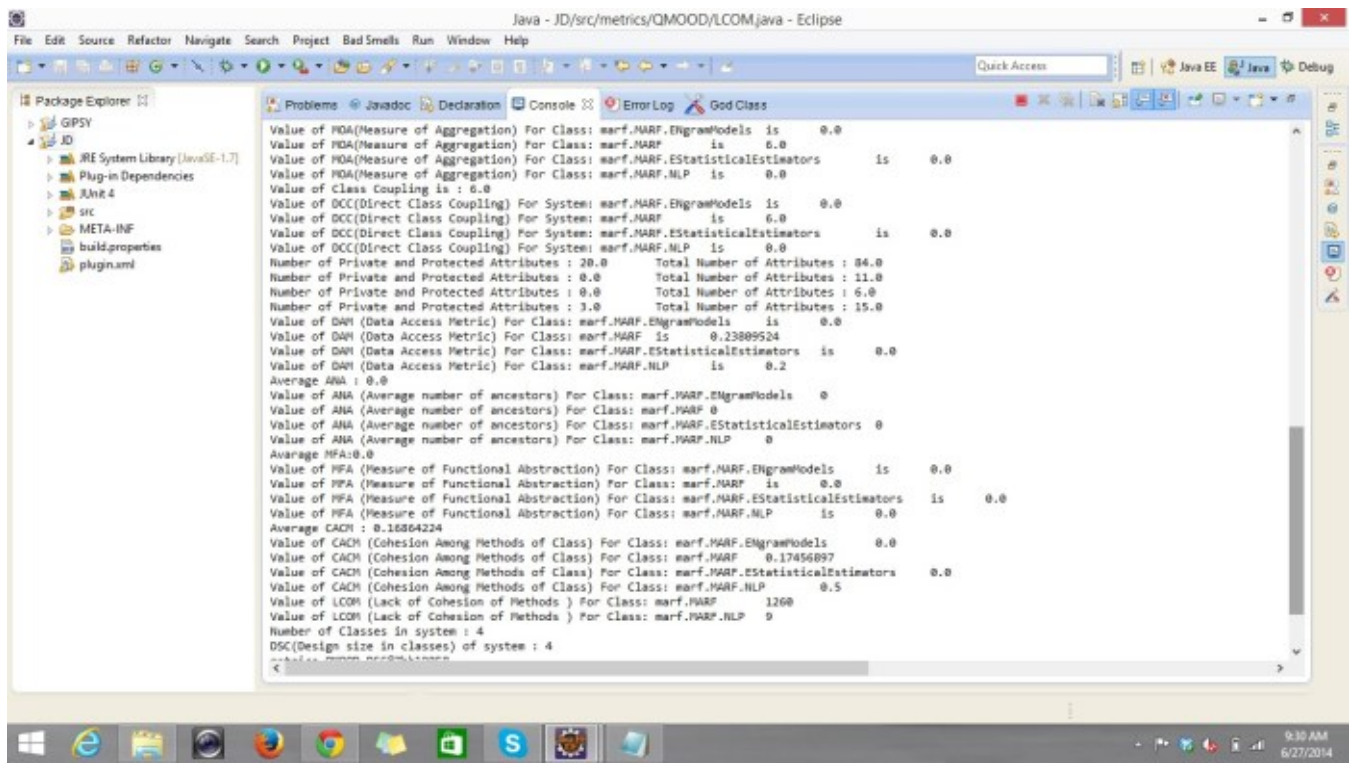

*Figure 52*

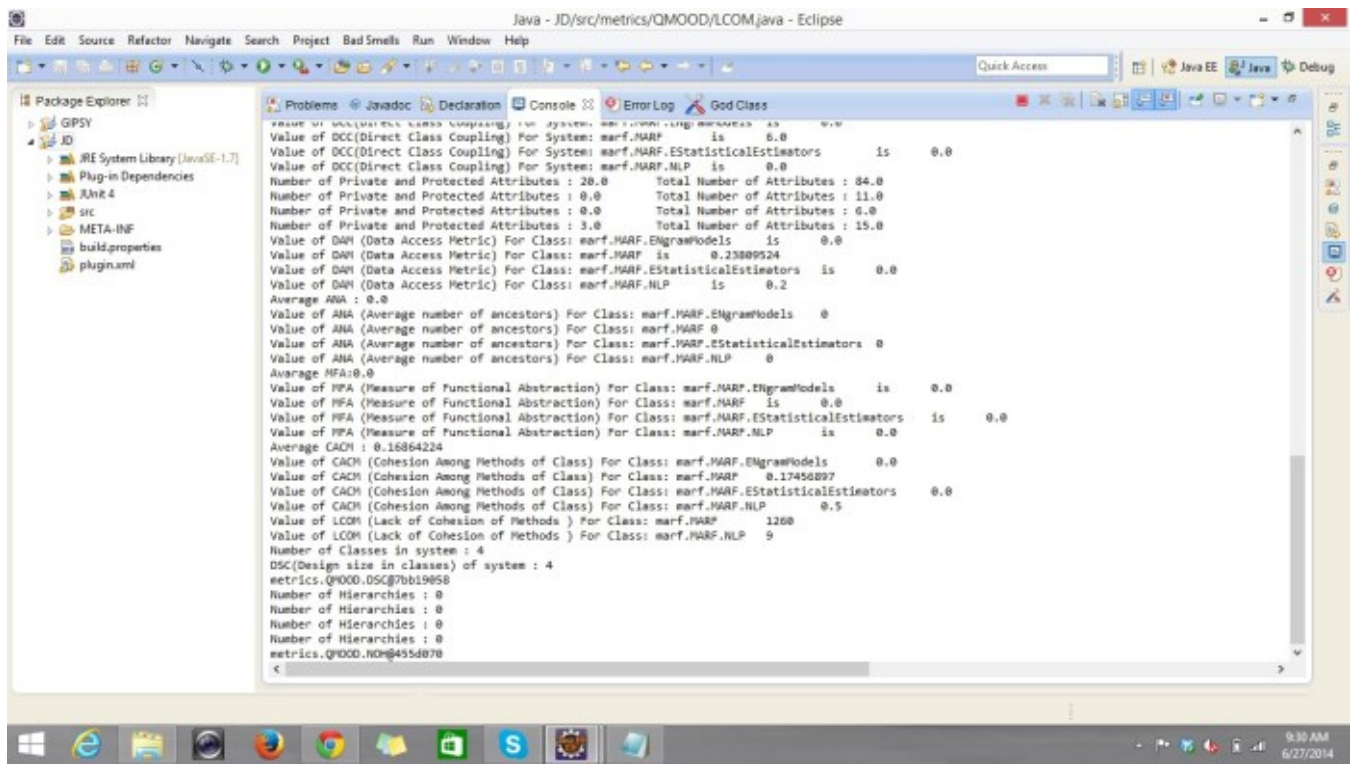

*Figure 53*

For MARF class2:
marf/src/marf/Classification/NeuralNetwork/NeuralNetwork.java

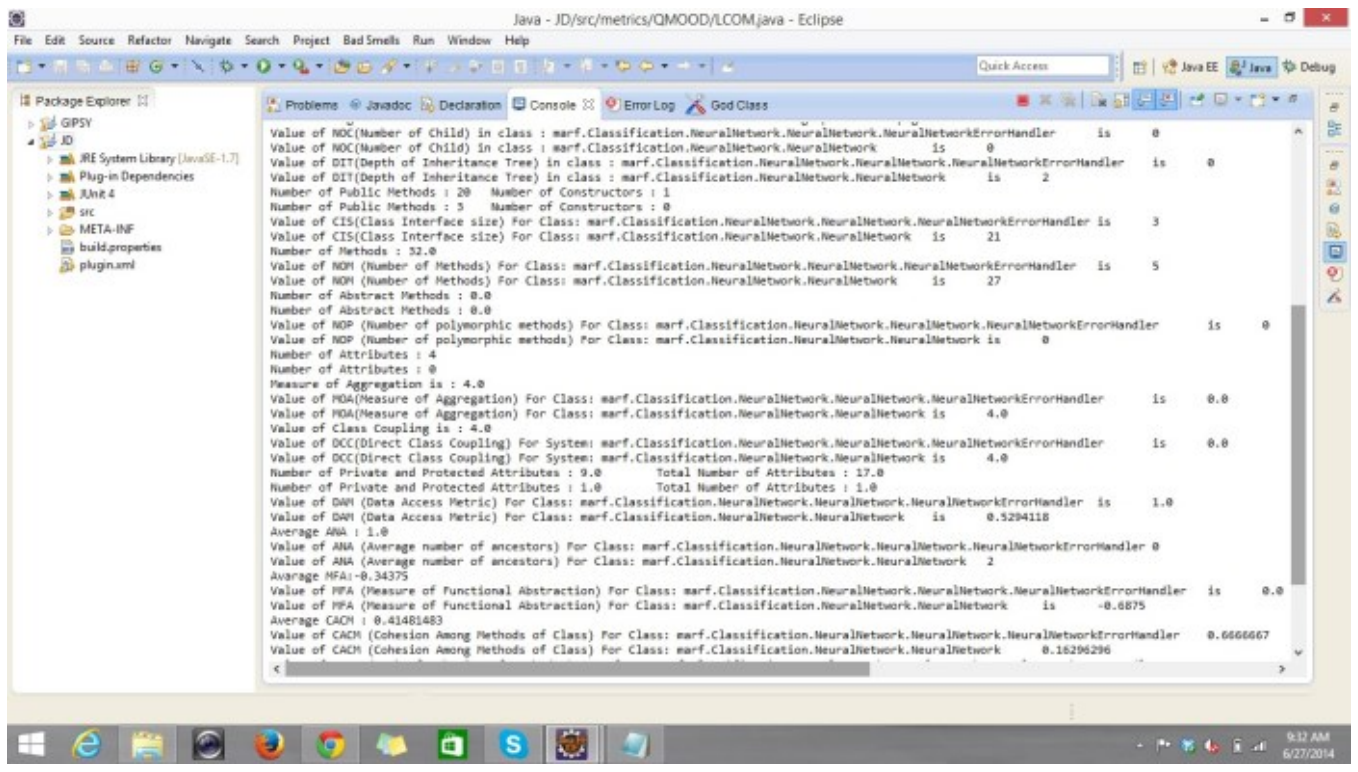

*Figure 54*

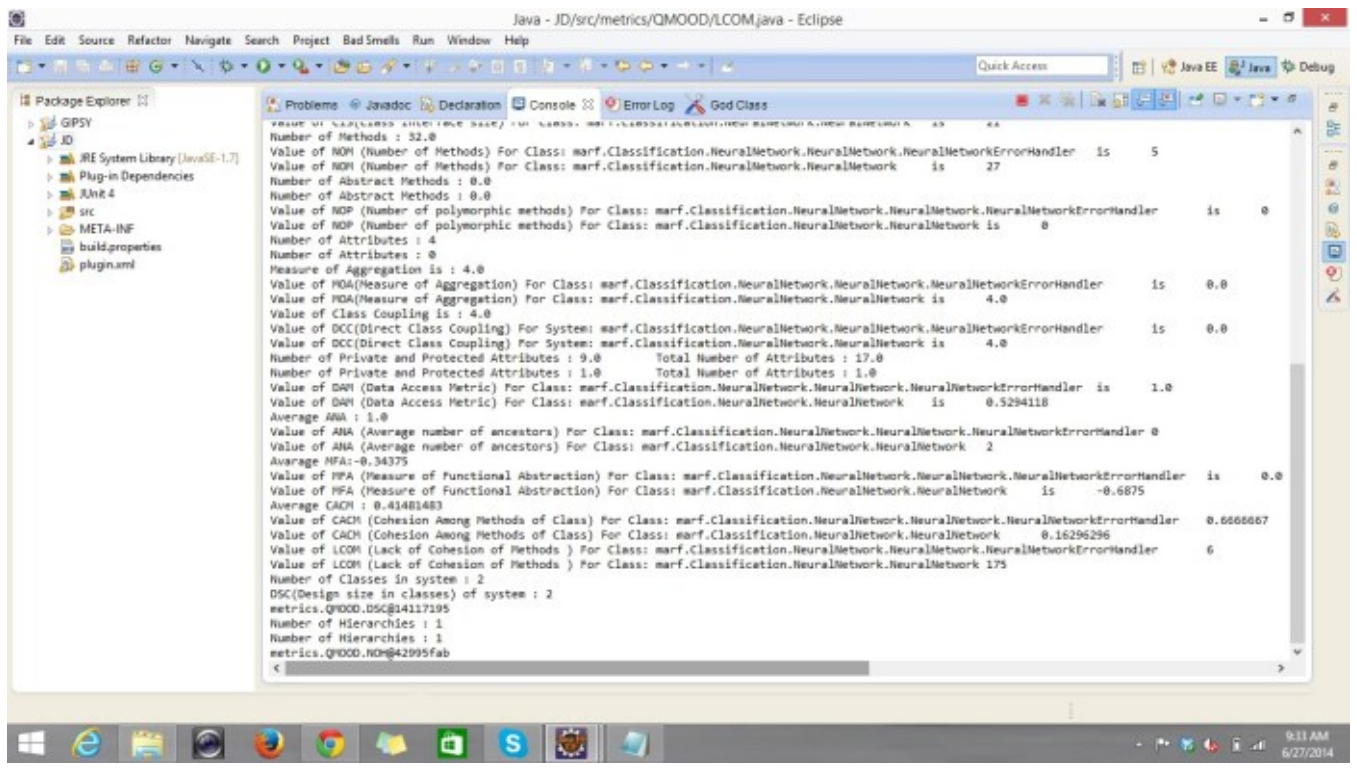

*Figure 55*

### c) Interpretation Problematic vs. less Problematic (In terms of packages)

From the metrics done for package the JDeodorant automatically highlight the problematic class and also the package of problematic class. The percentage of problematic classes in term of size(assuming line of code) is calculated by the tool and provides the approximate value. Here the percentage of the problematic class is always more than others classes which are in the respective package. Here have the reasons behind the problematic classes.

**Gipsy problematic class1:**
gipsy/src/gipsy/RIPE/editors/RunTimeGraphEditor/ui/GIPSYGMTOperator

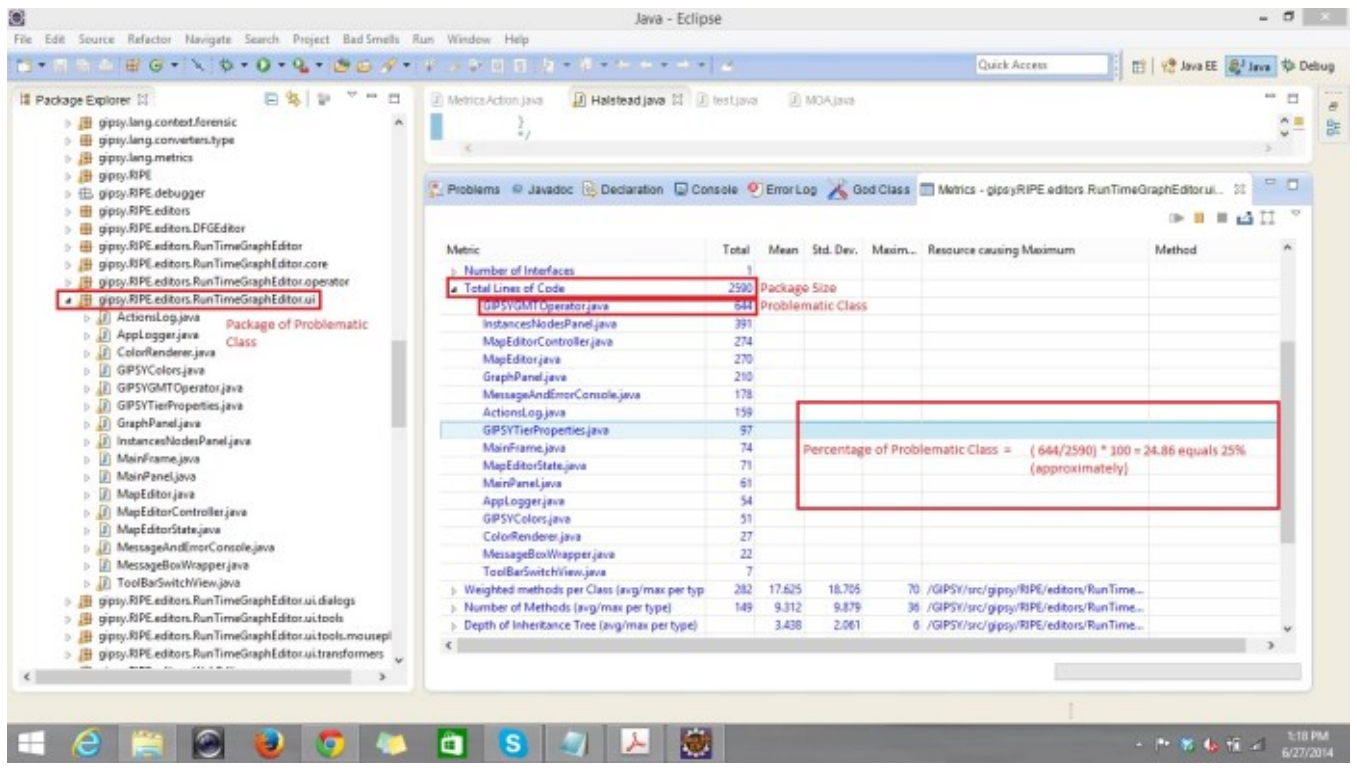

*Figure 56*

For gipsy problematic class the total lines of code is more than other classes in the selected package. And the percentage provided by tool was 25% approximately which means this class contains large number of code in it. The analysis for this problem is explained below

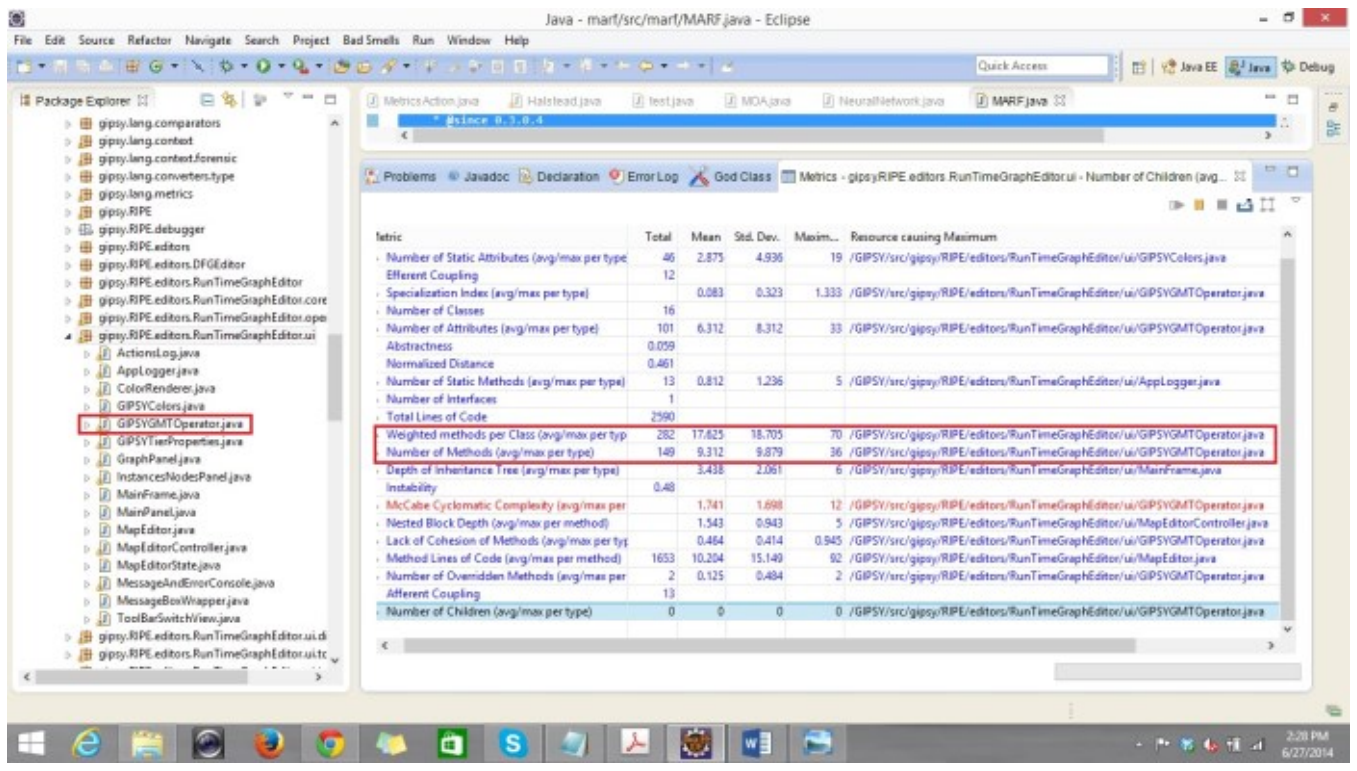

*Figure 57*

From the analysis the problem of large source code is due to large number of methods. For reference regarding the source code analysis is found in *APPENDIX*

**GIPSY Problematic class 2:**
gipsy/src/gipsy/GIPC/intensional/SIPL/ObjectiveLucid.ObjectiveGIPLParser

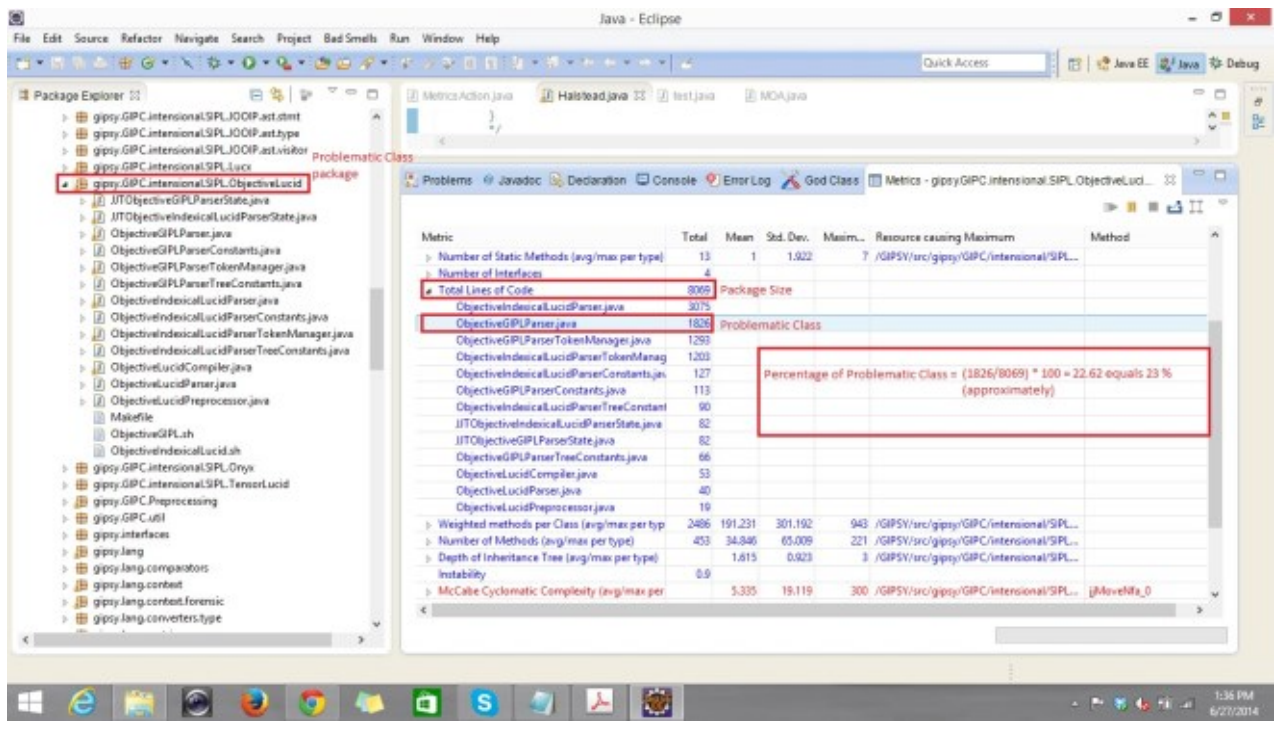

*Figure 58*

The size for problematic class is more compared to other classes. By comparing with the package size the percentage of problematic class size is 23% approximately provided by tool which is more than other classes in the package.

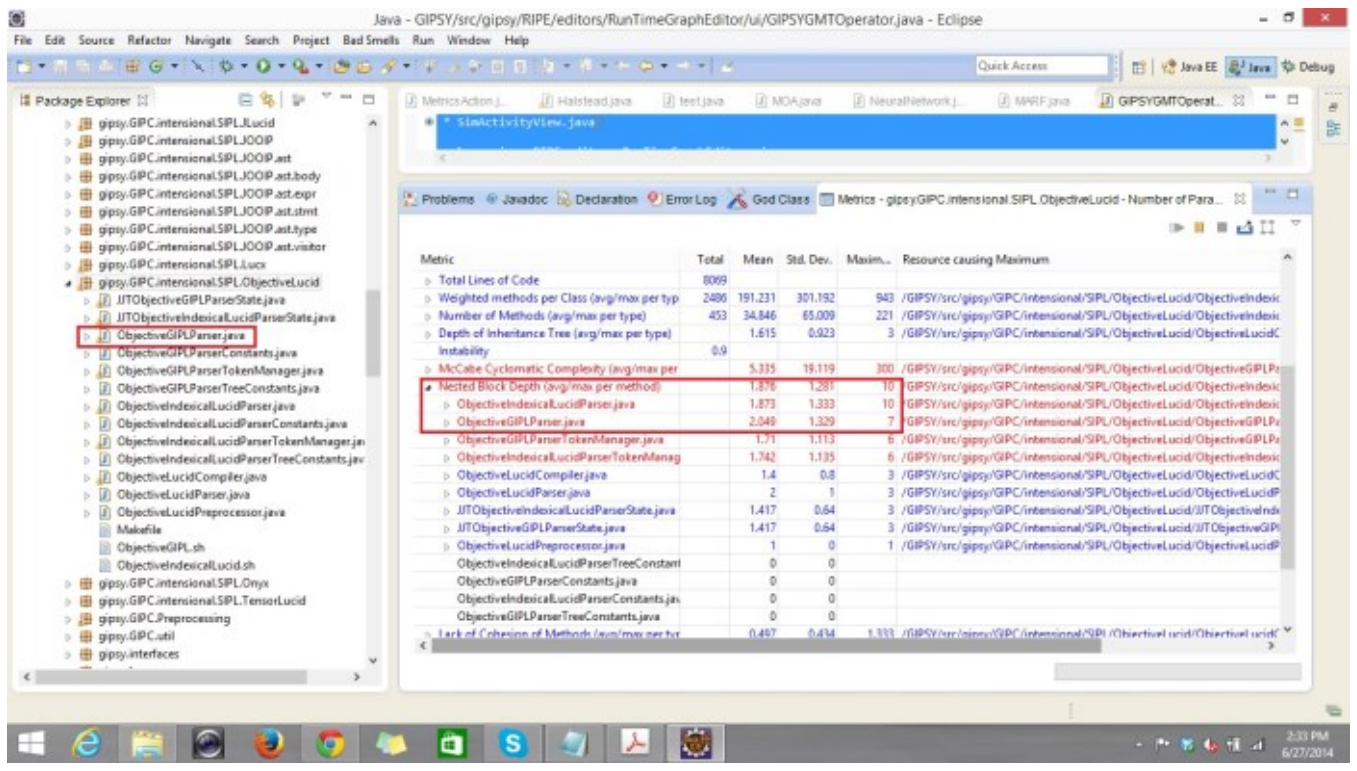

*Figure 59*

From the analysis the problem of large source code is due more number of decision predicates is 15 in that class. For reference regarding the source analysis is found in *APPENDIX*

**Problematic MARF Class1:**
marf/src/marf/MARF.java

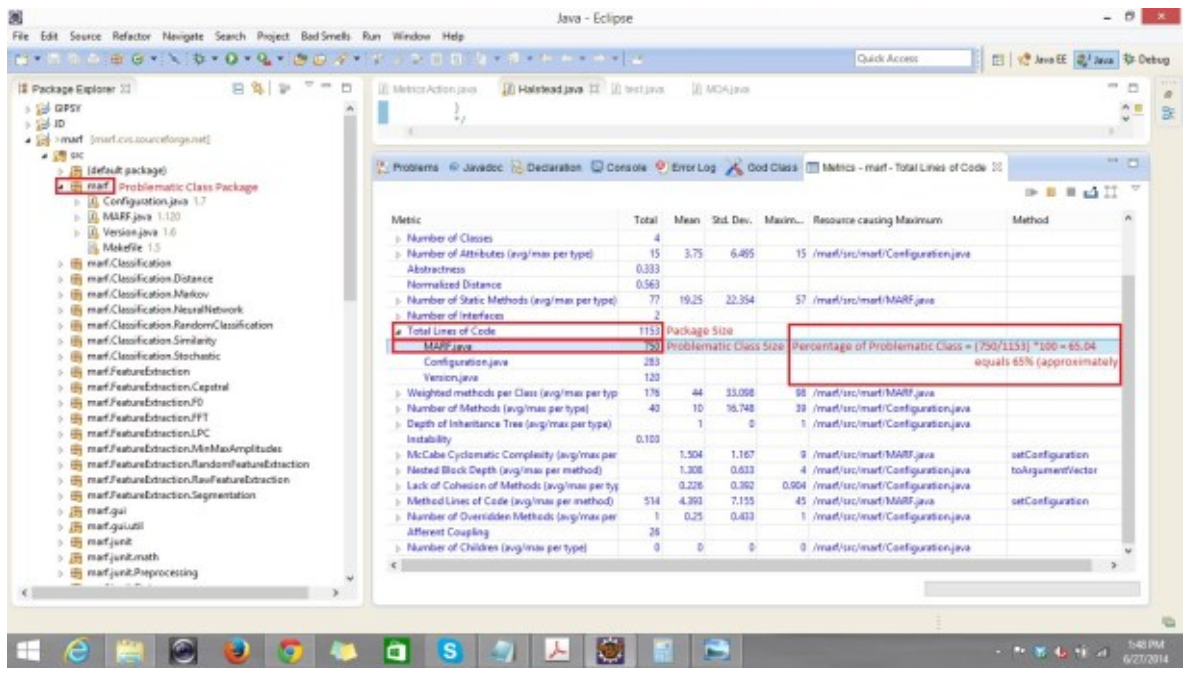

*Figure 60*

Here the lines of code for problematic class marf.MARF.java is more than other class in the respective package. The percentage of problematic class size to the package class size is 65% approximately taken from tool. This percentage for other classes in the same package will be less.

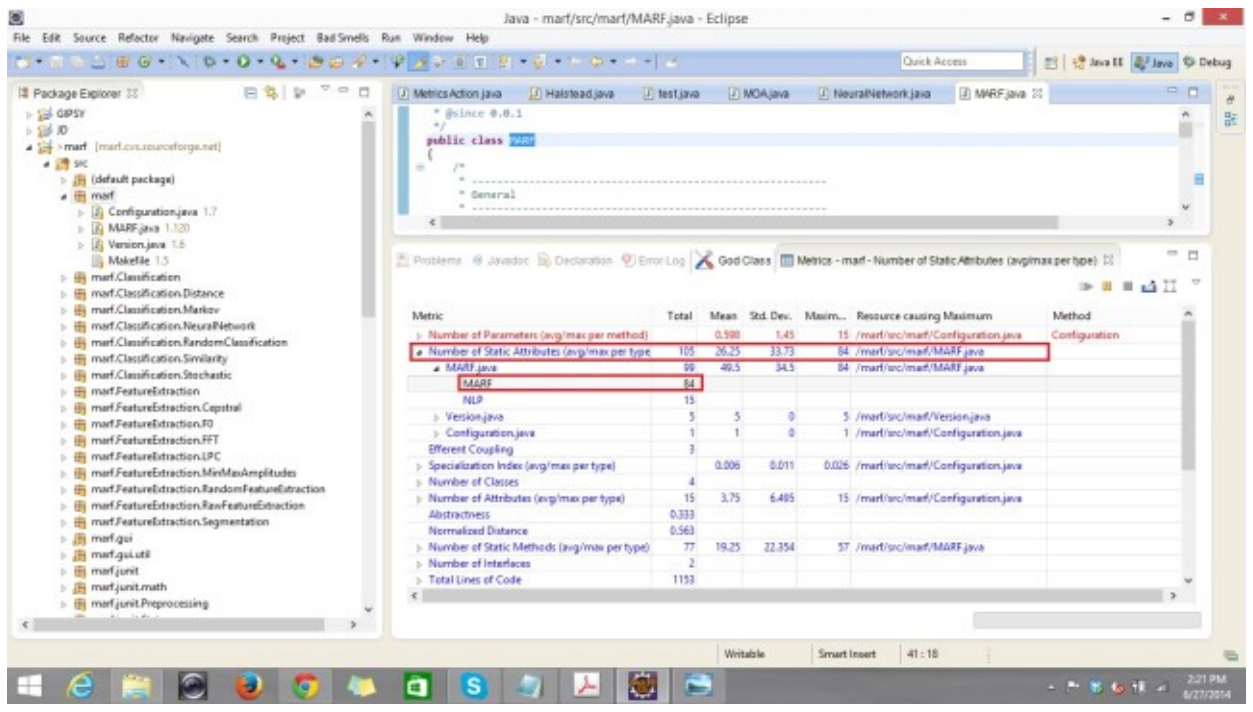

Here from the analysis the problem of large source code is due to large number of static attributes which is 84 . Reference for this analysis can be found in *APPENDIX*

MARF Problematic class2:
marf/src/marf/Classification/NeuralNetwork/NeuralNetwork.java

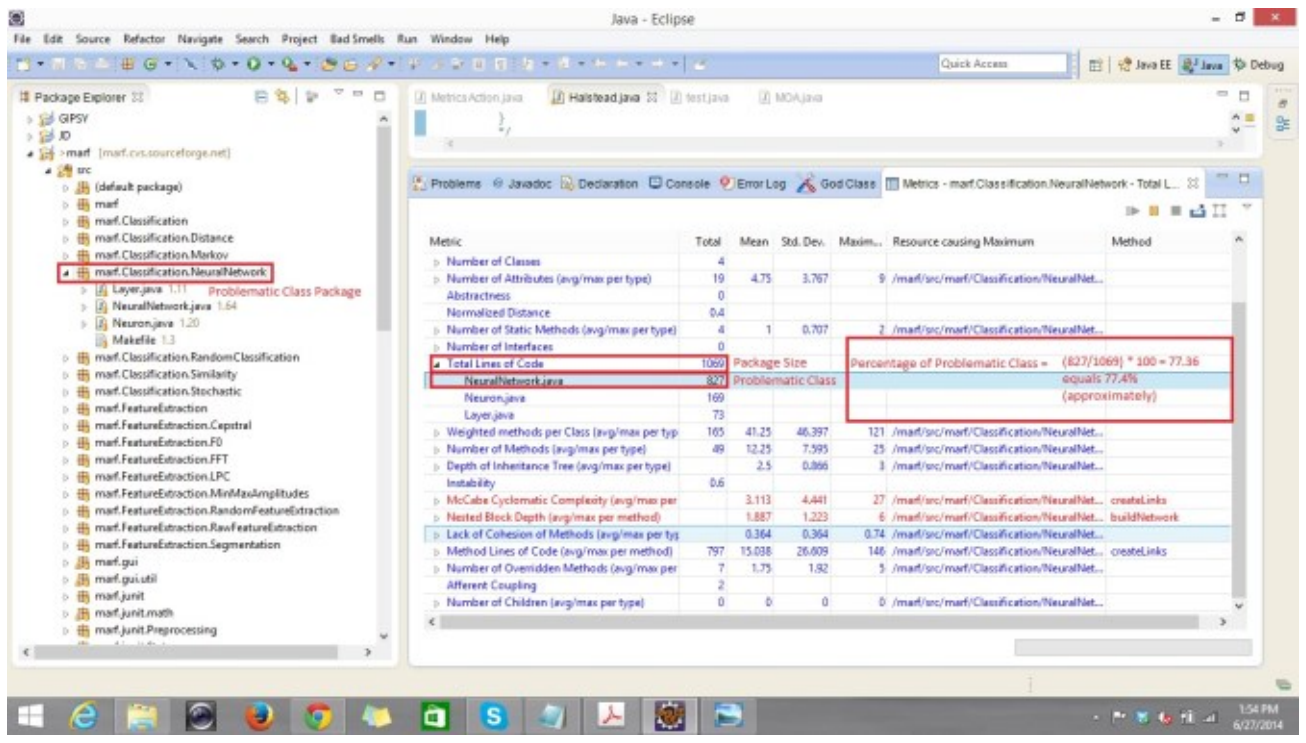

*Figure 61*

The size of this class is 827 and the percentage of problematic class size to the package is 77.4% approximately got from the tool.

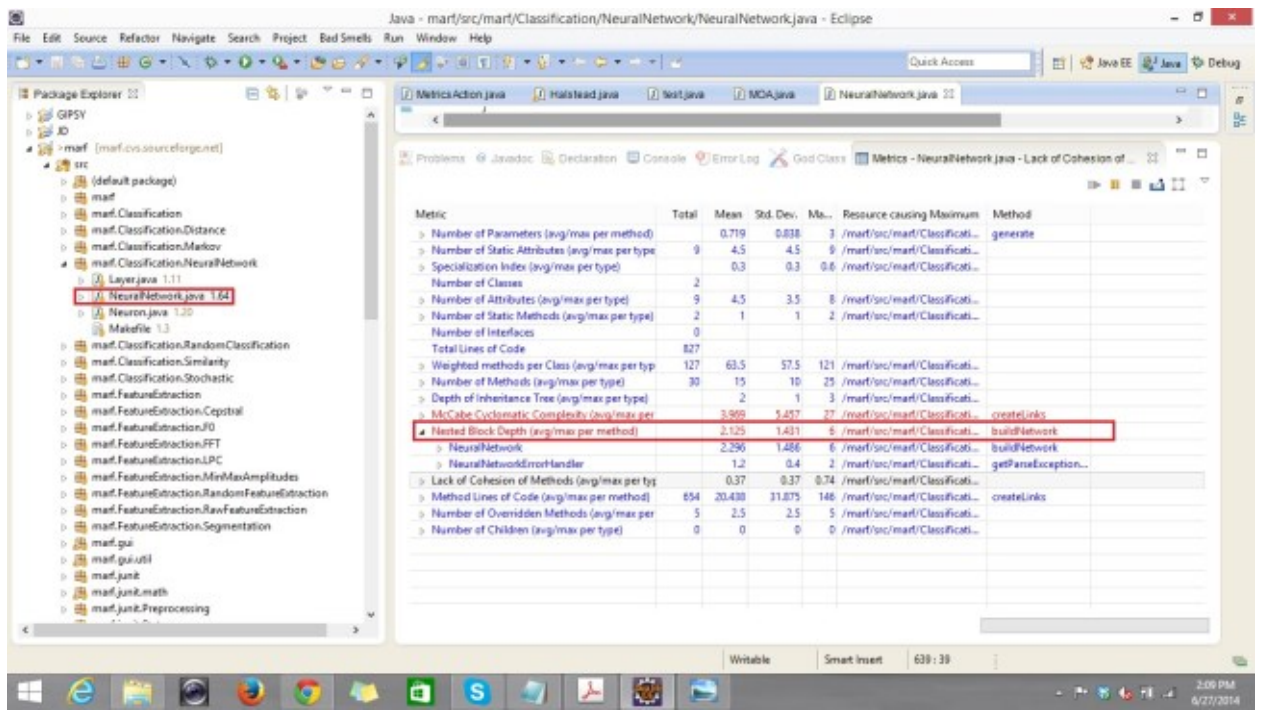

*Figure 62*

From analysis the problem of large source code is due to large number of decision predicate in the method which is 15.
Reference for this analysis can be found in *APPENDIX*

*d) Interpretation with Marfcat*

MARFCAT (scan all java files): MARFCAT is used to scan all the java files of MARF and GIPSY in order to check the vulnerability of classes. It is noticed that distance threshold value of all classes in GIPSY and MARF is 0.1, which makes warning to be reported false so there is no class in MARF and GIPSY that are highly vulnerable. Vulnerability of classes is high if the value of distance threshold is 0 and warring to be reported is true. One of the gipsy files is show in fig1 and MARF file in fig, Reference of all log files is in appendix.

```
                 File: ../test-cases/gipsy/bin/jini/start.sh
              Path ID: 2
               Config: -noprepre -raw -fft -cheb -flucid
      Processing time: 0d:0h:0m:0s:35ms:35ms
         Subject's ID: 61
   Subject identified: CVE-2007-1355
ResultSet: [suppressed; enable debug mode to show]
FileItem: strFileID: [2]
strPath: [../test-cases/gipsy/bin/jini/start.sh]
strFileType: [Bourne-Again shell script text executable]
bEmpty: [false]
oLocations: [[[]]]

       Second Best ID: 63
     Second Best Name: CVE-2006-7196
            Date/time: Wed Jun 25 22:07:14 EDT 2014
Outcome o (classifier-specific): 1885.2563697245303
             Distance threshold: 0.1
           Computed raw P = 1/o: 0.0
        Warning to be reported: false
         Computed normalized P: 0.0
```

Fig 63 :- marfcat parameter log file for all gipsy file

```
                 File: ../test-cases/marf/marf/build.xml
              Path ID: 2
               Config: -noprepre -raw -fft -cheb -flucid
      Processing time: 0d:0h:0m:0s:40ms:40ms
         Subject's ID: 49
   Subject identified: CVE-2008-1232
ResultSet: [suppressed; enable debug mode to show]
FileItem: strFileID: [2]
strPath: [../test-cases/marf/marf/build.xml]
strFileType: [XML 1.0 document text]
bEmpty: [false]
oLocations: [[[]]]

       Second Best ID: 39
     Second Best Name: CVE-2009-0783
            Date/time: Wed Jun 25 22:10:42 EDT 2014
Outcome o (classifier-specific): 847.6525089175706
             Distance threshold: 0.1
           Computed raw P = 1/o: 0.0
        Warning to be reported: false
         Computed normalized P: 0.0
```

Fig 64:- marfcat parameter log file for all gipsy file

*b) What probelemtic about probelmtic classes*

Going through code in the bad classes we find classes can be refactored based on the multiple attributes located in single class, we also verified this by JDeodrant Bad Smell Tool and God Class (violates, single responsibility principle and it control large number of object implementing different functionality the solution is to extract all the methods and fields which are related to specific functionality into separate class.)

Skimming through the lines of code problematic classes were found. To validate and make sure that the chosen classes were actually problematic JDedorant was used.

JDedorant Bad Smell Analysis

Path: marf/src/marf/MARF.java

voilates, single responsibility principle nd it control large number of object implemting different funclitonali the solution is to extract all the methods and fields which are realted to specific functionality itno separte class.

```
    /**
     * Gets outputs of a neural network run.
     * @return array of doubles read off the output layer's neurons
     */
    public double[] getOutputResults()
    {
        double[] adRet = new double[this.oOutputs.size()];

        for(int i = 0; i < this.oOutputs.size(); i++)
        {
            adRet[i] = this.oOutputs.get(i).dResult;
        }

        return adRet;
    }

    //---------- Methods for Outputting the NNet ----------------
```

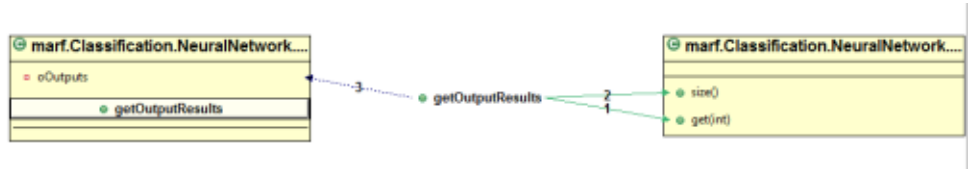

```
    /**
     * Sets inputs.
     * @param padInputs double array of input features
     * @throws ClassificationException if the input array's length isn't
     * equal to the size of the input layer
     */
    public final void setInputs(final double[] padInputs)
    throws ClassificationException
    {
        if(padInputs.length != this.oInputs.size())
        {
            throw new ClassificationException
            (
                "Input array size not consistent with input layer."
            );
        }

        for(int i = 0; i < padInputs.length; i++)
        {
            this.oInputs.get(i).dResult = padInputs[i];
        }
    }
```

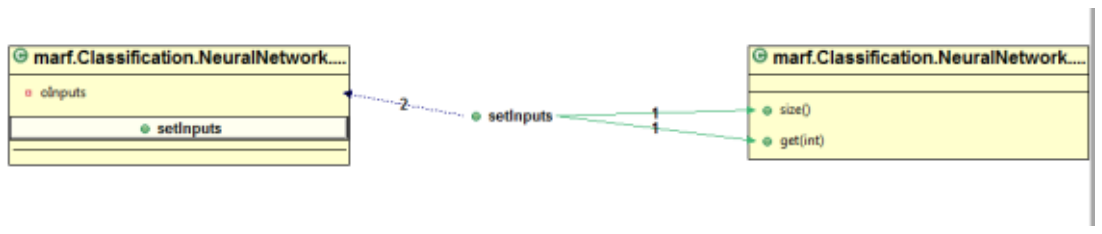

JDedorant GOD class

Path: marf/src/marf/Classification/NeuralNetwork/NeuralNetwork.java when a method references other class through methods and fields more often it reference its own class the solution is to refactor.

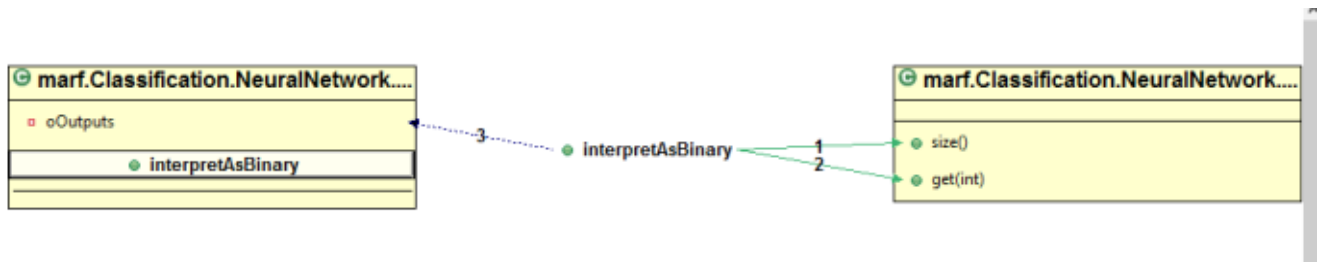

```java
/**
 * Interprets net's binary output as an ID for the final classification result.
 * @return ID, integer
 */
private final int interpretAsBinary()
{
    int iID = 0;

    for(int i = 0; i < this.oOutputs.size(); i++)
    {
        // Binary displacement happens to not have any
        // effect in the first iteration :-P
        iID *= 2;

        // Add 1 if the resulting weight is more than 0.5
        if(this.oOutputs.get(i).dResult > 0.5)
        {
            iID += 1;
        }

        Debug.debug(this.oOutputs.get(i).dResult + ",");
    }

    Debug.debug("Interpreted binary result (ID) = " + iID);

    return iID;
}
```

So is the case with gipsy problematic files, the class contains more than required static methods as detected by eclipse metric tool.

## IV. CONCLUSION

Based on project study we conclude that software quality can be measured for any system using different tools. We used tools like LOGISCOPE, McCabe, MARFCAT. According to the obtained results LOGISCOPE identifies problematic classes in any system. On the other hand MARFCAT identifies vulnerable and weak classes. We have carried out a systematic process to identify the weak classes. This was done using LOGISCOPE and the cylomatic Complexity was calculated using McCabe. To justify the selection of classes MARFCAT was used to identify the vulnerable classes. Dummy test cases were written to compare it with the metrics that were chosen earlier for the implementation. Therefore, it is clear that software quality can be measured for any system which dependent on the statistical values is obtained using different tools.

For Bonus implementation, there is extra implementation of QMood metrics, Marfcat .jar to eclipse plugin which are attached in source code for the project.

All the information about the source code and references are located in appendix section

## APPENDIX

TABLE I TERMINOLOGY

| | |
|---|---|
| MARF | Modular Audio Recognition Framework |
| NLP | Natural Language Processing |
| GIPSY | General Intensional Programming System |
| FFT | Fast Fourier transform |
| LPC | Linear predictive coding |
| CFE | Continuous Fraction Expansion |
| ASSL | Autonomic System Speciation Language |
| DMARF | Distributed Modular Audio Recognition Framework |
| ASIP | AS Interaction Protocol |
| AEIP | AE interaction protocol |
| AE | Autonomic Element |
| ADMARF | Autonomic Distributed Modular Audio Recognition Framework |
| GEE | General Education Engine |
| GIPC | General Intentional programming Language Compiler |
| RIPE | Runtime Programming Environment |
| IDS | Data Dependency Structure |
| IDP | Intentional Demand Propagator |
| IVW | Intentional value Warehouse |
| RIPE | Runtime Interactive Programming Environment |
| AC | Autonomic computing |
| AGIPSY | Autonomous GIPSY |
| GN | GIPSY Node |
| NM | Node Manager |
| GMT | Gipsy manager Tier |
| DST | Demand store Tier |
| DWT | Demand worker Tier |
| DMT | Demand manager Tier |
| TLOC | Total Lines Of Code Counts |
| OO | Object Oriented |
| AOP | Aspect Oriented Programming |
| AOSD | Aspect -Oriented Software Development |
| CF | Coupling Factor |
| LCOM | Lack of Cohesion of Methods |
| UP | Unified Process |
| SAIL | Simple Interpreted Language |
| WMC | Weighted methods per class |
| DIT | Depth of Inheritance Tree |
| NOC | Number of Children of a Class |
| CB0 | Coupling between Object classes |
| RFC | Response For a Class |
| QMOOD | Quality for Object Oriented Design |
| DSC | Design Class in Classes |
| NOH | Number of hierarchies |
| ANA | Average Number of ancestor |
| DAM | Data Access Metrics |
| DCC | Direct Class Coupling |
| CAM | Cohesion Among Methods of Class |
| MOA | Measure of Aggregation |
| MFA | Measure of Functional Abstraction |
| NOP | Class Interface Size |
| CIS | Class |
| NOM | Number of Methods |
| MFC | Microsoft Foundation Classes |
| OWL | Objects Window |
| HWMC | Hypothesis of Weighted Method per Class |
| HDIT | Hypothesis of Depth of Inheritance Tree of a class |
| HNOC | Hypothesis of Number Of Children of a Class |

TABLE II LIST OF CASE STUDIES & STUDENT NAME (MARF)

| S.No. | Name | Article |
|---|---|---|
| 1 | Aakash Parmar : | Study of best algorithm combinations for speech processing tasks in machine learning using median vs. mean clusters in MARF [2]. |
| 2 | Ajay Kumar Thakur (Team Leader) | Developing autonomic propertiesfor distributed pattern-recognition systems with ASSL [7] |
| 3 | Renuka Milkoori | A MARF approach to DEFT 2010 [6] |
| 4 | Biswajit Banik | Towards security hardening of scientific distributed demand-driven and pipelined computing systems [3] |
| 5 | Pankaj Kumar Pant | Choosing best algorithm combinations for speech processing tasks in machine learning using MARF [1] |
| 6 | Dhanashree Sankini | Writer identification using inexpensive signal processing techniques [4] |
| 7 | Dipesh Walia | Evolution of MARF, its NLP framework [5] |

TABLE III LIST OF CASE STUDIES & STUDENT NAME (GIPSSY)

| S.No. | Name | Article |
|---|---|---|
| 1 | Aakash Parmar : | A type system for hybrid intensional-imperative programming support in GIPSY. [11] |
| 2 | Ajay Kumar Thakur | An interactive graph-based automation assistant: A case study to manage the GIPSY's distributed multi-tier run-time system [14] |
| 3 | Renuka Milkoori | Advances in the design and implementation of a multi-tier architecture in the GIPSY environment. [13] |
| 4 | Biswajit Banik | The GIPSY architecture [8] |
| 5 | Pankaj Kumar Pant | Autonomic GIPSY [9] |
| 6 | Dhanashree Sankini | A type system for higher-order intensional logic support for variable bindings in hybrid intensional-imperative programs in GIPSY [12] |
| 7 | Dipesh Walia | Design and implementation of context calculus in the GIPSY environment [10] |

TABLE IV LIST OF CASE STUDIES & STUDENT NAME

| S.No. | Name | Article |
|---|---|---|
| 1 | Ajay Kumar Thakur (Team Leader) | Jagdish Bansiya and Carl G. Davis. A hierarchical model for object-oriented design quality assess- ment. IEEE Transactions on Software Engineering, 28(1):4–17, January 2002. |
| 2 | Renuka Milkoori | Rachel Harrison, Steve J. Counsell, and Reuben V. Nithi. An evaluation of the MOOD set of objectoriented software metrics. IEEE Transactions on Software Engineering, 24(6):491{496, June 1998.doi: 10.1109/32.689404. |
| 3 | Biswajit Banik | Victor R. Basili, Lionel C. Briand, and Walc elio L. Melo. A validation of object-oriented design metrics as quality indicators.IEEE Transactions on Software Engineering , 22(10):751{761, 1996. |
| 4 | Pankaj Kumar Pant | Measurement of Cohesion and Coupling in OO Analysis Model Based on Crosscutting Concerns |
| 5 | Dhanashree Sankini | A Unified Framework for Coupling Measurement in Object-Oriented Systems |
| 6 | Dipesh Walia | Towards a Simplified Implementation of Object-Oriented Design Metrics |

## SOURCE CODE REFERNCES

We have two folders in project folder

Analysis of MARCAT
JD Files (Implementation of Metrics)
Analysis of GIPSY MARF PROBELMETIC FILE
MARFCAT as Plugin

Under Analysis of MARFCAT we have log files

Analysis of MARFCAT 2 items

- marfcat--super-fast-test-quick-gipsy-cve.log
- marfcat--super-fast-test-quick-marf-cve.log

Under JD section we have all the required files MARFCAT

JD 8 items

- .settings
- bin
- META-INF
- src
- .classpath
- .project
- build.properties
- plugin.xml

MARFCAT as Plugin

Analysis of GIPSY MARF PROBELMETIC FILE

## METRICS RESULT SNAPSHOTS

### Number of languages (sloccount)

```
Totals grouped by language (dominant language first):
java:          98478 (97.53%)
ansic:          1631 (1.62%)
sh:              452 (0.45%)
xml:             368 (0.36%)
haskell:          44 (0.04%)
```

*Figure 34 Result Obtain from GIPSY Sloccount*

```
Totals grouped by language (dominant language first):
java:          24546 (88.45%)
sh:             1501 (5.41%)
perl:           1295 (4.67%)
xml:             409 (1.47%)
```

*Figure 35 Result Obtain from MARD Sloccount*

### Lines of text and classes (Metrics 1.3.6)

| Metric | Total | Mean | Std. Dev. | Maxim... | Resource causing Maximum | Method |
|---|---|---|---|---|---|---|
| Number of Overridden Methods (avg/max per | 439 | 0.757 | 1.761 | 21 | /GIPSY/src/gipsy/GIPC/intensional/SIPL/ForensicLuci... | |
| Number of Attributes (avg/max per type) | 1851 | 3.191 | 5.104 | 33 | /GIPSY/src/gipsy/RIPE/editors/RunTimeGraphEditor/... | |
| Number of Children (avg/max per type) | 236 | 0.407 | 1.812 | 21 | /GIPSY/src/gipsy/GIPC/intensional/SIPL/JOOIP/ast/ex... | |
| Number of Classes (avg/max per packageFragr | 580 | 5.472 | 5.572 | 36 | /GIPSY/src/gipsy/GIPC/intensional/SIPL/JOOIP/ast/ex... | |
| Method Lines of Code (avg/max per method) | 75262 | 12.44 | 40.732 | 933 | /GIPSY/src/gipsy/GIPC/SemanticAnalyzer.java | check |
| Number of Methods (avg/max per type) | 5746 | 9.907 | 33.947 | 555 | /GIPSY/src/gipsy/GIPC/intensional/SIPL/JOOIP/JavaP... | |
| Nested Block Depth (avg/max per method) | | 1.71 | 1.313 | 19 | /GIPSY/src/gipsy/GIPC/intensional/SIPL/Lucx/LucxPa... | jj_3_51 |
| Depth of Inheritance Tree (avg/max per type) | | 2.467 | 1.683 | 8 | /GIPSY/src/gipsy/GEE/multitier/GMT/demands/DWT... | |
| Number of Packages | 106 | | | | | |
| Afferent Coupling (avg/max per packageFragn | | 10.972 | 19.973 | 90 | /GIPSY/src/gipsy/GEE/IDP/demands | |
| Number of Interfaces (avg/max per packageFra | 75 | 0.708 | 1.244 | 6 | /GIPSY/src/gipsy/lang/context | |
| McCabe Cyclomatic Complexity (avg/max per | | 4.055 | 13.426 | 300 | /GIPSY/src/gipsy/GIPC/intensional/GIPL/GIPLParserT... | jjMoveNfa_0 |
| Total Lines of Code | 104073 | | | | | |
| Instability (avg/max per packageFragment) | | 0.597 | 0.361 | 1 | /GIPSY/src/gipsy/GEE/IDP/DemandGenerator/rmi | |
| Number of Parameters (avg/max per method) | | 0.794 | 1.099 | 11 | /GIPSY/src/gipsy/GIPC/intensional/SIPL/JOOIP/ast/b... | MethodDeclaration |
| Lack of Cohesion of Methods (avg/max per typ | | 0.243 | 0.361 | 1.625 | /GIPSY/src/gipsy/GIPC/util/Token.java | |
| Efferent Coupling (avg/max per packageFragm | | 4.377 | 4.721 | 34 | /GIPSY/src/gipsy/GIPC/intensional/SIPL/JOOIP/ast/ex... | |
| Number of Static Methods (avg/max per type) | 302 | 0.521 | 1.561 | 14 | /GIPSY/src/gipsy/GIPC/intensional/SIPL/JOOIP/ast/b... | |
| Normalized Distance (avg/max per packageFra | | � | � | 1 | /GIPSY/src/gipsy/RIPE/RuntimeSystem | |
| Abstractness (avg/max per packageFragment) | | � | � | 0.75 | /GIPSY/src/gipsy/GEE/IDP/DemandDispatcher | |
| Specialization Index (avg/max per type) | | 0.542 | 0.953 | 5.04 | /GIPSY/src/gipsy/tests/GEE/IDP/demands/DemandTe... | |
| Weighted methods per Class (avg/max per typ | 24533 | 42.298 | 185.189 | 2856 | /GIPSY/src/gipsy/GIPC/intensional/SIPL/JOOIP/JavaP... | |
| Number of Static Attributes (avg/max per type | 836 | 1.441 | 4.938 | 76 | /GIPSY/src/gipsy/GIPC/intensional/SIPL/JOOIP/JavaP... | |

*Figure 36 Result obtained using eclipse with respect to GIPSY (metrics as a plug in)sss*

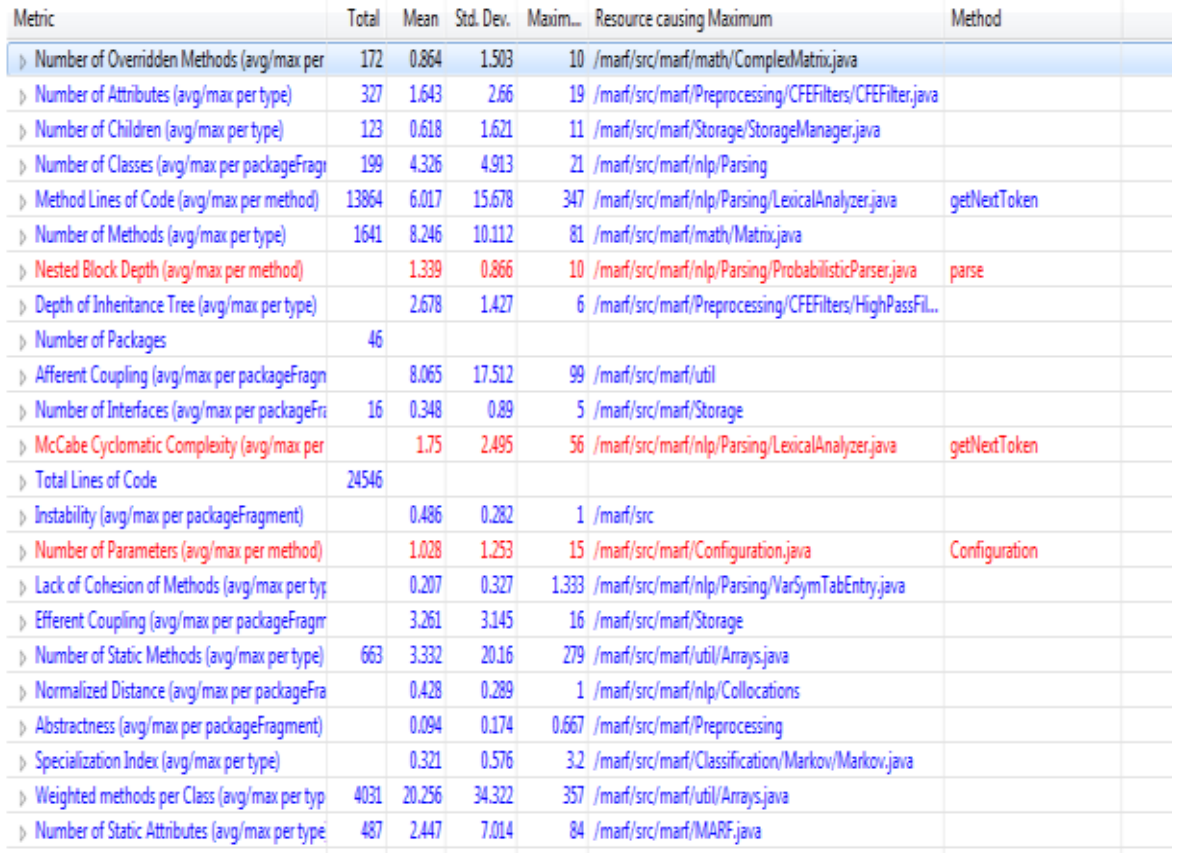

| Metric | Total | Mean | Std. Dev. | Maxim... | Resource causing Maximum | Method |
|---|---|---|---|---|---|---|
| Number of Overridden Methods (avg/max per | 172 | 0.864 | 1.503 | 10 | /marf/src/marf/math/ComplexMatrix.java | |
| Number of Attributes (avg/max per type) | 327 | 1.643 | 2.66 | 19 | /marf/src/marf/Preprocessing/CFEFilters/CFEFilter.java | |
| Number of Children (avg/max per type) | 123 | 0.618 | 1.621 | 11 | /marf/src/marf/Storage/StorageManager.java | |
| Number of Classes (avg/max per packageFragı | 199 | 4.326 | 4.913 | 21 | /marf/src/marf/nlp/Parsing | |
| Method Lines of Code (avg/max per method) | 13864 | 6.017 | 15.678 | 347 | /marf/src/marf/nlp/Parsing/LexicalAnalyzer.java | getNextToken |
| Number of Methods (avg/max per type) | 1641 | 8.246 | 10.112 | 81 | /marf/src/marf/math/Matrix.java | |
| Nested Block Depth (avg/max per method) | | 1.339 | 0.866 | 10 | /marf/src/marf/nlp/Parsing/ProbabilisticParser.java | parse |
| Depth of Inheritance Tree (avg/max per type) | | 2.678 | 1.427 | 6 | /marf/src/marf/Preprocessing/CFEFilters/HighPassFil... | |
| Number of Packages | 46 | | | | | |
| Afferent Coupling (avg/max per packageFragn | | 8.065 | 17.512 | 99 | /marf/src/marf/util | |
| Number of Interfaces (avg/max per packageFra | 16 | 0.348 | 0.89 | 5 | /marf/src/marf/Storage | |
| McCabe Cyclomatic Complexity (avg/max per | | 1.75 | 2.495 | 56 | /marf/src/marf/nlp/Parsing/LexicalAnalyzer.java | getNextToken |
| Total Lines of Code | 24546 | | | | | |
| Instability (avg/max per packageFragment) | | 0.486 | 0.282 | 1 | /marf/src | |
| Number of Parameters (avg/max per method) | | 1.028 | 1.253 | 15 | /marf/src/marf/Configuration.java | Configuration |
| Lack of Cohesion of Methods (avg/max per typ | | 0.207 | 0.327 | 1.333 | /marf/src/marf/nlp/Parsing/VarSymTabEntry.java | |
| Efferent Coupling (avg/max per packageFragm | | 3.261 | 3.145 | 16 | /marf/src/marf/Storage | |
| Number of Static Methods (avg/max per type) | 663 | 3.332 | 20.16 | 279 | /marf/src/marf/util/Arrays.java | |
| Normalized Distance (avg/max per packageFra | | 0.428 | 0.289 | 1 | /marf/src/marf/nlp/Collocations | |
| Abstractness (avg/max per packageFragment) | | 0.094 | 0.174 | 0.667 | /marf/src/marf/Preprocessing | |
| Specialization Index (avg/max per type) | | 0.321 | 0.576 | 3.2 | /marf/src/marf/Classification/Markov/Markov.java | |
| Weighted methods per Class (avg/max per typ | 4031 | 20.256 | 34.322 | 357 | /marf/src/marf/util/Arrays.java | |
| Number of Static Attributes (avg/max per type | 487 | 2.447 | 7.014 | 84 | /marf/src/marf/MARF.java | |

*Figure 37 Result obtained from eclipse with respect to MARF system*